\documentclass[12pt, a4paper]{article}

\usepackage{rotating}
\usepackage{amsmath}
\usepackage{mathtools}
\usepackage{amsthm}
\usepackage{amsfonts}
\usepackage[dvips]{epsfig}
\usepackage{graphicx}
\usepackage{psfrag}

\usepackage{amssymb}
\usepackage{cancel}
\usepackage{axodraw4j}
\usepackage{pstricks} 
\usepackage{lscape}
\usepackage{cancel}
\usepackage{slashed}
\usepackage{pstricks} 
\usepackage{lscape}
\usepackage{color}
\usepackage{cite}
\usepackage{url}
\usepackage{caption}
\usepackage{bbm}
\usepackage{bm}
\usepackage{subfig}
\usepackage{bbold}
\usepackage{caption}

\newcommand{\beq}{\begin{equation}}
\newcommand{\eeq}{\end{equation}}
\newcommand{\ga}{\lower.7ex\hbox{$\;\stackrel{\textstyle>}{\sim}\;$}}
\newcommand{\la}{\lower.7ex\hbox{$\;\stackrel{\textstyle<}{\sim}\;$}}

\newcommand{\kahler}{K\"ahler }

\makeatletter
\newcommand{\Cen}[2]{%
  \ifmeasuring@
    #2%
  \else
    \makebox[\ifcase\expandafter #1\maxcolumn@widths\fi]{$\displaystyle#2$}%
  \fi
}
\makeatother

\begin{document}

\begin{flushright}
{\tt KCL-PH-TH/2017-17}, {\tt CERN-PH-TH/2017-071}  \\
{\tt UT-17-16, ACT-03-17, MI-TH-1752} \\
{\tt UMN-TH-3623/17, FTPI-MINN-17/08} \\
\end{flushright}

\vspace{0.5cm}
\begin{center}
{\bf {\large Starobinsky-like Inflation, Supercosmology  \\ 
\vspace {0.05in}
and Neutrino Masses in No-Scale Flipped SU(5)}}
\end{center}

\vspace{0.05in}

\begin{center}{
{\bf John~Ellis}$^{a}$,
{\bf Marcos~A.~G.~Garcia}$^{b}$,
{\bf Natsumi Nagata}$^{c}$, \\
{\bf Dimitri~V.~Nanopoulos}$^{d}$ and
{\bf Keith~A.~Olive}$^{e}$
}
\end{center}

\begin{center}
{\em $^a$Theoretical Particle Physics and Cosmology Group, Department of
  Physics, King's~College~London, London WC2R 2LS, United Kingdom;\\
Theoretical Physics Department, CERN, CH-1211 Geneva 23,
  Switzerland}\\[0.2cm]
  {\em $^b$Physics \& Astronomy Department, Rice University, Houston, TX
 77005, USA}\\[0.2cm] 
  {\em $^c$Department of Physics, University of Tokyo, Bunkyo-ku, Tokyo
 113--0033, Japan}\\[0.2cm] 
{\em $^d$George P. and Cynthia W. Mitchell Institute for Fundamental
 Physics and Astronomy, Texas A\&M University, College Station, TX
 77843, USA;\\ 
 Astroparticle Physics Group, Houston Advanced Research Center (HARC),
 \\ Mitchell Campus, Woodlands, TX 77381, USA;\\ 
Academy of Athens, Division of Natural Sciences,
Athens 10679, Greece}\\[0.2cm]
{\em $^e$William I. Fine Theoretical Physics Institute, School of
 Physics and Astronomy, University of Minnesota, Minneapolis, MN 55455,
 USA}

\end{center}

\bigskip

\centerline{\bf ABSTRACT}

\noindent  
{\small We embed a flipped ${\rm SU}(5) \times {\rm U}(1)$ GUT model in
a no-scale supergravity framework, and discuss its predictions for
cosmic microwave background observables, which are similar to those of
the Starobinsky model of inflation. Measurements of the tilt in the
spectrum of scalar perturbations in the cosmic microwave background,
$n_s$, constrain significantly the model parameters. We also discuss the
model's predictions for neutrino masses, and pay particular attention to the
behaviours of scalar fields during and after inflation, reheating and
the GUT phase transition. We argue in favor of strong reheating in order
to avoid excessive entropy production which could dilute the generated
baryon asymmetry.}

\vspace{0.2in}

\begin{flushleft}
April 2017
\end{flushleft}
\medskip
\noindent

\newpage

\section{Introduction}

One of the biggest issues in particle physics is how to construct a
testable theory of unification that goes beyond the Standard Model and
also incorporates our phenomenological knowledge of neutrino masses and
mixing. In parallel, one of the key issues in cosmology is how to
construct a model of cosmological inflation that accommodates the
current experimental constraints and relates it to particle physics in a
testable way, {\it e.g.}, by making specific predictions for reheating
after inflation.

Flipped ${\rm SU}(5) \times {\rm U}(1)$ \cite{Barr,DKN,flipped2,AEHN} offers a
promising framework for supersymmetric grand unification that offers
resolutions of several important phenomenological issues in particle
physics. For example, in addition to accommodating small neutrino masses
\cite{flipped2,eln,Ellis:1992nq,Ellis:1993ks}, it provides a minimal
mechanism for splitting the masses of the triplet and doublet components
of the fiveplets of GUT Higgs fields \cite{flipped2}. Moreover, flipped ${\rm SU}(5) \times {\rm U}(1)$ can be
extracted from string theory \cite{AEHN,cehnt}. 

In parallel, a very attractive
framework for constructing models of cosmological inflation
\cite{ENO6,ENO7,ENO8,EGNO4,EGNO5,EGNO6,EGNOP} is provided by no-scale
supergravity \cite{no-scale,LN}, which offers a positive semi-definite
potential that accommodates naturally an asymptotically-flat direction
that makes predictions similar to the Starobinsky model
\cite{Staro,MC,Staro2} that is highly consistent with the available
cosmological data \cite{planck15}. The next frontier in the
phenomenology of Starobinsky-like models is to construct a model of
post-inflationary reheating \cite{EGNOP}, which is testable in principle
by a precise measurement of the tilt in the scalar perturbation
spectrum, $n_s$. We addressed this issue recently in the framework of an
SO(10) model of grand unification \cite{EGNNO}: here we revisit
Starobinsky-like inflation in the framework of the supersymmetric flipped
${\rm SU}(5) \times {\rm U}(1)$ GUT.

Working within such a specific framework enables---indeed, requires---us
to address a wide range of related issues. For example, in connection
with particle physics, one must check consistency with the available
information on neutrino masses and mixing \cite{flipped2,AEHN,abel,art,eln,Ellis:1992nq,Ellis:1993ks,lv,rt},
proton stability \cite{eln1,lt,Li:2010dp,Li:2009fq}
and provide for the cold dark matter of the Universe \cite{ehkno,dt,dm}. 
Also, in connection with
cosmology, one must check the evolution in the early Universe of the
various scalar fields that necessarily appear in any more complete model
of inflation \cite{fi}. Finally, one should aim at a successful scenario for
generating the cosmological baryon asymmetry \cite{cehno2,Ellis:1988qe}.

In this paper we address these issues in the supersymmetric flipped ${\rm SU}(5) \times {\rm U}(1)$
GUT framework. We consider various possible identifications of the
inflaton, and analyze the circumstances under which they can reproduce
successful Starobinsky-like predictions for the tensor-to-scalar
perturbation ratio, $r$, as well as the scalar tilt, $n_s$. We also
study numerically the cosmological evolutions of the various GUT-singlet
and massive sneutrino fields in the theory. Concerning reheating, we
find a non-trivial link between the reheating temperature, and hence
$n_s$, and the values of the neutrino masses. In the contexts of two
specific scenarios for the masses and couplings of the singlet fields in
the model, we show that experimental measurements of $n_s$ constrain
significantly the model parameters. We also consider the GUT phase
transition in the supersymmetric flipped ${\rm SU}(5) \times {\rm U}(1)$ model,
following~\cite{Ellis:1992nq,Campbell:1987eb}.  We argue that 
so long as the inflationary reheat temperature is sufficiently high (higher
than the strong coupling scale associated with SU(5)), excessive
entropy release can be avoided, and it is relatively easy to obtain
an adequate cosmological baryon asymmetry.

The layout of this paper is as follows. In Section~\ref{sec:model} we
introduce the flipped ${\rm SU}(5) \times {\rm U}(1)$ GUT model that we study,
Section~\ref{sec:inflation} discusses cosmological inflation in this
model, considering various possible inflaton assignments and the
corresponding constraints on the model parameters that yield
Starobinsky-like inflation. This Section also contains our numerical
analysis of the behaviours of the various scalar fields during and after
inflation. Neutrino masses and the decays of scalar fields are discussed
in Section~\ref{sec:reheating}, and the GUT transition and scenarios for baryogenesis are discussed in
Section~\ref{sec:phasetr}. Finally, Section~\ref{sec:summary} summarizes
and discusses our results.

\section{The No-Scale Flipped ${\rm SU}(5) \times {\rm U}(1)$ GUT model}
\label{sec:model}

The field content of the flipped ${\rm SU}(5) \times {\rm U}(1)$ GUT we
consider~\cite{Barr,DKN,flipped2,AEHN} consists of three generations of Standard Model (SM)
matter fields, each with the addition of a right-handed neutrino, arranged in
a $\mathbf{10}$, $\bar{\mathbf{5}}$, and $\mathbf{1}$ of SU(5) with the
right-handed electrons and neutrinos, as well as the up- and down-type
right-handed quarks, ``flipped'' with respect to a
standard SU(5) assignment. The ${\rm SU}(5) \times {\rm U}(1)$ GUT group is
subsequently broken to the SM group via
$\mathbf{10}+\overline{\mathbf{10}}$ representations of SU(5), and
subsequently to the ${\rm SU}(3) \times {\rm U}(1)$ symmetry via electroweak doublets
in $\mathbf{5}+\bar{\mathbf{5}}$ representations. Our notations for the
fields and their gauge representations are as follows:
\begin{alignat}{3}
&F_i && = ({\bf 10},1)_i  && \ni \; \left\{d^c,Q,\nu^c\right\}_i~,\nonumber\\
&\bar{f}_i &&=(\bar{\bf 5},-3)_i && \ni \; \{u^c,L\}_i~,\nonumber\\
&\ell^c_i &&=({\bf 1},5)_i && \ni \; \{e^c\}_i~,\nonumber\\
&H &&=({\bf 10},1)~,\nonumber\\
&\bar{H} &&=(\overline{\bf 10},-1)~,\nonumber\\
&h &&=({\bf 5},-2)~,\nonumber\\
&\bar{h} &&=(\bar{\bf 5},2) \, ,
\label{eq:charges}
\end{alignat}
where the subscripts $i = 1, 2, 3$ are generation indices that we
suppress for clarity when they are unnecessary. 
Following the notation of~\cite{flipped2}, the states in $H$ will be
labeled by the same symbols as in the $F_i$, but with an additional subscript:
$d_H^c, Q_H,\ldots$, and states in $\bar{H}$ are similarly labelled
including bars. States in $h$ are denoted by $(D, D, D, h^- , h^0)$ and
in $\bar{h}$ by $(\bar{D}, \bar{D}, \bar{D}, h^+ , \bar{h}^0)^T$. With
these charge assignments, the hypercharge $Y$ is given by a linear
combination of the SU(5) generator $T_{24} \equiv {\rm
diag}(2,2,2,-3,-3)/\sqrt{60}$ and the U(1) charge $Q_X$ as
\begin{equation}
 Y = \frac{1}{\sqrt{15}} T_{24} +\frac{1}{5} Q_X ~.
\end{equation}
Hence, the hypercharge $Y$ is not traceless in this model, contrary to
a conventional SU(5) GUT.

The model
also employs four singlet fields, which have no U(1) charges and are
denoted by $\phi_a=({\bf 1},0)$, $a=0,\ldots,3$.  As we comment below, it is also
sufficient to consider only 3 singlets. In this case, 
the inflaton (associated with one combination of the singlets)
participates in the neutrino mass matrix as in the SO(10) model discussed in 
Ref. \cite{EGNNO}.

The generic form for the superpotential of the theory can be written
as\footnote{Note that these couplings are exactly what would be allowed
by SO(10). In the case where the $\text{SU}(5) \times \text{U}(1)$ gauge
group is embedded into SO(10), the couplings $\lambda_1$, $\lambda_2$,
and $\lambda_3$ are unified to a single Yukawa coupling. Moreover, in
this case, additional chiral superfields need to be introduced so that
$H$ and $\bar{H}$ are embedded into SO(10) representations. }
\begin{align} \notag
W &=  \lambda_1^{ij} F_iF_jh + \lambda_2^{ij} F_i\bar{f}_j\bar{h} +
 \lambda_3^{ij}\bar{f}_i\ell^c_j h + \lambda_4 HHh + \lambda_5
 \bar{H}\bar{H}\bar{h}\\ 
&\quad   + \lambda_6^{ia} F_i\bar{H}\phi_a + \lambda_7^a h\bar{h}\phi_a
 + \lambda_8^{abc}\phi_a\phi_b\phi_c + \mu^{ab}\phi_a\phi_b\,, 
\label{Wgen} 
\end{align}
where the indices $i,j$ run over the three fermion families, for
simplicity we have suppressed gauge group tensor indices, and we impose
a $\mathbb{Z}_2$ symmetry  
\beq
H\rightarrow -H \, , 
\label{eq:z2h}
\eeq
that prevents the mixing of SM matter fields with Higgs colour triplets and
members of the Higgs decuplets. This symmetry also suppresses the
supersymmetric mass term for $H$ and $\bar{H}$, which has the advantage
of suppressing the dangerous dimension-five proton decay operators
as we discuss below. Expansion of the superpotential (\ref{Wgen}) in
component fields reveals the following couplings 
\beq\label{Wsinglets}
W\supset \mu^{ab}\phi_a\phi_b + \lambda_8^{abc}\phi_a\phi_b\phi_c + \lambda_6^{ia}\nu^c_i\nu^c_{\bar{H}}\phi_a\,,
\eeq
for the SM gauge singlets $\phi_a$. 

The \kahler potential for the model is assumed to have the no-scale
form
\beq
K = -3\ln\left[T+\bar{T}-\frac{1}{3}\left(|\phi_a|^2+|\ell^c|^2 +
f^{\dagger}f + h^{\dagger}h + \bar{h}^{\dagger}\bar{h} + F^{\dagger}F +
H^{\dagger}H+ \bar{H}^{\dagger}\bar{H} \right)\right]\,. 
\eeq
Therefore, in the absence of any moduli dependence of the gauge kinetic
function, the scalar potential will have the form 
\beq
V=e^{2K/3}\left(|W_i|^2 + \frac{1}{2}D^aD^a\right)\,,
\eeq
where the $D$-term part of the potential in the limit of vanishing SM
non-singlets has the form\footnote{We can always rotate the vacuum
expectation values (vevs) of $H$ and $\bar{H}$ into the $\tilde\nu_H^c$
and $\tilde\nu_{\bar{H}}^c$ directions, respectively, via SU(5) gauge
transformations. } 
\begin{equation}
 D^aD^a = \left(\frac{3}{10}g_5^2 +
\frac{1}{80} g_X^2\right)\,\left(|\tilde{\nu}^c_i|^2
+|\tilde{\nu}^c_{H}|^2-|\tilde{\nu}^c_{\bar{H}}|^2\right)^2\,,
\end{equation}
where we have rescaled the U(1) gauge coupling by a factor of
$\sqrt{40}$; namely, the U(1) charges in \eqref{eq:charges} are
expressed in units of $1/\sqrt{40}$.
The ${\rm SU}(5) \times {\rm U}(1)$ GUT symmetry is therefore broken
along the $F$- and $D$-flat direction $\langle
\tilde{\nu}^c_H\rangle=\langle \tilde{\nu}^c_{\bar{H}}\rangle \neq 0$. These vevs, which can naturally be large
thanks to the $F$- and $D$-flatness, are induced by the soft
supersymmetry-breaking masses. The resultant symmetry-breaking pattern
is 
\begin{equation}
 {\rm SU}(5) \times {\rm U}(1) \to {\rm SU}(3)_C \times {\rm SU}(2)_L
  \times {\rm U}(1)_Y ~.
\end{equation}
Notice that this symmetry-breaking pattern is unique, contrary to the
case of an ordinary supersymmetric SU(5) GUT, which has degenerate vacua, so that
SU(5) may be broken into other gauge groups such as ${\rm
SU}(4) \times {\rm U}(1)$. We also note that this model is free from any
monopole problem, since the ${\rm SU}(5) \times {\rm U}(1)$ gauge group
is not simple \cite{DKN}.

After $H$ and $\bar{H}$ develop vevs, thirteen gauge fields (out of the
twenty-five in ${\rm SU}(5) \times {\rm U}(1)$) acquire masses of order the
GUT scale by absorbing  the corresponding Nambu--Goldstone chiral superfields
in $H$ and $\bar{H}$. The remaining seven chiral superfields in $H$ and
$\bar{H}$ appear as physical states: one is a SM singlet and the others
are the $d_H^c$ and $d^c_{\bar{H}}$. The former, which is a linear
combination of $\nu_H^c$ and $\nu_{\bar H}^c$, is massless in the
supersymmetric limit due to the presence of an $F$- and $D$-flat
direction in the potential, and has a mass of order the soft supersymmetry-breaking mass
scale; we denote this combination by $\Phi$, and refer to it as the flaton. On
the other hand, the $d_H^c$ and $d^c_{\bar{H}}$ are combined with the $D$ and
$\bar{D}$ in $h$ and $\bar{h}$ via the couplings $\lambda_4$ and
$\lambda_5$, respectively,  have GUT-scale masses. The minimal
supersymmetric Standard Model (MSSM) Higgs
multiplets $h_d$ and $h_u$ in $h$ and $\bar{h}$, respectively, do not
acquire masses through the vevs of $\nu_H^c$ and $\nu_{\bar H}^c$, and
thus remain light. This realizes the so-called missing-partner mechanism
\cite{flipped2, Masiero:1982fe}, which solves naturally the
doublet-triplet splitting problem. We note that the flat direction is
expected to be lifted by a higher order operator of the form $(H
\bar{H})^n/M_P^{2n-3}$ in the superpotential~\footnote{We use natural units with $M_P^{-2} = 8\pi G_N
\equiv 1$ throughout this paper.}.  In order to obtain
a GUT scale vev, we should have $n\geq 4$. As a result, we expect flaton and flatino 
masses to be of order $M_{\rm GUT}^6/M_P^5$, \textit{i.e.}, of order 
the supersymmetry-breaking scale, facilitating their decays into lighter MSSM particles.

In order to achieve successful
electroweak symmetry breaking, we need a $\mu$-term for $h$ and
$\bar{h}$ of order the supersymmetry-breaking scale. This can be generated via the
Giudice--Masiero (GM) mechanism \cite{gm} or through the coupling
$\lambda_7^a$ with supersymmetry-breaking scale vevs of $\phi_a$~\cite{flipped2},
as in the next-to-minimal supersymmetric Standard Model (NMSSM). 
In order for $\phi_a$ to develop a TeV-scale vev, its supersymmetric
mass term should be $\lesssim {\cal O}(1)$~TeV. In this case,
a supersymmetry-breaking soft mass for $\phi_a$, which is driven
negative by renormalization-group effects, allows $\phi_a$ to acquire
a vev of order the soft mass scale, which can naturally explain the origin
of the TeV-scale MSSM $\mu$-term. When only three singlets are included in the model,
a $\mu$ term generated by the GM mechanism is necessary.

As we have mentioned above, a superpotential $\mu$-term for $H$ and $\bar{H}$ is
suppressed by the $\mathbb{Z}_2$ symmetry, and thus the chirality
flip between the color-triplet Higgs multiplets can occur only via
the $\mu$-term for $h$ and $\bar{h}$~\footnote{It is also possible that
a GM term for  $H$ and $\bar{H}$ could be generated in the K\"ahler
potential through loop corrections accompanied by an explicit
$\mathbb{Z}_2$-symmetry-breaking effect. Such a term would also
contribute to the mass of the flaton and flatino. In addition, such an
explicit $\mathbb{Z}_2$-symmetry-breaking term can prevent the generation of
domain walls when the field $H$ acquires a vev. This
$\mathbb{Z}_2$-symmetry-breaking effect would also generate a dimension-five
proton-decay operator, but its contribution to the proton decay rate is
suppressed by a factor of $(\mu_H/M_{H_C})^2$ (where $\mu_H$ is the induced
$\mu$-term for $H$ and $\bar{H}$), and thus does not lead to a proton decay problem. } 
Since the dimension-five
proton-decay process through the color-triplet Higgs exchange requires a
chirality flip, this rate is suppressed by a factor of
$(\mu/M_{H_C})^2$, where $M_{H_C}$ denotes the color-triplet Higgs
mass. As a consequence, this model can easily avoid the dimension-five
proton decay limit from the $p \to K^+ \bar{\nu}$ mode, $\tau (p\to K^+
\bar{\nu}) > 6.6\times 10^{33}$~yrs \cite{Takhistov:2016eqm}, without
relying on multi-TeV scale sfermions. This can enlarge the MSSM
parameter space where both the correct dark matter density and the 125
GeV Higgs boson mass are obtained \cite{Ellis:2017djk}.

In an ${\rm SU}(5) \times {\rm U}(1)$ GUT, the SU(3)$_C$ and SU(2)$_L$
gauge couplings unify at a high scale, $M_{32}\equiv M_{\rm GUT}$, into
a single SU(5) gauge coupling $\alpha_5$:
\begin{equation}
 \alpha_3 (M_{32}) = \alpha_2 (M_{32}) = \alpha_5 (M_{32}) = 0.0374 ~,
 \label{alphagut}
\end{equation}
where $M_{32} = 1.2 \times 10^{16}$~GeV when we use $\alpha_3(M_Z) =
0.1181$ \cite{Olive:2016xmw}, the SM beta functions below 10~TeV, and
the MSSM beta functions above 10~TeV---both at two-loop level---and
neglect threshold corrections. At this scale, the hypercharge gauge
coupling $\alpha_1 \equiv 5\alpha_Y/3$ is matched onto the U(1) gauge
coupling $\alpha_X$ as  
\begin{equation}
 \frac{25}{\alpha_1 (M_{32})} = \frac{1}{\alpha_5 (M_{32})} 
+ \frac{24}{\alpha_X (M_{32})} ~. 
\end{equation} 
We see in these equations that
the U(1) gauge coupling $\alpha_X$ is not necessarily equal to
the SU(5) gauge coupling $\alpha_5$ at $M_{32}$. These couplings may
unify at a higher scale, which is required if the ${\rm SU}(5) \times
{\rm U}(1)$ gauge group is embedded into a simple group such as SO(10) at
high energies, and in string constructions.

Since the unification of $\alpha_3$ and $\alpha_2$ should occur below
the scale of complete unification, the first unification scale $M_{32}$
is expected to be smaller than the unification scale in the minimal
SU(5) GUT \cite{Ellis:1995at,
Ellis:2002vk}. This indicates that the proton decay rate of the $p\to
e^+ \pi^0$ channel in this model, which is induced by the exchange of
the SU(5) gauge bosons with masses around $M_{32}$, may be larger
than that in the ordinary SU(5) GUT. The proton lifetime in the
supersymmetric flipped $\text{SU}(5) \times \text{U}(1)$ model was
evaluated in Ref.~\cite{Li:2010dp} as~\footnote{The lifetime would be
much shorter if there were additional vector-like multiplets at
the TeV scale \cite{Li:2009fq}, since they would give a positive contribution to
the gauge coupling beta functions, making the GUT-scale gauge
coupling larger. In general, such extra matter multiplets at low energies can enhance
proton decay rate considerably \cite{Hisano:2012wq}.}:
\begin{equation}
 \tau_p =4.6\times 10^{35} \times \left(\frac{M_{32}}{10^{16}~\text{GeV}}\right)^4
  \times \left(\frac{0.0374}{\alpha_5(M_{32})}\right)^2 ~\text{yrs}~.
\end{equation}
This may be compared with the current experimental limit on the $p\to e^+
\pi^0$ channel given by the Super-Kamiokande collaboration, which is $\tau
(p\to e^+ \pi^0) > 1.6\times 10^{34}$~yrs \cite{Miura:2016krn}. 
This sets a lower limit on the scale $M_{32}$: 
\begin{equation}
 M_{32} > 4.3 \times 10^{15}~\text{GeV} \times 
\left(\frac{\alpha_5(M_{32})}{0.0374}\right)^{\frac{1}{2}} ~,
\label{eq:protlim}
\end{equation}
which is generically satisfied when the low-energy matter content is the
MSSM \cite{Ellis:1995at, Ellis:2002vk}. 
A part of the predicted range of proton lifetimes
may be within the reach of future proton decay experiments, such as
the Hyper-Kamiokande experiment \cite{Abe:2011ts}.

\section{Inflation}
\label{sec:inflation}

The inflaton, which we denote by $S$, is in general a linear
combination of the singlet fields $\phi_a$, $a= 0,1,2,3$\footnote{The discussion in this section
is not affected by the number of singlets, and models with just three singlets with $a = 0, 1, 2$ are also possible.}. 
The asymptotically-flat Starobinsky potential is realized for $S$ if its
superpotential takes the form~\footnote{An alternative choice of
superpotential involving the moduli is $W = S(T-1/2)$
\cite{Cecotti}. Families of superpotentials that lead to the Starobinsky
potential were discussed in \cite{ENO7}.} \cite{ENO6}
\beq\label{WZW}
W\supset m\left( \frac{S^2}{2}  - \frac{S^3}{3\sqrt{3}}\right) \,,
\eeq
with $m\simeq 10^{-5}$ set by the measured primordial power spectrum
amplitude.
One would expect that the dimensionful couplings $\mu^{ab}$ in
(\ref{Wsinglets}) would naturally be around the GUT scale $M_{\rm
GUT}\sim 10^{-2}$, which is a few orders of magnitude above the
intermediate scale set by the magnitude of $m$. Two scenarios for these
couplings are possible: (1) the light state $S$  appears along with
three heavy states upon diagonalizing a nearly-degenerate matrix
$\mu^{ab}$, and (2) through an unspecified mechanism, some or all of the
couplings are $\mu^{ab} \lesssim 10^{-5}$ at the inflation scale,
implying that there may be more than one light scalar field, and in
principle any of the $\phi_a$ could be identified with the inflaton. We
consider both possibilities in what follows, and derive the
corresponding phenomenological constraints on the model parameters.

\subsection{Scenario (1): hierarchy of singlet masses with one  light
  state}
\label{sec:singlst}

The simplest realization of (\ref{WZW}) corresponds to the assumption
that the state $S$ is the light eigenstate of a nearly-degenerate mass matrix $\mu^{ab}$. We identify
$S$ with the rotated field $\phi^D_0$ in the diagonal basis denoted by
subscripts $D$, where 
\beq
\mu^{ab}_D = {\rm
diag}\left(m/2,\mu_D^{11},\mu_D^{22},\mu_D^{33}\right)\,,\qquad
\mu_D^{ab}\leq M_{\rm GUT}\,, 
\eeq
with $m \simeq 10^{-5}$. Such a light eigenstate exists if ${\rm
det}\,\mu^{ab}\ll M_{\rm GUT}^4$. 

In order to realize successful Starobinsky-like inflation as in
(\ref{WZW}), in the diagonal basis the Yukawa coupling must satisfy 
\beq\label{StaroCond1}
-3\sqrt{3}\, \lambda_{8,D}^{000} = m\,.
\eeq
For the remainder of this (sub)section we drop the index $D$, assuming
implicitly that we refer to rotated fields and couplings. 

We now investigate the conditions for sufficient inflation. Expanding
the ($F$-term) scalar potential for the singlet fields reveals that
large masses for the fields $\tilde{\nu}^c$ and $\tilde{\nu}_{\bar{H}}^c$ may be
induced during inflation: 
\begin{align} \notag
 V_F &\;=\;  e^{2K/3}\left[ m^2|S-S^2/\sqrt{3}|^2 + \sum_i
 \left(|\lambda_6^{i0}\tilde{\nu}_{\bar{H}}^c S|^2 + |\lambda_6^{i0}
 \tilde{\nu}_{i}^c S|^2\right) + \cdots \right] \\  \notag
&\;\simeq\; \frac{3}{4}m^2\left(1- e^{-\sqrt{2/3}\,s}\right)^2 +
 \frac{3}{4}\sinh^2(\sqrt{2/3}\,s) \sum_i |\lambda_6^{i0}|^2 \left(
 |\tilde{\nu}_{\bar{H}}^c |^2 + |\tilde{\nu}_{i}^c |^2 \right)\\
 & \qquad + \frac{1}{8}m^2e^{\sqrt{2/3}s}\left(|\tilde{\nu}^c_{\bar{H}}|^2 + \sum_i |\tilde{\nu}^c_i|^2\right) + \cdots \,. 
\end{align}
where $s=\sqrt{6}\tanh^{-1}(S/\sqrt{3})$ denotes the
canonically-normalized inflaton and the index $i = 1,2,3$. In the second
expression for $V_F$, we see the standard Starobinsky potential, followed by
correction terms. At large $s$, we see the origin of the large masses
for $\tilde{\nu}^c$ and $\tilde{\nu}_{\bar{H}}^c$. Therefore, for generic
couplings, $\tilde{\nu}^c$ and the GUT-breaking field $\tilde{\nu}_H^c$ will be
driven to zero in about one Hubble time during inflation, leaving the
Universe in the symmetric phase at the end of inflation. A subsequent
phase transition driven by the renormalization-group (RG) flow of the
soft masses of $H$ and $\bar{H}$ via the couplings $\lambda_{4,5,6}$ can
lead to the breaking of the ${\rm SU}(5) \times {\rm U}(1)$
symmetry~\cite{flipped2}~\footnote{We note that entropy could be released during this
transition~\cite{Ellis:1992nq,Campbell:1987eb}, whose amount and potential danger we discuss in
Section~\ref{sec:phasetr}.}. No constraints on the couplings
$\lambda_6^{i0}$ are necessary for a successful inflationary phase, and
we assume from now on that $\tilde{\nu}^c=\tilde{\nu}_H^c=0$ during
inflation.

Another source for a deformation of the inflationary potential is the
coupling of $S$ with the other $\phi_i$ fields. In order to determine
its effect, we evaluate the gradient of the scalar potential during
inflation: 
\beq\label{dVdphiInf}
e^{-2K/3}\frac{\partial V}{\partial \bar{\phi}^a} \;=\; \sum_{b}W^b
\left(\frac{2}{3}K_a\bar{W}_b + \bar{W}_{ab}\right)\,. 
\eeq
During inflation, the fields $\phi_j$ (as well as any other scalars) get large masses,
\beq
\frac{\partial^2V}{\partial\phi_i\partial \bar{\phi}^j} \;=\; \frac{2}{3}e^{K}m^2|S-S^2/\sqrt{3}|^2\delta_j^i + \cdots  \;\simeq\; \frac{1}{8}m^2e^{\sqrt{2/3}s}\delta_j^i + \cdots \;\gg\; H^2\,,
\label{heavy}
\eeq 
and hence fluctuations displacing them from the origin can be neglected.
Since all non-singlet fields vanish, and we assume we are in the
$\mu$-diagonal basis, the superpotential derivatives that appear in this
expression correspond to 
\begin{align} \label{Wbsing1} \displaybreak[0]
W^i &= 3\lambda_8^{00i}S^2  + 2 \sum_{j} (\mu^{ij}+ 3\lambda_8^{0ij}S)
 \phi_j + 3 \sum_{j,k} \lambda_8^{ijk}\phi_j \phi_k\,,\\
 \displaybreak[0] 
W^0 &= m(S-S^2/\sqrt{3})  + 6 S \sum_{j} \lambda_8^{00j} \phi_j + 3
 \sum_{j,k} \lambda_8^{0jk}\phi_j \phi_k\,,\\ \displaybreak[0] 
\bar{W}_{ab} &= 2\bar{\mu}_{ab} + 6 \bar{\lambda}_{8\,0ab} \bar{S} + 6 \sum_{j}
 \bar{\lambda}_{8\,abj}\bar{\phi}^j\,. 
\end{align}
We notice that, unless $\lambda_8^{00i}=0$, (\ref{dVdphiInf})
implies that the singlet fields will relax to non-zero values during
inflation. The scenario in which $\phi_i=0$ is possible if $\mu^{ab}$
and $\lambda_8^{0ab}$ can be diagonalized simultaneously. In that case,
the couplings $\lambda_8^{00i}$ {\it are} all absent and  substituting
$\phi_i=0$ into the effective potential yields 
\beq\label{staropot}
V_{\rm inf} = \frac{3}{4}m^2\left(1- e^{-\sqrt{2/3}\,s}\right)^2\,,
\eeq
{\it i.e.}, simply the Starobinsky potential. 

\begin{figure}[t!]
\centering
   \subfloat{\scalebox{0.5}{\includegraphics{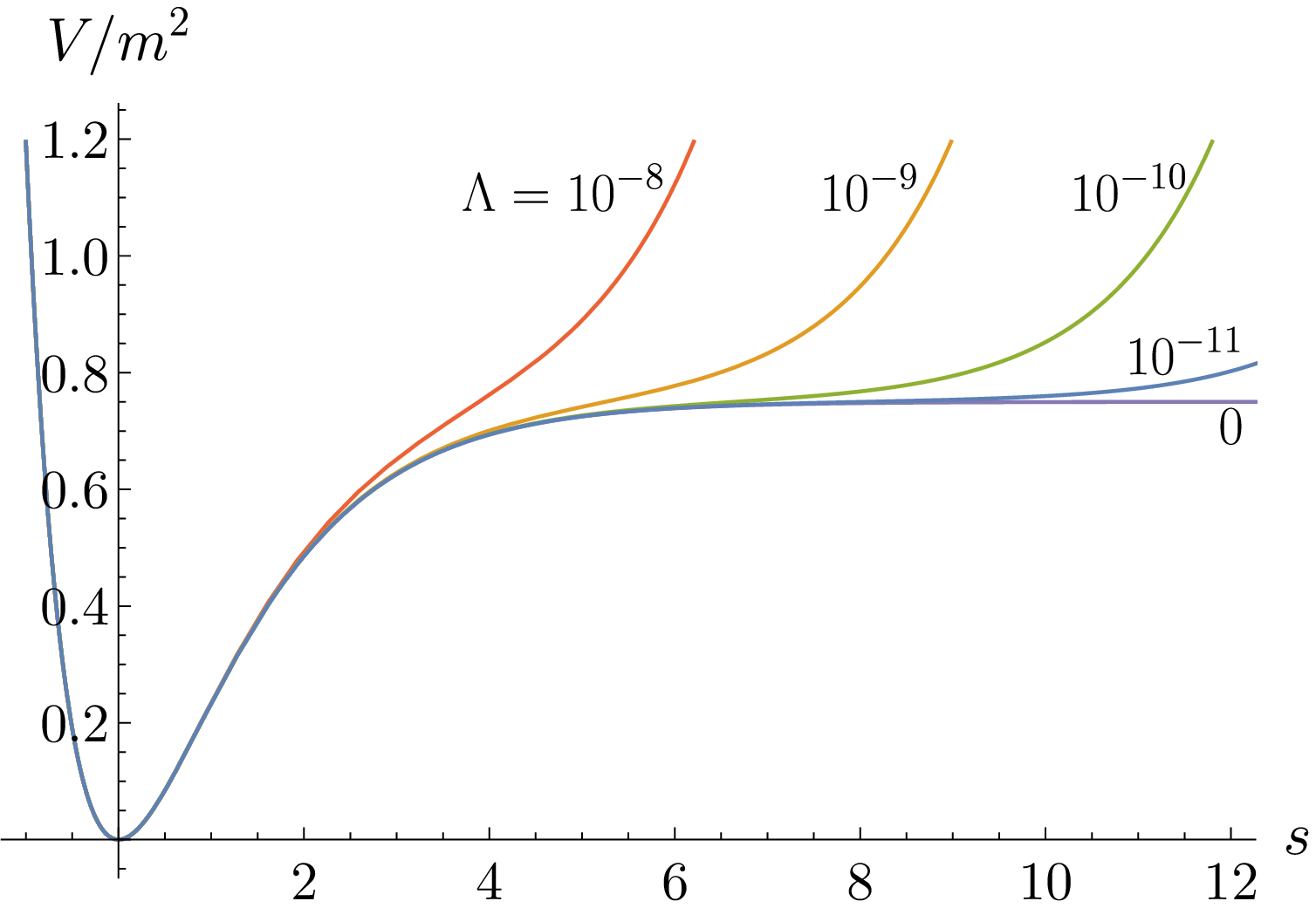}}}
    \hfill
    \subfloat{\scalebox{0.5}{\includegraphics{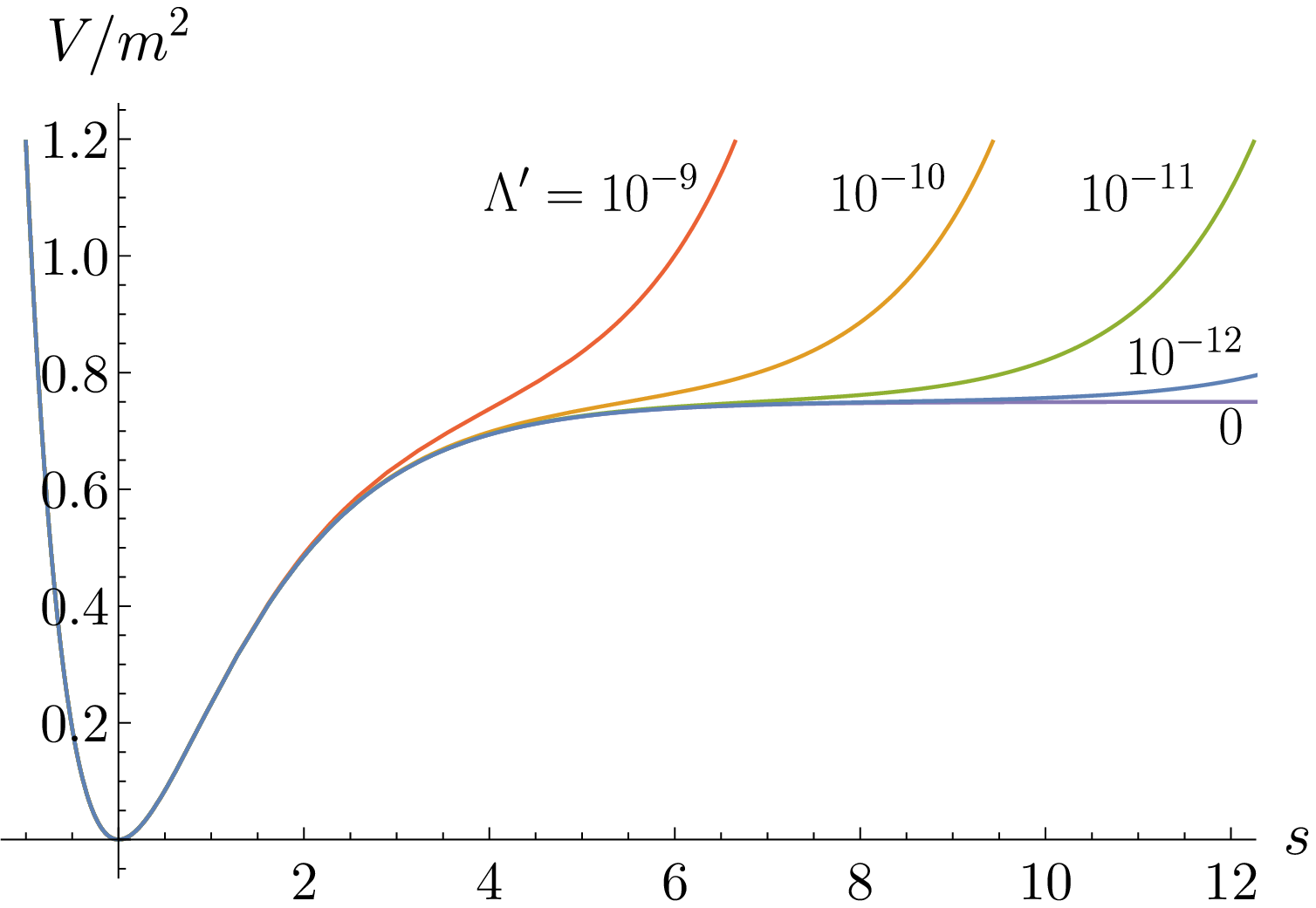}}}
    \caption{\em The effective inflationary potentials (\ref{LinearPot1}) and (\ref{LinearPot2}) for different values of 
    $\Lambda$ (left) and $\Lambda'$ (right), with $m=10^{-5}$. The curves labeled $\Lambda,\Lambda' = 0$ 
    correspond to the Starobinsky potential (\ref{staropot}). 
        }
    \label{fig:Vcorr0}
\end{figure}

If $\mu^{ab}$ and $\lambda_8^{0ab}$ are {\it not} simultaneously
diagonalizable, then $\lambda_8^{00i}$ may not vanish and thus $\phi_i$
may be non-zero during inflation. Since $m\simeq 10^{-5}$, and during
inflation when $s$ is large,  $S\sim \sqrt{3}$, we can disregard the
contribution proportional to $W^0$ in (\ref{dVdphiInf}) for the solution
of the values of $\phi_i$, if in addition $\lambda_8^{00i} \ll
\lambda_8^{0ij}$. This would imply that the instantaneous singlet vevs
correspond approximately to the solutions of the equations $W^i=0$. The
scalar potential during inflation then takes the form 
\beq\label{Veff1}
V \simeq e^{2K/3}\Big|m(S-S^2/\sqrt{3}) + 6S \sum_{i} \lambda_8^{00i}
\phi_i + 3\sum_{j,k} \lambda_8^{0jk} \phi_j \phi_k\Big|^2\,. 
\eeq
For $\lambda_8^{00a}\ll \lambda_8^{0ij}$, the singlet vevs are
approximately given by the solution of the system of equations 
\beq \label{LinearSol1}
 3\lambda_8^{00i}S^2  + 2 \sum_{j} (\mu^{ij}+ 3\lambda_8^{0ij}S) \phi_j
 \simeq 0\,, 
\eeq
and the effective potential (\ref{Veff1}) can be approximately written as
\begin{align} \notag
V_{\rm inf} &\simeq \frac{3}{4}m^2\left(1-e^{-\sqrt{2/3}\,s}\right)^2\\
 &+\frac{3\sqrt{3}\, m
 \sinh(\sqrt{2/3}\,s)}{2(1+\tanh(s/\sqrt{6}))} \left[ 2 \sqrt{3}
 \tanh{(s/\sqrt{6})} \sum_{i} \lambda_8^{00i} \phi_i + \sum_{i,j}
 \lambda_8^{0ij} \phi_i \phi_j \right] + {\rm h.c.}\, 
\end{align}
In general, the singlet vevs may be obtained by inverting the matrix
$(\mu^{ij}+3\lambda_8^{0ij}S)$, but the resulting general expressions
are not particularly enlightening. Instead, let us consider two limiting
cases. First, let us assume that the couplings $\lambda_8^{0ij}\gtrsim
\mu^{ij}$. As $S = {\cal O}(1)$ during inflation, in this case we can
disregard the $\mu^{ab}$ term in (\ref{LinearSol1}), and the effective
potential (\ref{Veff1}) takes the approximate form 
\begin{align} 
V_{\rm inf} &\simeq \frac{3}{4}m^2\left(1-e^{-\sqrt{2/3}\,s}\right)^2
 +\frac{27\, m \sinh^2(s/\sqrt{6})}{2(1+\tanh(s/\sqrt{6}))} \left(
 \sum_{i} \lambda_8^{00i} \phi_i + {\rm h.c.}  \right) \\
 \label{LinearPot1} 
& \simeq \frac{3}{4}m^2\left(1-e^{-\sqrt{2/3}\,s}\right)^2  +
 \frac{27\sqrt{3}}{4}\, m \Lambda \, e^{-s/\sqrt{6}}\sinh^3(s/\sqrt{6})
 \,. 
\end{align}
where we have defined
\beq\label{lambdadef}
\Lambda \equiv - \sum_{i,j} (\lambda_8^{0ij})^{-1} \lambda_8^{00i}
\lambda_8^{00j} + {\rm h.c.} \,  
\eeq
The left panel of Fig.~\ref{fig:Vcorr0} shows the form of the scalar
potential (\ref{LinearPot1}) as a function of $\Lambda$.  

\begin{figure}[t!]
\centering
   \subfloat{\scalebox{0.584}{\includegraphics{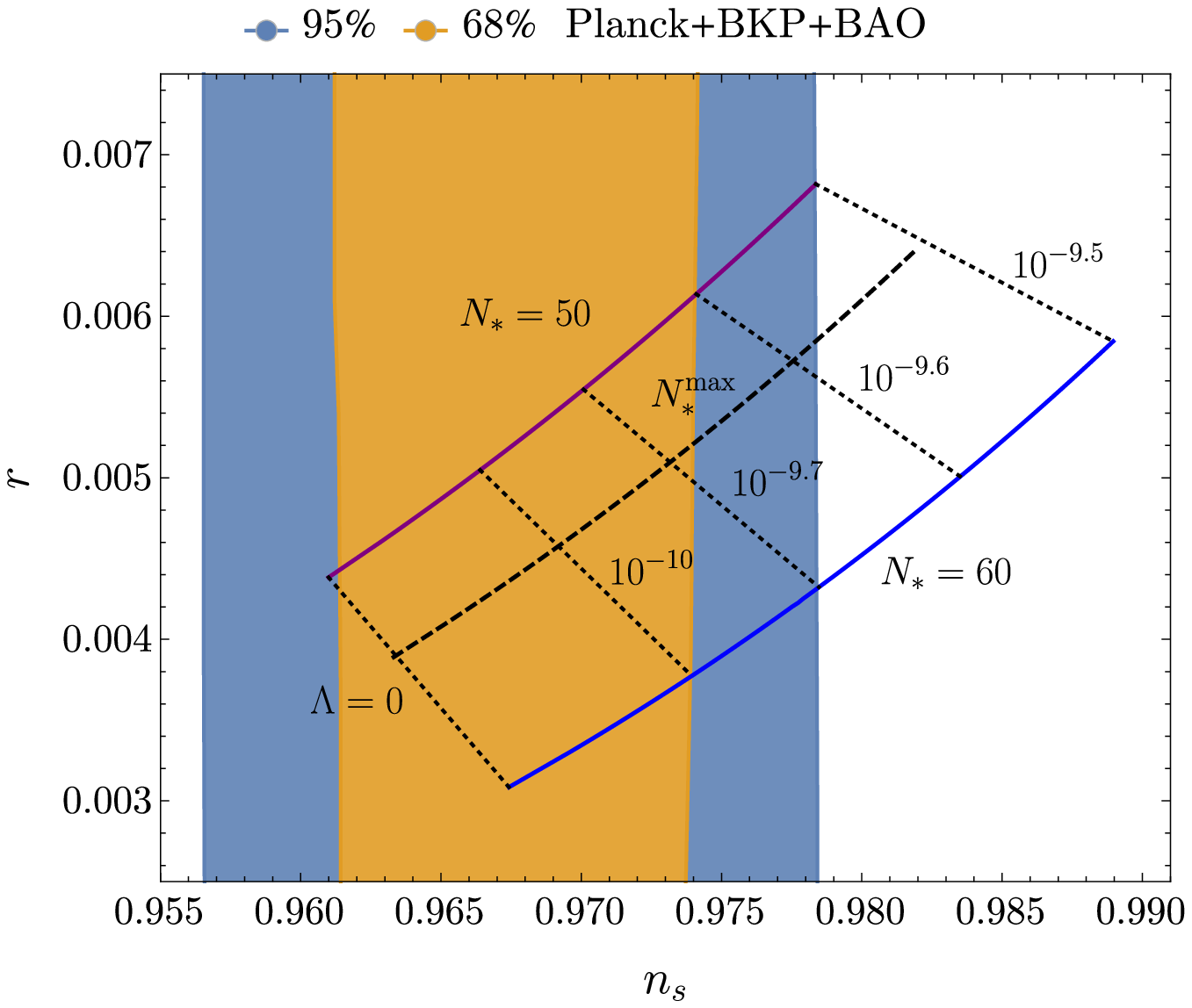}}}
    \hfill
    \subfloat{\scalebox{0.584}{\includegraphics{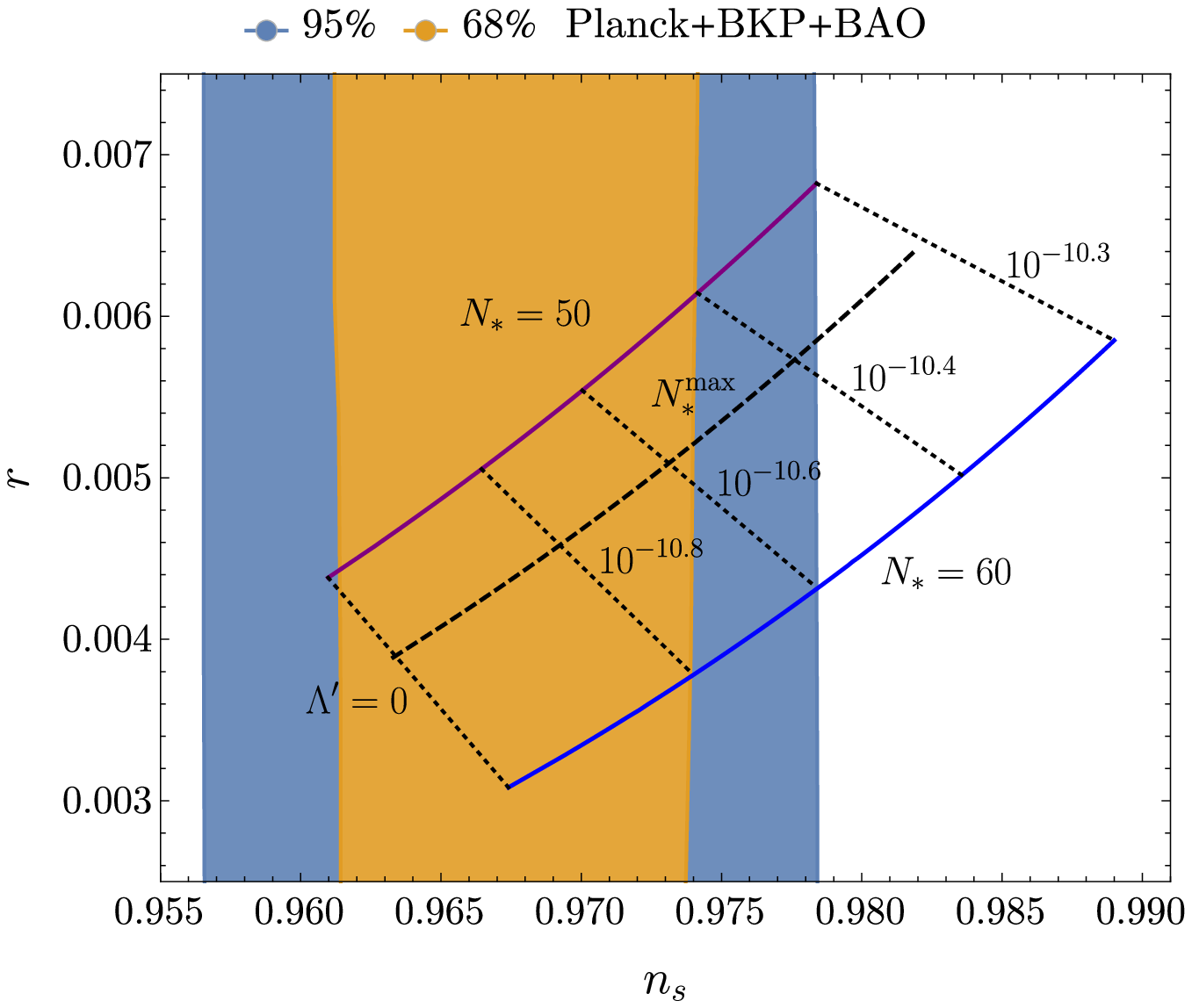}}}
    \caption{\em Parametric $(n_s, r)$ curves as functions of $\Lambda$ (left) and $\Lambda'$ (right) for $N_* = 50,60$ and $N_*^{\rm max}\simeq 53.3$, 
    with the 68\% and 95\% CL constraints from Planck and other data~\protect\cite{planck15} shown in the background. The solid curves show the parametric dependence 
    using the analytical approximations (\ref{LinearPot1}) and (\ref{LinearPot2}). The dashed curve uses these analytical approximations together with (\ref{howmany}) to determine the dependence at $N_*^{\rm max}$. The dotted curves are for illustrative values of $\Lambda,\Lambda'$.
        }
    \label{fig:nsrcurve0}
\end{figure}

Starobinsky-like inflation with a total number of $e$-folds $N>60$ is
realized only if $\Lambda\lesssim 10^{-10}$, which corresponds,
schematically, to $\lambda_8^{00i}\lesssim 10^{-5}
(\lambda_8^{0ij})^{1/2}$~\footnote{In general, as we see also in later examples,
Starobinsky-like inflation occurs if the deviation from the minimal
Starobinsky potential is small for values of the inflaton field that are
$\lesssim 6$.}. The left panel of Fig.~\ref{fig:nsrcurve0} shows the
parametric dependence of the scalar tilt $n_s$ and the tensor-to-scalar
ratio $r$ on $\Lambda$, and compares them to the 68\% and 95\% CL
constraints from Planck and other data~\cite{planck15}. We see that the model predicts $r \lesssim 0.007$ for the
number of $e$-folds to the end of inflation, $N_* > 50$, within a factor
$\sim 2$ of the Starobinsky prediction and far below the current
experimental upper limit. On the other hand, for either $N_*=50$ or $N_*^{\rm max}$, Planck
compatibility with the 95\% CL range of $n_s$ is lost for
$\lambda_8^{00i}\gtrsim 10^{-4.8} (\lambda_8^{0ij})^{1/2}$, and for $N_*=60$ this occurs for $\lambda_8^{00i}\gtrsim 10^{-4.9} (\lambda_8^{0ij})^{1/2}$. Here $N_*^{\rm max}$ is defined as the maximum number of e-folds after horizon crossing, which is compatible with the bound on the reheating temperature due to thermal production of gravitinos (see section~\ref{sec:entropygrav} for a detailed discussion). $N_*^{\rm max}$ is a function of the energy density at horizon crossing $V_*$, and therefore it is dependent on $\Lambda$; the curve shown in Fig.~\ref{fig:nsrcurve0} takes into account this dependence, which is very weak, merely a $\lesssim 0.3\%$ overall variation with respect to the Starobinsky limit $N_*^{\rm max}\simeq 53.3$.

If instead $\lambda_8^{0ij}\lesssim \mu^{ij}$, as would be the case for
a strongly-segregated inflaton sector, the effective potential can be
approximated by  
\beq \label{LinearPot2}
V_{\rm inf} \simeq \frac{3}{4}m^2\left(1-e^{-\sqrt{2/3}\,s}\right)^2 +
81\, m  \, \sinh^4(s/\sqrt{6})\left(\tanh(s/\sqrt{6})-1\right) \sum_{i}
\left[   \mu_i^{-1}(\lambda_8^{00i})^2  + {\rm h.c.} \right]\, , 
\eeq
where $\mu^a \equiv \mu^{aa}$ since we have already assumed a basis
where $\mu$ is diagonal. This potential is shown in the right panel of
Fig.~\ref{fig:Vcorr0}, where we have denoted  
\beq
\Lambda'= -\sum_{a} \mu_i^{-1}(\lambda_8^{00i})^2 + {\rm h.c.} .
\eeq 
In this case, Starobinsky-like inflation is obtained for 
$\Lambda'\lesssim10^{-11}$, or $\lambda_8^{00i}\lesssim 10^{-5.5}
(\mu^i)^{1/2}\sim 10^{-6.5}$. As shown in the right panel of
Fig.~\ref{fig:nsrcurve0}, whereas the model prediction for $r$ is again
similar to the Starobinsky prediction, compatibility with the 95\%
Planck range of $n_s$ is lost for $\lambda_8^{00i}\gtrsim 10^{-6.2}$
when $N_* =50$ or $N_*^{\rm max}$, and for $\lambda_8^{00i}\gtrsim 10^{-6.3}$ when $N_*=60$.

We now discuss the dynamics of the scalar fields subsequent to
inflation. If $\mu^{ab}$ and $\lambda_8^{0ab}$ are simultaneously
diagonalizable, the potential gradient for the non-inflaton singlet
fields vanishes, implying they are not excited during reheating. If
$\lambda_8^{0ab}$ is not diagonal in the rotated basis, the singlets
start oscillations about their minima, as they have non-zero vevs during
inflation. Moreover, these oscillations are forced, driven by the
oscillating inflaton $S$. To see this, let us investigate the potential
gradient with respect to $\bar{\phi}^i$: 
\begin{align}\notag
\frac{\partial V}{\partial \bar{\phi}^i} &\simeq  \sum_{j} \Big[
 4\Big(|\mu^i|^2\delta_i^j + 3\bar{\mu}_i \lambda_8^{0ij}S +
 3\bar{\lambda}_{8\,0ij}\mu^j \bar{S} +  9 |S|^2
 \sum_{k}\lambda_8^{0jk}\bar{\lambda}_{8\,0ik} \Big)\phi_j + 18
 \bar{\lambda}_{8\,0ij}\lambda_8^{00j} |S|^2 S  \Big]\\ 
&\qquad  + 6\bar{\mu}_i\lambda_8^{00i}S^2 +  \cdots ~.
\label{VgradC2}
\end{align}
Since we assume that $\mu^i\gg m$ and $\langle\phi_i\rangle_{\rm
inf}\ll1$, the fields $\phi_i$ will track quasi-statically the solution
of the vev condition, $\partial V/\partial \bar{\phi}^i =0$. Naively,
this implies that during reheating the singlets will remain small,
$\phi_i \sim \lambda_8^{00i}$. In general, however, if
$\lambda_8^{0ij}\gtrsim \mu^{ij}$, this may only be true during the
first oscillation(s) of the inflaton $S$, as the approximation
(\ref{VgradC2}) breaks down if  
\beq\label{detcond}
\det\Big(|\mu^i|^2\delta_i^j +
3\bar{\mu}_i\lambda_8^{0ij}S+3\bar{\lambda}_{8\,0ij}\mu^j \bar{S} + 9
|S|^2 \sum_{k}\lambda_8^{0jk}\bar{\lambda}_{8\,0ik} \Big) = 0\,. 
\eeq
If a real solution exists, the $\phi_i$ exhibit resonant behaviour for
$S\simeq 0$~\footnote{Equation (\ref{detcond}) is a polynomial equation of
the form $\sum_{n=0}^{6} a_n S^n = 0$, with $a_0\ll a_1 \ll \cdots \ll
a_6$, for which a solution (if it is real) is given by $S_0\simeq
-a_0/a_1\ll 1$. As $S$ oscillates about the origin with an initial
amplitude $S\sim \mathcal{O}(1)$, it crosses this point at least
once.}. Therefore, in this case, numerical integration of the equations
of motion is necessary, as the singlets can in principle drain the
energy density from the inflaton, and thereby be responsible for the
eventual reheating of the Universe. We explore this effect in
Section~\ref{sec:numerinf}.

If no real solution for (\ref{detcond}) exists, for
$\lambda_8^{0ij}\gtrsim \mu^{ij}$ the singlets evolve adiabatically with
the inflaton oscillation. Modeling the inflaton oscillation as $s\simeq
s_0 \sin(mt)/mt$ and $H\simeq 2/3t$, with $s_0\simeq 0.6$~\cite{EGNO5},
the time evolution of $\phi_i$ previous to the decay of $S$ can be
approximated by 
\beq\label{phisol01}
\phi_i(t) \simeq - \frac{1}{2\sqrt{2}}\sum_{j}  (\lambda_8^{0ij})^{-1}\lambda_8^{00j}s_0
\left( \frac{\sin mt }{m t}\right) \,,\qquad t_{\rm end}\lesssim t
\lesssim t_{\rm reh}\,, 
\eeq
disregarding corrections due to the finite mass of $\phi_a$ at the
points for which $S=0$, which slightly overdamp the amplitude of the
oscillations (see below). The timescales $t_{\rm end}$ and $t_{\rm reh}$
denote the times when inflation has ended and reheating has taken place
respectively~\footnote{See \cite{EGNO5} for more precise definitions of these quantities.}. (\ref{phisol01}) implies that the ratio of the
energy densities stored in the fields is given approximately by 
\beq\label{rhorat1}
\frac{\rho_{\phi_i}}{\rho_s} \sim
\left(\frac{\lambda_8^{00i}s_0}{m}\right)^2 \left( \frac{\sin mt }{m
t}\right)^2\,. 
\eeq
Therefore, for $\lambda_8^{00i}\lesssim 10^{-5}$, the energy density
during reheating is always dominated by the oscillating inflaton,
meaning that any phenomenological constraints related to the decay of
the singlet fields, such as gravitino overproduction, can be reduced to
the usual discussion for reheating bounds. Of course, this is the case
as long as the singlet excitations decay sufficiently rapidly to avoid a
matter-dominated era after reheating, which needs to be checked on a
case-by-case basis (see Section~\ref{sec:singletdecay}). We further note
that this limit on $\lambda_8^{00i}$ is less constraining than the
previous limit from sufficient inflation.

When the strong segregation condition $\lambda_8^{0ij}\lesssim \mu^i$ is
satisfied, the right-hand side of \eqref{VgradC2} is simply given by 
\beq
\frac{\partial V}{\partial \bar{\phi}^i} \simeq 4|\mu^i|^2\phi_i +
6\bar{\mu}_i \lambda_8^{00i}S^2 + 18 \sum_{j}
\bar{\lambda}_{8\,0ij}\lambda_8^{00j} |S|^2 S + \cdots ~,
\eeq
and thus the evolution is always adiabatic. Under the same assumptions
as in the previous case, the time evolution of $\phi_i$ previous to the
decay of $S$ can be approximated by 
\beq
\phi_i(t) \simeq -\frac{3\lambda_8^{00i}s_0^2}{4\mu^i}\left( \frac{\sin
mt }{m t}\right)^2 \,,\qquad t_{\rm end}\lesssim t \lesssim t_{\rm
reh}\,. 
\eeq
Substitution shows that the ratio of the energy densities is also given
by (\ref{rhorat1}). Thus, in this case reheating through $S$ decay also
occurs, given that $\lambda_8^{00i}\lesssim 10^{-5}$.

\subsection{Scenario (2): multiple light singlet states}\label{sec:nondiaginf}

\begin{figure}[t!]
\centering
    \scalebox{0.5}{\includegraphics{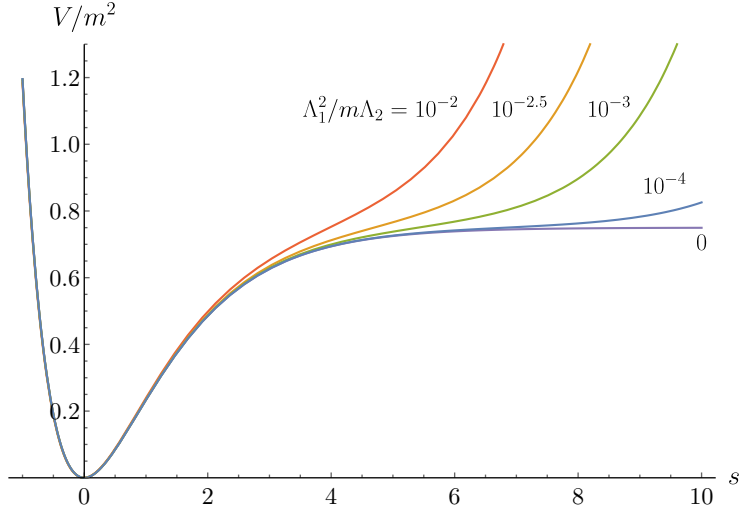}}
    \caption{\em The effective inflationary potential (\ref{deltaV}) for different values of $\Lambda_1^2/m\Lambda_2$. 
    The curve labeled $\Lambda_1^2/m\Lambda_2 = 0$ is the Starobinsky potential. 
        }
    \label{fig:Vcorr1}
\end{figure}

\begin{figure}[h!]
\centering
    \scalebox{0.76}{\includegraphics{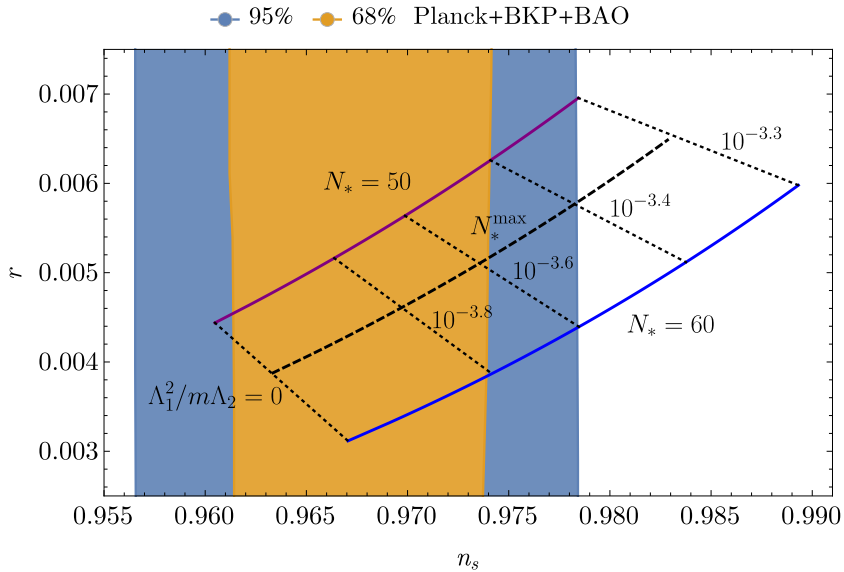}}
    \caption{\em Parametric $(n_s, r)$ curves as functions of
 $\Lambda_1^2/m\Lambda_2$ for $N_* = 50,60$ and $N_*^{\rm max}\simeq 53.3$, with the 68\% and 95\% CL constraints from Planck and other data~\protect\cite{planck15} shown in the background. The solid curves illustrate the 
    parametric dependence using the analytical approximation (\ref{deltaV}). The dashed curve uses the analytical approximation (\ref{deltaV}) together with (\ref{howmany}) to determine the dependence at $N_*^{\rm max}$.
    The dotted curves are for illustrative values of $\Lambda_1^2/m\Lambda_2$.
        }
    \label{fig:nsrcurve}
\end{figure}

We now consider the case with multiple light singlet states, and assume
that the relation  
\beq\label{StaroCond}
-3\sqrt{3}\, \lambda_8^{000} = 2\mu^{00} =  m
\eeq
is realized off-diagonally, {\it i.e.}, in a basis where $\phi_0$ is not
a mass eigenstate. In this case, the superpotential parameters
$\mu^{0i}$ and $\lambda_8^{0ij}$ must be constrained in order to allow
for Starobinsky-like inflation. Despite the increased number of
problematic parameters, the analysis is analogous to that in the
previous Section. The singlet fields $\phi_i$ develop non-vanishing vevs
during inflation, which can be found by solving (\ref{dVdphiInf}), where
in this case the superpotential derivatives are given by 
\begin{align} \displaybreak[0]
W^i &=  2\mu^{0i}S + 3\lambda_8^{00i}S^2  + 2 \sum_{j} (\mu^{ij}+
 3\lambda_8^{0ij}S) \phi_j + 3 \sum_{jk} \lambda_8^{ijk}\phi_j
 \phi_k\,,\\ \displaybreak[0] 
W^0 &= m(S-S^2/\sqrt{3})  + 2  \sum_{j} ( \mu^{0j} +3\lambda_8^{00j}S)
 \phi_j + 3 \sum_{jk} \lambda_8^{0jk}\phi_j \phi_k\,,\\ \displaybreak[0] 
\bar{W}_{ab} &= 2\bar{\mu}_{ab} + 6 \bar{\lambda}_{8\,0ab} \bar{S} + 6 \sum_{j}
 \bar{\lambda}_{8\,abj}\bar{\phi}^j\,. 
\end{align}
The effective potential during inflation then takes the form
\begin{align} \displaybreak[0]
V &= e^{2K/3}\Big|m(S-S^2/\sqrt{3}) + 2 \sum_{i}
 (\mu^{0i}+3\lambda_8^{00i}S) \phi_i + 3 \sum_{i,j} \lambda_8^{0ij}
 \phi_i \phi_j\Big|^2\\ \notag 
&\simeq \frac{3}{4}m^2\left(1-e^{-\sqrt{2/3}\,s}\right)^2\\  \label{LinearPot}
&\qquad +\frac{\sqrt{3}\, m
 \sinh(\sqrt{2/3}\,s)}{2(1+\tanh(s/\sqrt{6}))} \left[ 2 \sum_{i}
 (\mu^{0i}+3\sqrt{3}\lambda_8^{00i} \tanh{(s/\sqrt{6})}) \phi_i +
 3\sum_{i,j} \lambda_8^{0ij} \phi_i \phi_j \right] + {\rm h.c.}\,, 
\end{align}
where in the second line we have assumed that
$\mu^{0i},\lambda_8^{00i}\ll1$, in which case the singlet vevs are
approximately given by the solution of the system of equations 
\beq \label{LinearSol}
2\mu^{0j}S + 3\lambda_8^{00j}S^2  + 2 \sum_{k} (\mu^{jk}+
3\lambda_8^{0jk}S) \phi_k \simeq 0\,. 
\eeq
As was done in the previous Section, the system of equations
(\ref{LinearSol}) can be formally solved and substituted into
(\ref{LinearPot}) in order to obtain the $S$-dependent effective
inflationary potential. However, once again the resulting expression is
not particularly illustrative. Let us write schematically
$\lambda_8^{00i}S \sim \mu^{0i}\sim  \Lambda_1$ and $\lambda_8^{0ij}S
\sim \mu^{ij} \sim \Lambda_2$, so that $\langle \phi_i \rangle_{\rm inf}
\sim \Lambda_1/\Lambda_2$, and 
\beq\label{deltaV}
\Delta V_{\inf} \sim \frac{\sqrt{3}\, m
\sinh(\sqrt{2/3}\,s)}{2(1+\tanh(s/\sqrt{6}))}\,\frac{\Lambda_1^2}{\Lambda_2}
\sim m\frac{\sqrt{3} \Lambda_1^2}{8 \Lambda_2}e^{\sqrt{2/3}\,s}\,. 
\eeq
Fig.~\ref{fig:Vcorr1} shows the form of the scalar potential
(\ref{LinearPot}) as a function of $\Lambda_1^2/m\Lambda_2$,
demonstrating that the mixing parameters $\mu^{0i},\lambda_8^{00i}$ need
only be small compared to $(m\Lambda_2)^{1/2}$ in order to allow for
inflation. Fig.~\ref{fig:nsrcurve} displays the corresponding CMB
parameters $n_s$ and $r$, and compares them with the 68\% and 95\% CL
limits from Planck and other data~\cite{planck15}. As in the previous cases, we see that $r$ lies
within a factor $\sim 2$ of the Starobinsky prediction, far below the
current upper limit, whereas $n_s$ lies beyond the Planck 95\% CL range
for $\Lambda_1^2/m\Lambda_2 < \{10^{-3.3},10^{-3.4},10^{-3.5}\}$ for $N_* = \{50,N_*^{\rm max},60\}$. 


\begin{figure}[t!]
\centering
    \scalebox{0.5}{\includegraphics{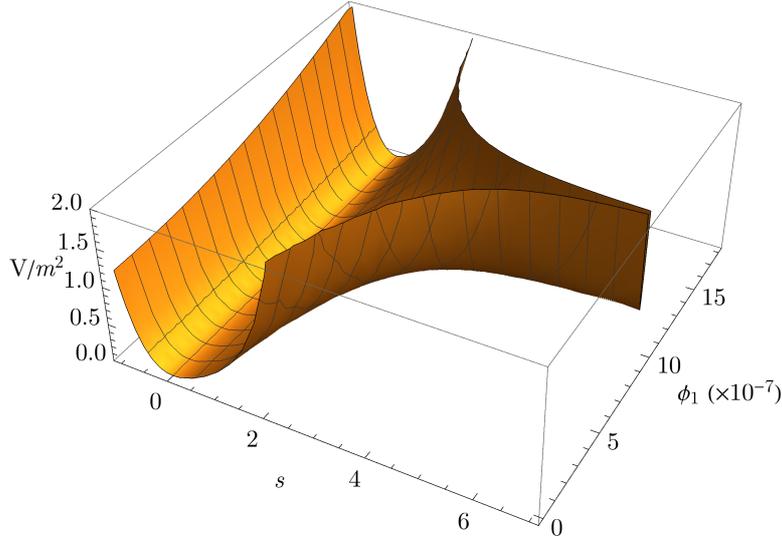}}
    \caption{\em The effective inflationary potential for a representative set of
    parameters with $\mu^{ij} \sim (0.1- 0.8) M_{\rm GUT}$,
 $\lambda_8^{0ij}$ and $\lambda_8^{ijk} \sim \pm(0.1 - 1)$ and
 $\mu^{0i}=\lambda_8^{00i}=10^{-6}$. The singlets $\phi_2$ and $\phi_3$
 are `integrated out' numerically for every value of $s$.   
  The evolution is seen to proceed to $s = 0$,
   yielding Starobinsky-like inflation. 
        }
    \label{fig:VcorrF}
\end{figure}

The curves shown in Fig.~\ref{fig:Vcorr1} must be taken with a pinch of
salt, as the approximation (\ref{deltaV}) is only an order-of-magnitude
estimate for the shape of the inflationary potential. In general, the
couplings $\mu^{ab}$ and $\lambda_8^{abc}$ will be unrelated, and the
corresponding effective potential can acquire a more complicated
structure. The potential can for example, begin to rise exponentially,
or develop a secondary minimum thus preventing the successful
realization of Starobinsky inflation.  In Fig.~\ref{fig:VcorrF}, we show
the form of the effective potential for an acceptable set of parameters
where we have taken $\mu^{0i}=\lambda_8^{00i}=10^{-6}$ and a representative set
of parameters in the range $\mu^{ij} \sim (0.1- 0.8) M_{\rm GUT}$,
$\lambda_8^{0ij}$ and $\lambda_8^{ijk} \sim \pm(0.1 - 1)$. We see a
simple valley structure in which the evolution of $s$ will lead to a
standard Starobinsky-like inflationary behavior, as discussed in more
detail in the next subsection.  Note that the evolution in
Fig.~\ref{fig:VcorrF} would appear to involve a change in direction of
the fields in field space. In principle, this could have a significant
effect on the final values of the anisotropy parameters $n_s$ and $r$
\cite{egno3}. However, the field remains
closely aligned with the instantaneous minimum and inflation proceeds as
in the single field case, although this is not apparent in the figure, because of the range of scales plotted. There is no significant production of isocurvature perturbations, even for singlets $\phi_i$ lighter than the inflaton, as the no-scale structure naturally constrains the width of the inflationary valley so that during inflation, their masses are much larger than the Hubble scale as we already saw in Eq. (\ref{heavy}) and
as illustrated in Fig.~\ref{fig:VcorrF}.

After inflation ends, $S$ and the other singlets $\phi_i$ undergo damped
oscillations. Assuming that during these oscillations the $\phi_i$
remain small, the evolution will track the instantaneous solution to the
system of equations 
\begin{align}\notag
\frac{\partial V}{\partial \bar{\phi}^i} &\simeq  4\sum_{j,k}
 \Big(\bar{\mu}_{ik}\mu^{kj} + 3\bar{\mu}_{ik} \lambda_8^{kj0}S +
 3\bar{\lambda}_{8\,ik0}\mu^{jk} \bar{S} + 9 |S|^2
 \lambda_8^{0jk}\bar{\lambda}_{8\,0ik} \Big)\phi_j  \\ \label{VgradCx} 
&+ 2S \sum_{j}\Big(2\bar{\mu}_{ij}\mu^{j0} +
 3\bar{\mu}_{ij}\lambda_8^{00j}S  + 6 \mu^{0j} \bar{\lambda}_{8\, 0ij} \bar{S}
+ 9\bar{\lambda}_{8\,0ij}\lambda_8^{00j} |S|^2 \Big)+  \cdots \, .
\end{align}
Similarly to the case studied in the previous Section, this
approximation may break down for $\lambda_8^{0ij}\gtrsim \mu^{ij}$. When
this is not the case, and if for simplicity we assume that
$\mu^{0i}\gtrsim \lambda_8^{00i}$~\footnote{Otherwise, if
$\mu^{0i}\lesssim \lambda_8^{00i}$, the result reduces to
(\ref{rhorat1}).}, the equations (\ref{VgradCx}) can be solved
trivially, resulting in the ratio 
\beq
\frac{\rho_{\phi_i}}{\rho_s} \sim m^{-2} \sum_{j}\bar{\mu}_{0j}\mu^{0j}\,.
\eeq
Therefore, reheating through inflaton decay requires $\mu^{0i} < 10^{-5}$.

\subsection{Numerical results}\label{sec:numerinf}

In subsections \ref{sec:singlst} and \ref{sec:nondiaginf} we have
discussed the analytical constraints on the superpotential parameters
$\mu^{0i}$ and $\lambda_8^{00i}$ that are imposed by the requirement of
successful Starobinsky-like inflation. We now show numerical results
that support these conclusions.  

\begin{figure}[!p]
\centering
	\hspace{0.3cm}\scalebox{0.68}{\includegraphics{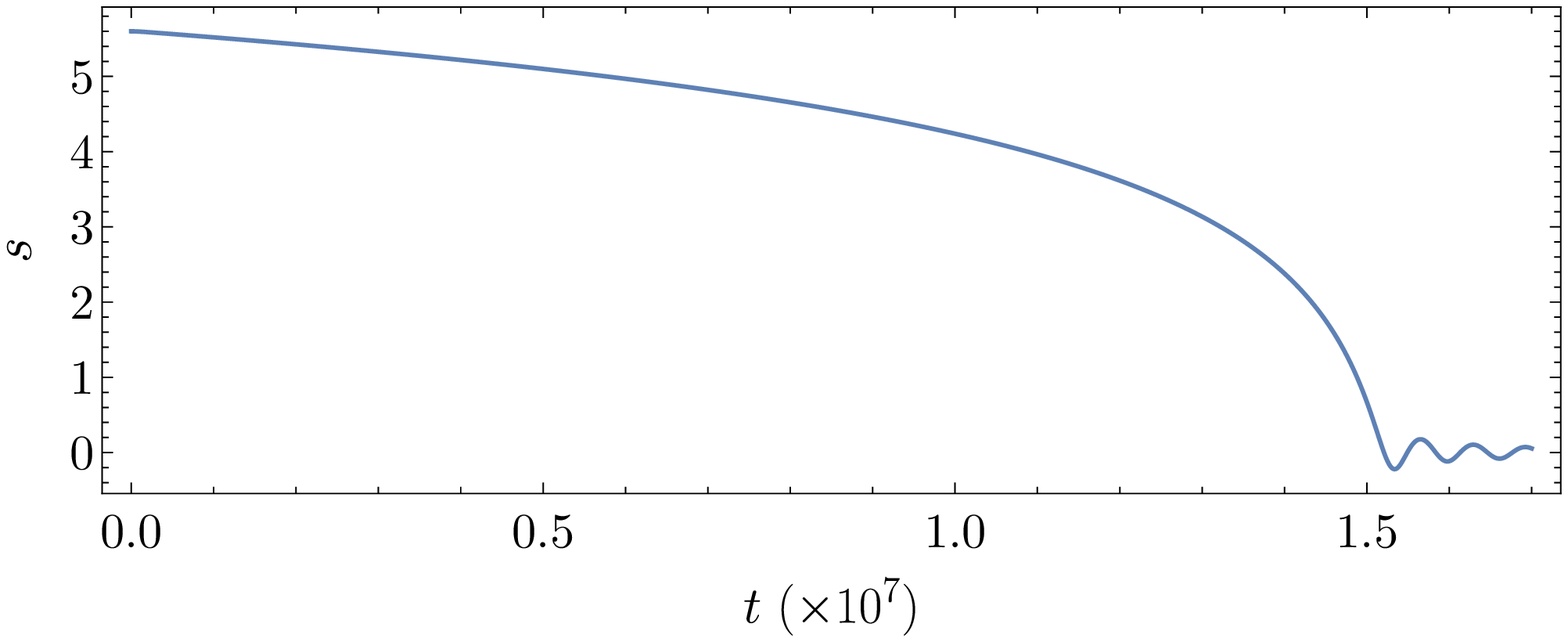}} \vspace{10pt}
	\vspace{5pt}
	\scalebox{0.70}{\includegraphics{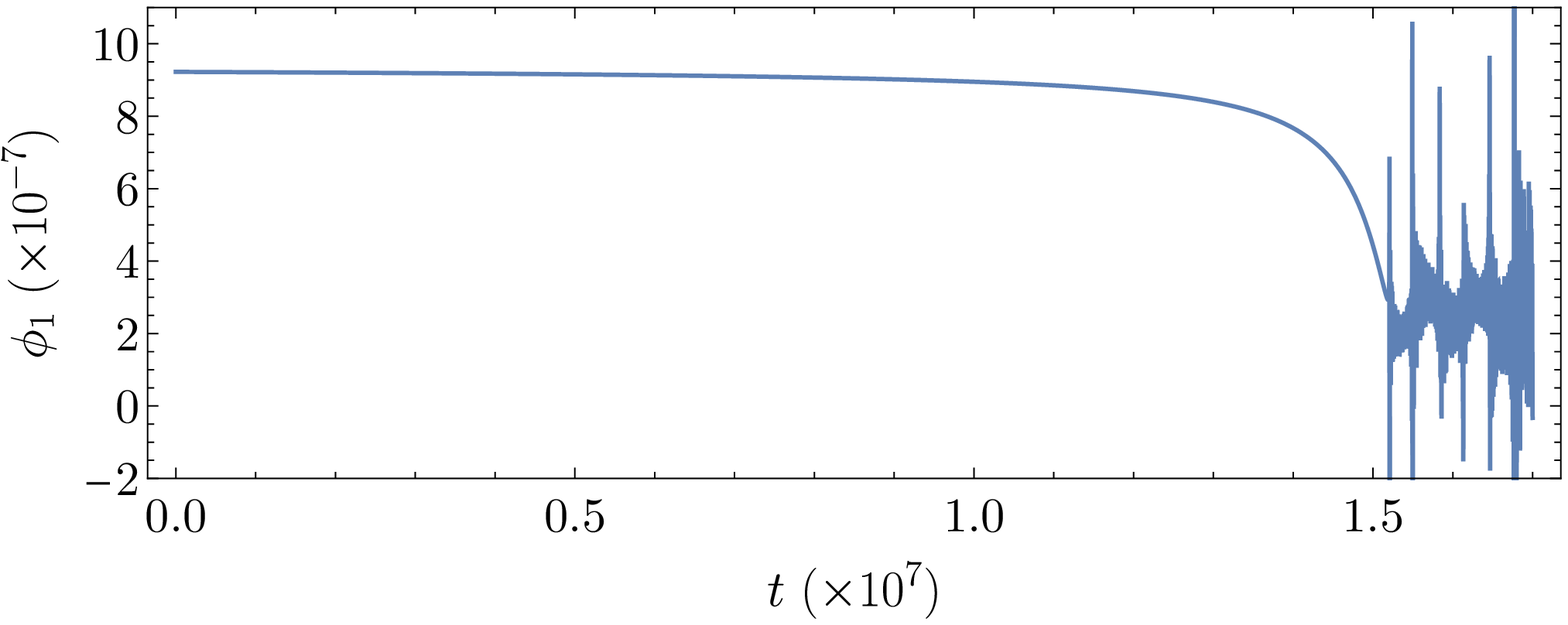}} \vspace{10pt}
	\vspace{5pt}
	\scalebox{0.70}{\includegraphics{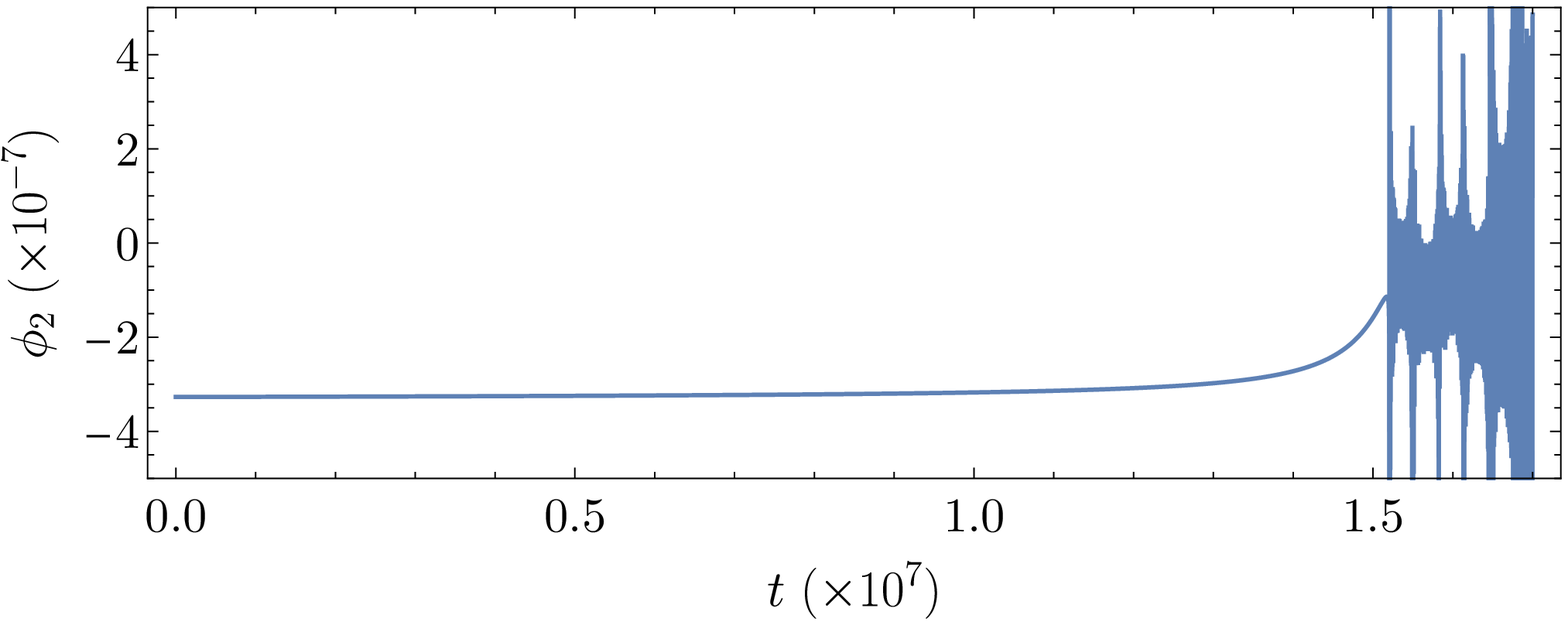}} 
	\caption{\it Evolution of the canonically-normalized inflaton
 $s$ and the SM singlet fields $\phi_1$ and $\phi_2$ during inflation,
 for $\mu^{0i}=\lambda_8^{00i}=10^{-6}$ and the same representative set of
 parameters as used in Fig.~\ref{fig:VcorrF}. The fields are assumed to
 start along the inflationary trajectory.}  
	\label{fig:sol1_1}
\end{figure} 

\begin{figure}[!p]
\centering
	\hspace{0.3cm}\scalebox{0.68}{\includegraphics{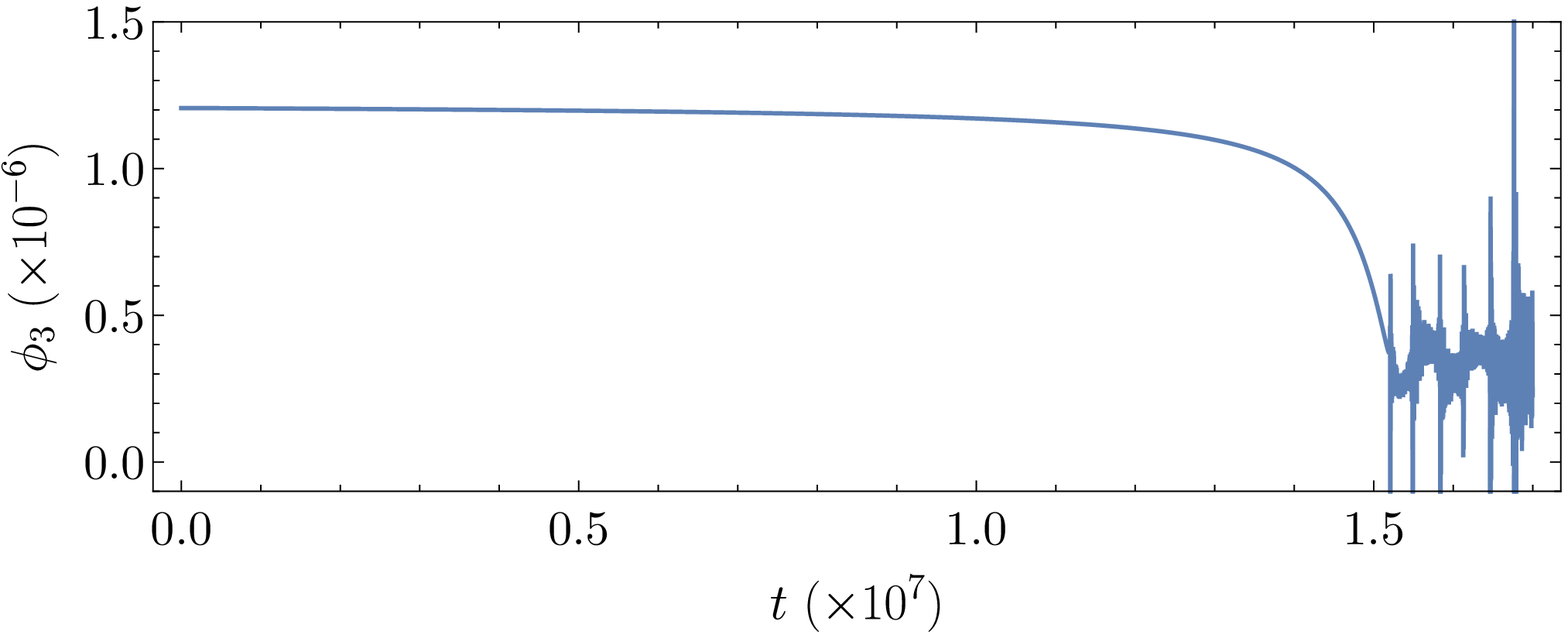}} \vspace{10pt}
	\vspace{5pt}
	\scalebox{0.70}{\includegraphics{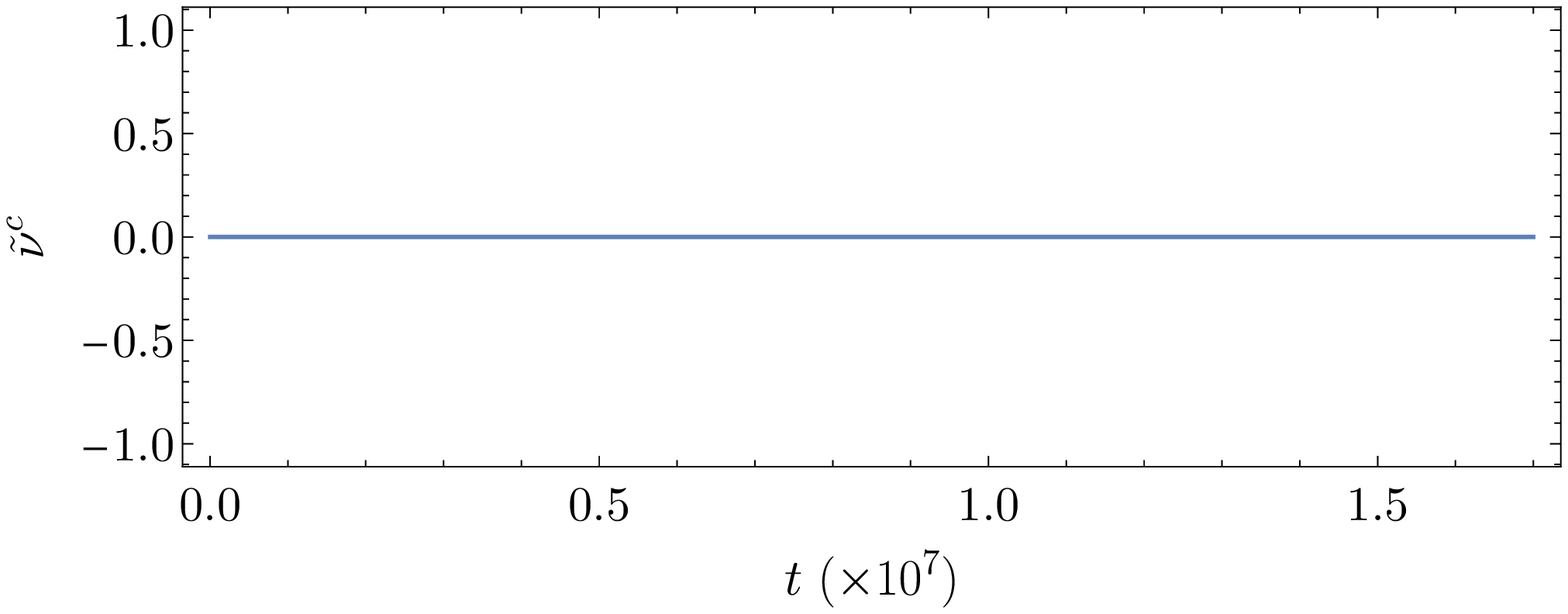}} \vspace{10pt}
	\vspace{5pt}
	\scalebox{0.70}{\includegraphics{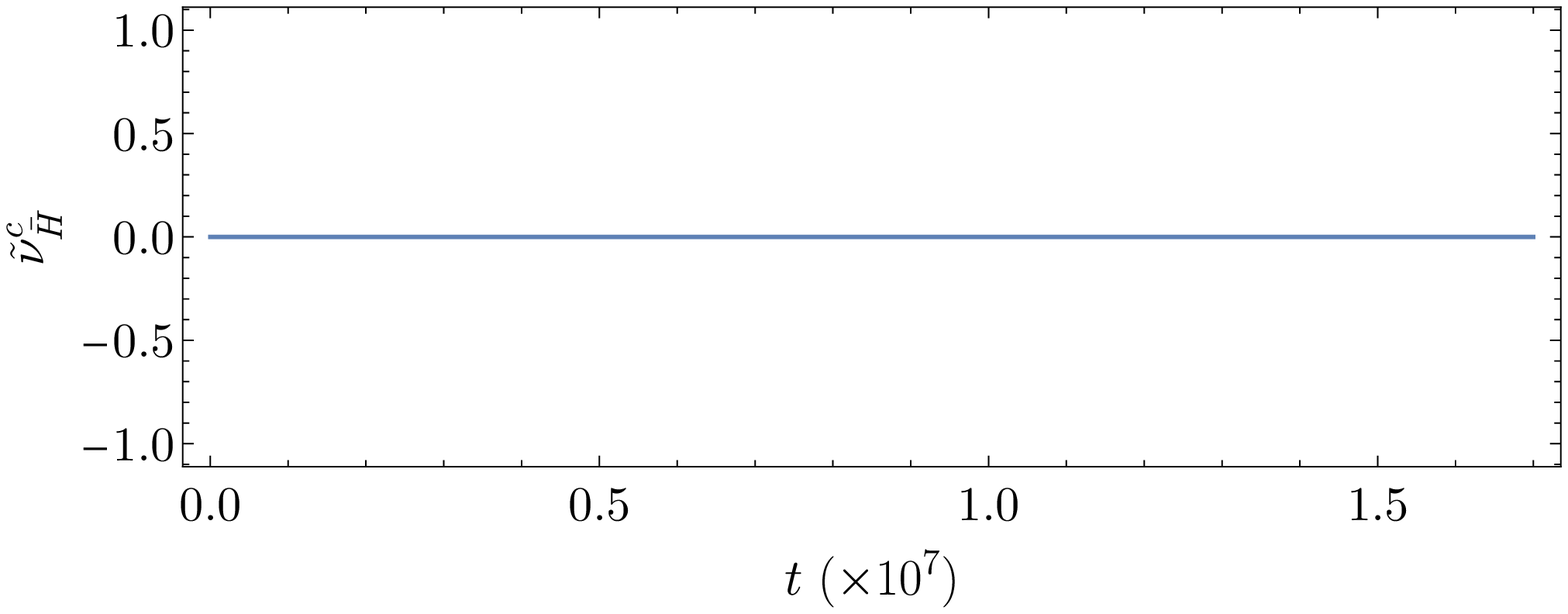}} 
	\caption{\it Evolution of the fields $\phi_3$, $\tilde{\nu}^c$
 and $\tilde{\nu}_{\bar{H}}^c$ during inflation, also for
 $\mu^{0i}=\lambda_8^{00i}=10^{-6}$ and the same representative set of
 parameters as used in Fig.~\ref{fig:VcorrF}. The fields are again
 assumed to start along the inflationary trajectory.} \label{fig:sol1_2} 
\end{figure}

\begin{figure}[t!]
\centering
	\scalebox{0.68}{\includegraphics{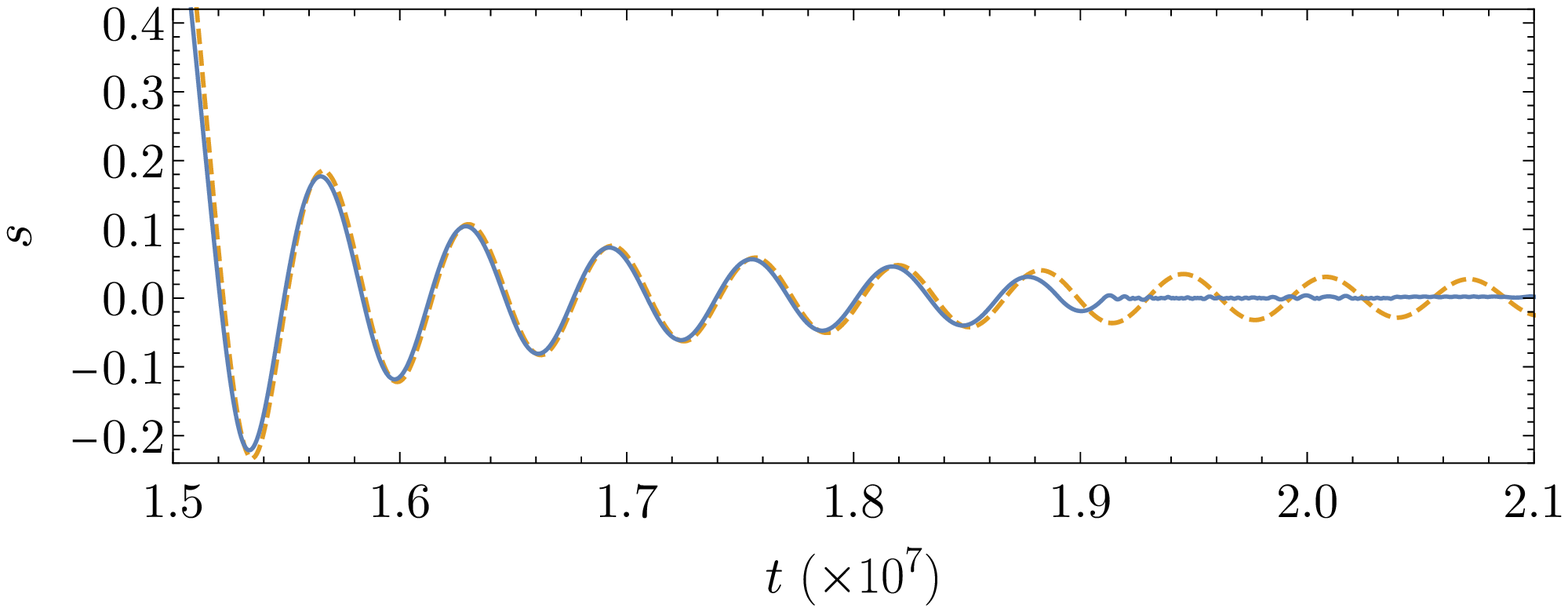}} \vspace{10pt}
	\subfloat{\scalebox{0.63}{\includegraphics{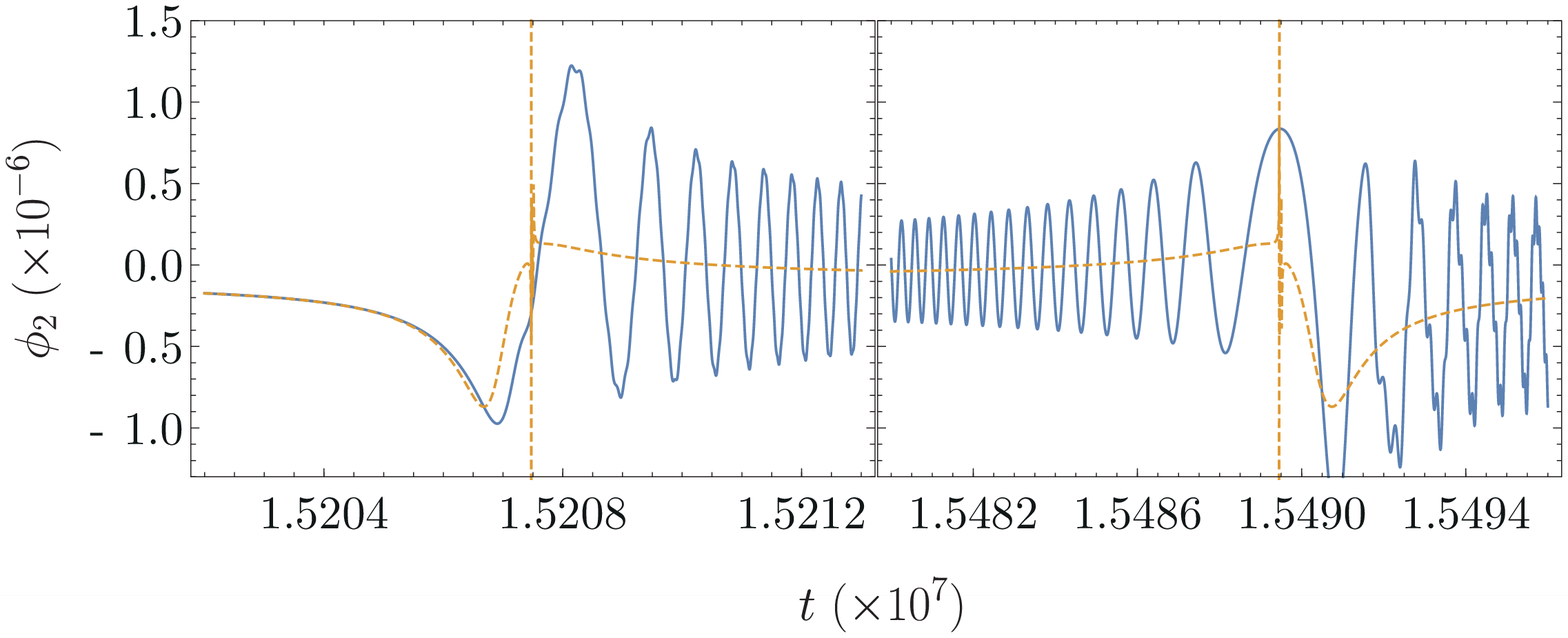}}}
	\caption{\it Evolution of the canonically-normalized inflaton
 $s$ and the SM singlet $\phi_2$ during reheating, for
 $\mu^{0i}=\lambda_8^{00i}=10^{-6}$ and the same representative set of
 parameters as used in Fig.~\ref{fig:VcorrF}. Upper panel: evolution
 of the inflaton field $s$ (blue, continuous), compared to the pure
 Starobinsky case (orange, dashed). Lower panel: evolution of the
 singlet field $\phi_2$ (blue, continuous), compared to the
 instantaneous solution of the equations $\partial_{\phi_i}V=0$ (orange,
 dashed). Notice the different horizontal scales in the two panels.}
 \label{fig:sol1_3} 
\end{figure}

Figures \ref{fig:sol1_1} and \ref{fig:sol1_2} show the time evolutions
of all the scalar singlet fields and $\tilde{\nu}^c$,
${\tilde\nu}^c_{\bar{H}}$ for $\mu^{0i}=\lambda_8^{00i}=10^{-6}$ and the same
representative set of parameters  as used in Fig.~\ref{fig:VcorrF}, assuming that
the fields start along the bottom of the inflationary `valley' in field
space. These results were obtained by numerical integration of the
supergravity equations of motion: 
\beq\label{eomfull}
\ddot{\Psi}^\alpha + 3H\dot{\Psi}^\alpha + \Gamma^\alpha_{\beta
\gamma}\dot{\Psi}^\beta\dot{\Psi}^\gamma + K^{\alpha
\bar{\beta}}\frac{\partial V}{\partial \bar{\Psi}^{\bar{\beta}}} \;=\;
0\,. 
\eeq
Here the indices run over all field components, with
$\Psi^\alpha\equiv\{T,\phi_a,\tilde{\nu}^c, \tilde{\nu}^c_{\bar{H}},\cdots\}$,
$K^{\alpha \bar{\beta}}$ denotes the inverse \kahler metric, and the
connection coefficients are given by  
\beq
\Gamma^\alpha_{\beta \gamma} = K^{\alpha\bar{\delta}}\partial_\beta K_{\gamma\bar{\delta}}\,.
\eeq
It is clear from the evolution of $s$ in Fig.~\ref{fig:sol1_1} that
Starobinsky-like inflation is realized. Figs.~\ref{fig:sol1_1} and
\ref{fig:sol1_2} show that after the end of inflation the singlets
$\phi_i$ undergo forced oscillations, with amplitudes much larger than
the expected values assuming an adiabatic tracking of the inflaton
value. This is an illustration of the phenomenon discussed around
(\ref{detcond}), namely that adiabaticity is violated when relatively
large values for the parameters $\lambda_8^{0ij}$ are chosen. As
Fig.~\ref{fig:sol1_3} demonstrates, after a few oscillations a
significant fraction of the energy stored in the inflaton oscillations
is transferred to the singlets $\phi_i$. The lower panel of
Fig.~\ref{fig:sol1_3} exhibits resonant enhancement of the $\phi_2$
amplitude when the solution of $\partial_{\phi_i}V=0$ is divergent, {\it
i.e.}, around the points where (\ref{VgradCx}) cannot be inverted. As
expected, the enhancement occurs when  
$s\ll 1$. This result confirms that strong segregation is a sufficient
condition for reheating to occur through the decay of $s$.

\begin{figure}[!p]
\centering
	\hspace{0.3cm}\scalebox{0.68}{\includegraphics{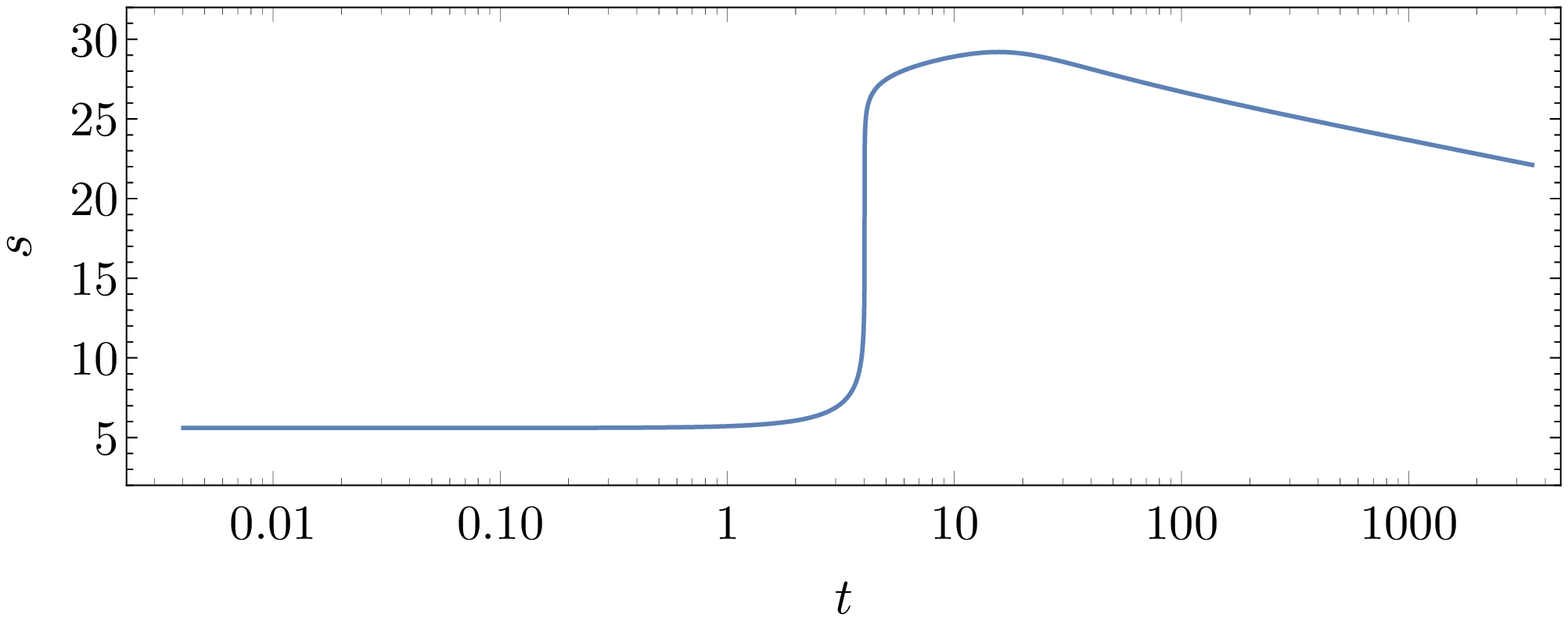}} \vspace{10pt}
	\vspace{5pt}
	\scalebox{0.70}{\includegraphics{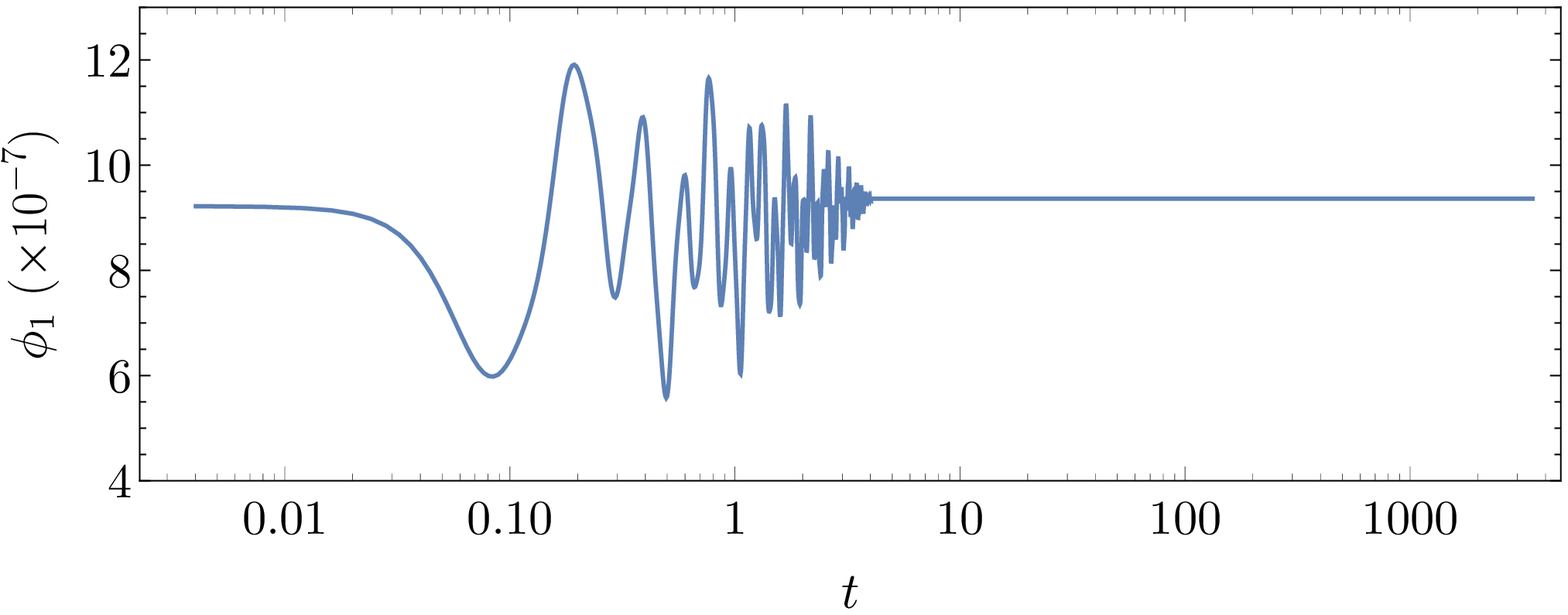}} \vspace{10pt}
	\vspace{5pt}
	\scalebox{0.70}{\includegraphics{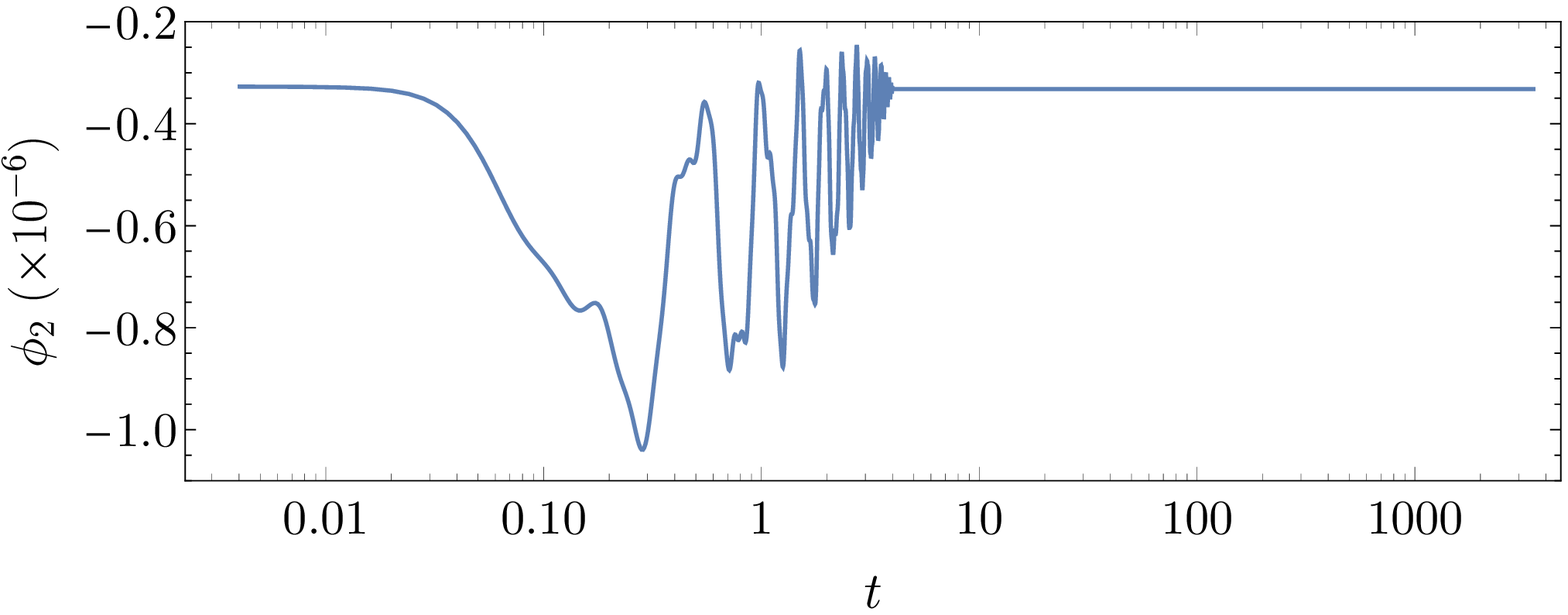}} 
	\caption{\it Evolution of the canonically-normalized inflaton
 $s$ and the SM singlets $\phi_1$ and $\phi_2$ during inflation, for
 $\mu^{0i}=\lambda_8^{00i}=10^{-6}$ and the same representative set of
 parameters as used in Fig.~\ref{fig:VcorrF}. The perturbed initial
 condition $\tilde{\nu}^c_{\bar{H}}=5\times10^{-3}$ is assumed.}
 \label{fig:sol2_1} 
\end{figure}

\begin{figure}[!p]
\centering
	\hspace{0.3cm}\scalebox{0.68}{\includegraphics{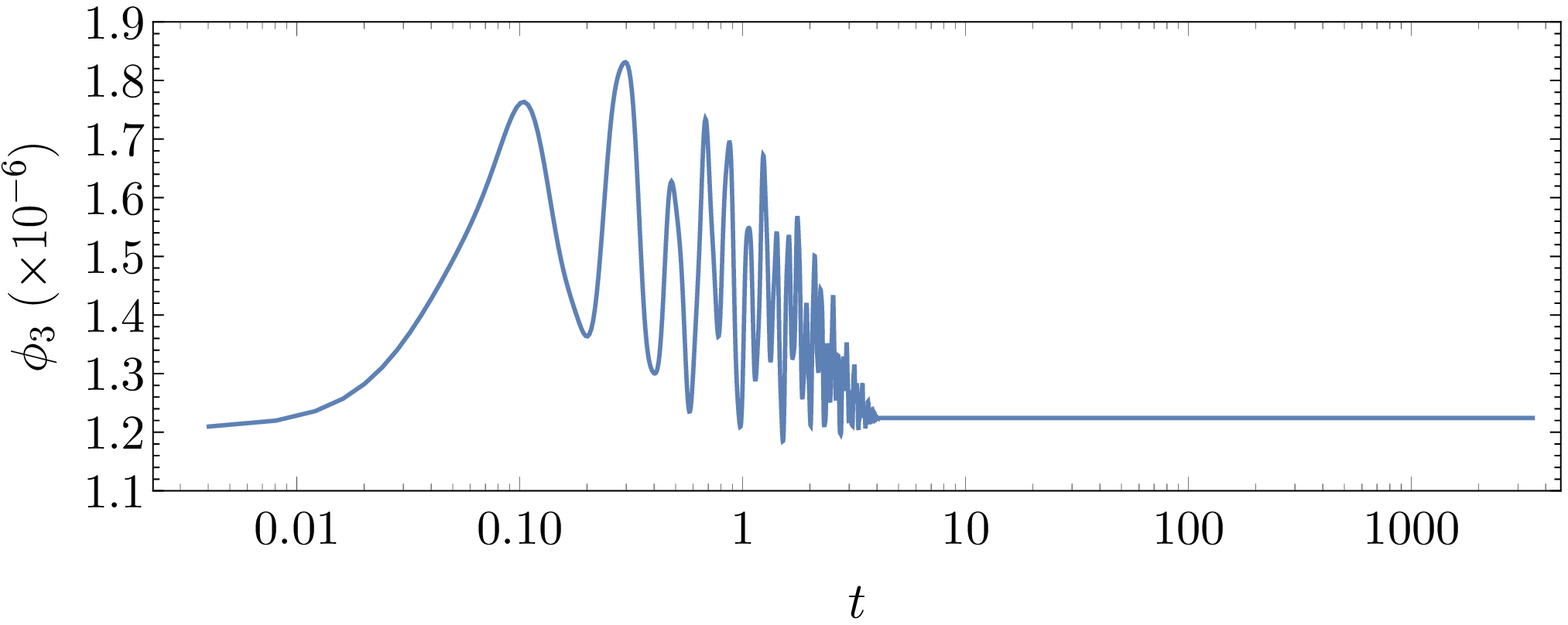}} \vspace{10pt}
	\vspace{5pt}
	\scalebox{0.70}{\includegraphics{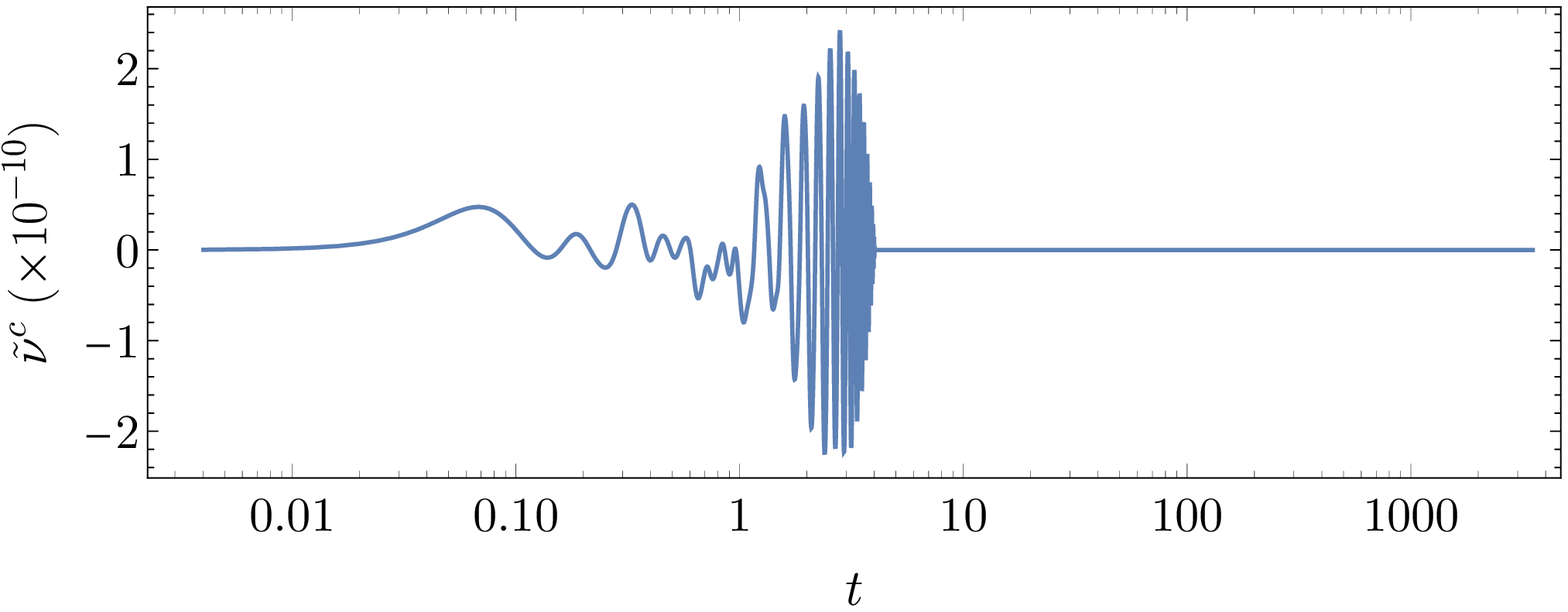}} \vspace{10pt}
	\vspace{5pt}
	\scalebox{0.70}{\includegraphics{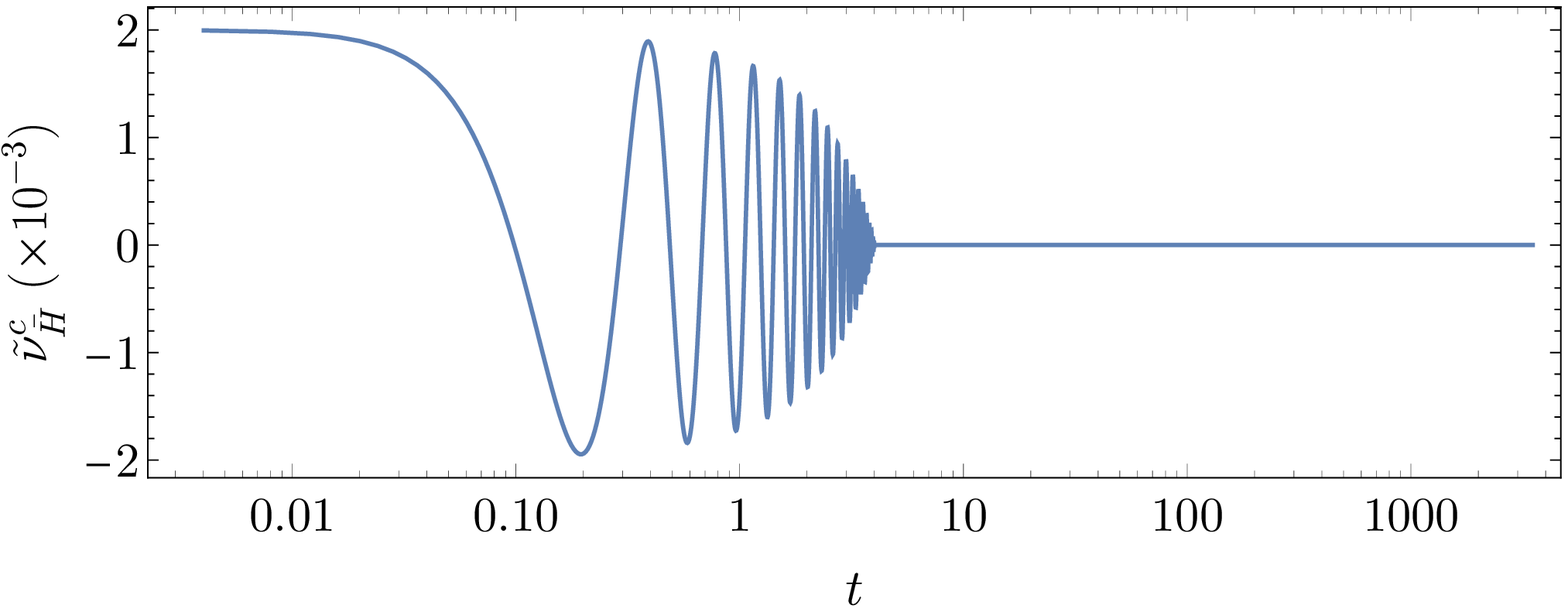}} 
	\caption{\it Evolution of the fields $\phi_3$, $\tilde{\nu}^c$
 and $\tilde{\nu}_{\bar{H}}^c$ during inflation, also for
 $\mu^{0i}=\lambda_8^{00i}=10^{-6}$ and the same representative set of
 parameters as used in Fig.~\ref{fig:VcorrF}. The perturbed initial
 condition $\tilde{\nu}^c_{\bar{H}}=5\times10^{-3}$ is again assumed.}
 \label{fig:sol2_2} 
\end{figure}

Figs.~\ref{fig:sol2_1} and \ref{fig:sol2_2} show the corresponding numerical
results for a solution with the same parameters, but with a perturbed
initial condition $\tilde{\nu}^c_{\bar{H}}=5\times10^{-3}$. This perturbation
seeds the oscillations of the remaining SM singlets, and drives an
uphill roll of the inflaton due to the connection-dependent terms in
(\ref{eomfull}), namely:  
\beq
\Gamma^S_{\alpha\beta}\dot{\Psi}^\alpha\dot{\Psi}^\beta \simeq
-\frac{1}{2\sqrt{3}}\,\sinh(\sqrt{2/3}\,s) \left(\dot{\phi}_1^2+
\dot{\phi}_2^2 + \dot{\phi}_3^2 + (\dot{\tilde{\nu}}^c)^2 +
(\dot{\tilde{\nu}}^c_{\bar{H}})^2 \right) + \cdots \, .  
\eeq
As the value of $s$ increases, the oscillations of the fields are
rapidly damped, and the subsequent evolution resembles that shown in
Fig.~\ref{fig:sol1_1} and Fig.~\ref{fig:sol1_2}, though with an
increased total number of $e$-folds, and Starobinsky-like values of $(n_s,r)$.

\begin{figure}[ht]
\centering
    \subfloat[$\mu^{ij}\lesssim 10^{-2},\ \lambda_8^{ijk}\lesssim 1$]{\scalebox{0.73}{\includegraphics{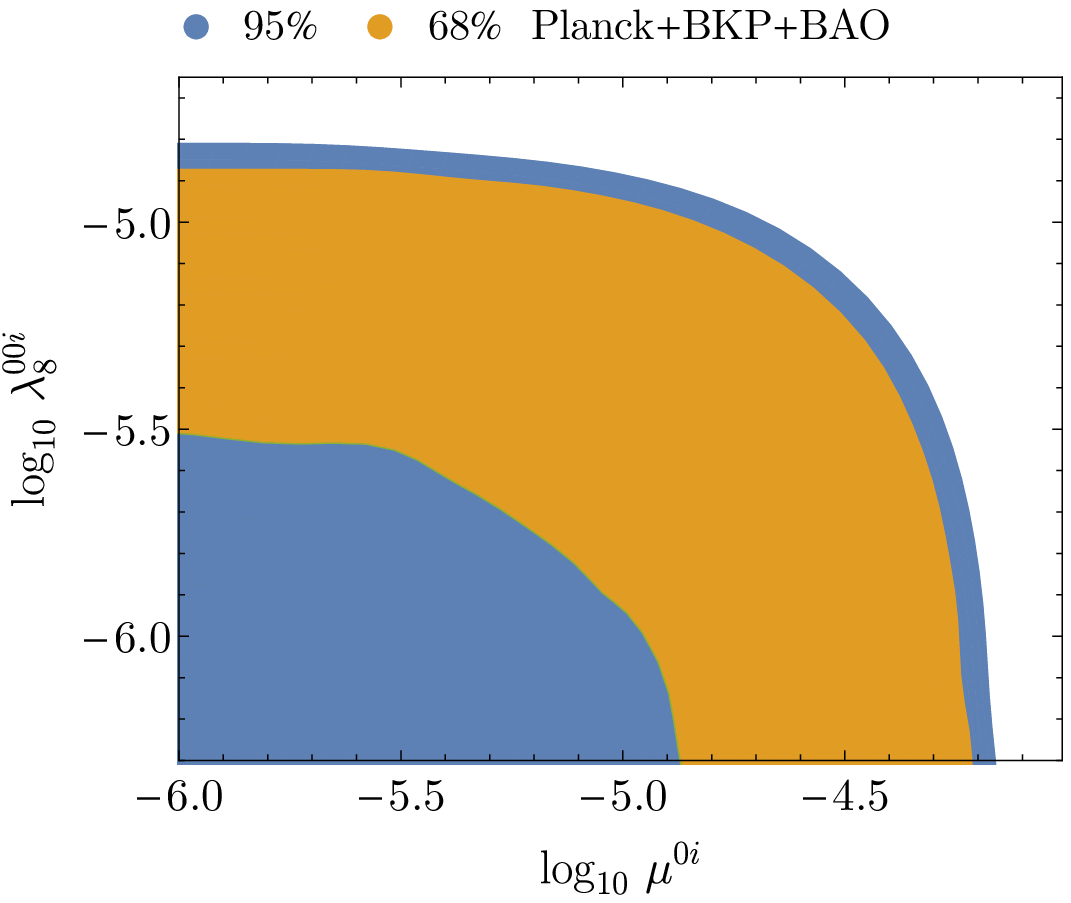}}\label{fig:f1}}
    \hfill
    \subfloat[$\mu^{ij}, \lambda_8^{ijk} \lesssim 10^{-2}$]{\scalebox{0.73}{\includegraphics{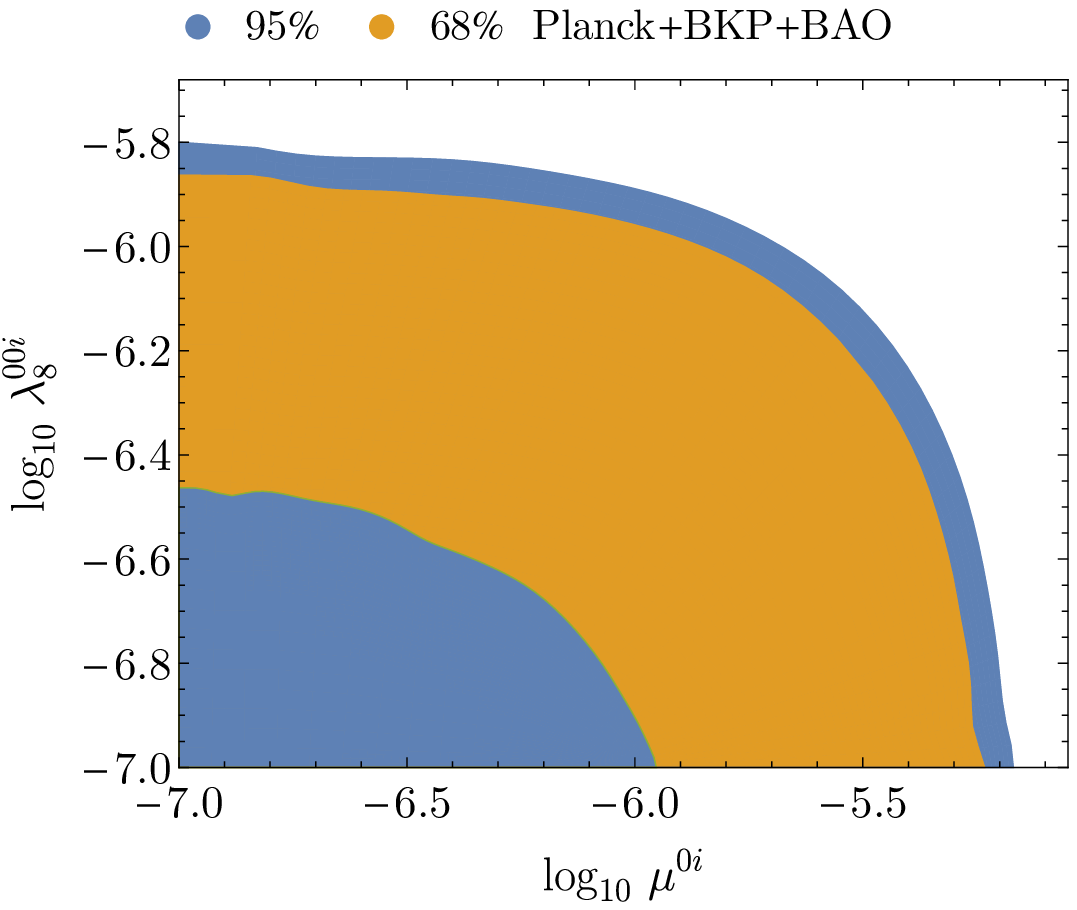}}\label{fig:f2}}
    \caption{\em The numerically-calculated 68\% and 95\% CL regions in the $(\mu^{0i},\lambda_8^{00i})$ plane at $N_*=50$ for the 
    no-scale model with superpotential (\ref{Wsinglets}). Here for simplicity we have assumed that $\mu^{0i}=\mu^{0i'}$ for any 
    $i,i'$, and similarly $\lambda_8^{00i} = \lambda_8^{00i'}$. 
        }
    \label{fig:Planck1}
\end{figure}

Fig.~\ref{fig:Planck1} shows the constraints from Planck and other data~\cite{planck15} on the model in
the $(\mu^{0i},\lambda_8^{00i})$ plane, calculated in a fully numerical
fashion. Due to the large number of independent parameters, we have made
the simplifying assumption $\mu^{0i}=\mu^{0i'}$ for any $i,i'$ (and
similarly $\lambda_8^{00i} = \lambda_8^{00i'}$). In the left panel the
constraints are calculated for the same representative set of parameters as used in
Fig.~\ref{fig:VcorrF}. We note that, {\it e.g.}, the upper limit for
$\mu^{0i}$ is reasonably close to the  analytical approximation
$\Lambda_1\sim(10^{-3.3}m\Lambda_2)^{1/2}$ obtained in
Fig.~\ref{fig:nsrcurve}. In the right panel, the constraints correspond
to the same representative set of parameters, but with the
$\lambda_8^{ijk} \rightarrow  10^{-2}\lambda_8^{ijk}$. Here the solution
is reasonably close to the analytical approximation
(\ref{deltaV}). We have also verified that the 68\% and 95\% CL contours
are unchanged under the assumption of multiple light states with
$\mu^{ij}\lesssim 10^{-5}$. 

\subsection{A symmetry argument for segregation}
\label{sec:rparity}

As we have seen above, the parameters that cause mixing between the
inflaton field $S$ and other singlet fields $\phi_i$, such as
$\lambda_8^{00i}$ and $\mu^{0i}$, must be strongly suppressed in order
to achieve successful inflation. In fact, such a suppression can
naturally be obtained if one adopts an appropriate definition of
$R$-parity. Let us assign odd $R$-parity to the matter superfields $F$,
$\bar{f}$, and $\ell^c$, and  even $R$-parity  to the Higgs multiplets $h$
and $\bar{h}$, as usual. We also assign $R$-parity even to $H$,
$\bar{H}$, and $S$, while $\phi_i$ are assumed to be $R$-parity
odd. With this assignment as well as the $\mathbb{Z}_2$ parity defined
in \eqref{eq:z2h}, we can forbid the following couplings:
\begin{equation}
 \lambda_6^{i0} = \lambda_7^i = \lambda_{8}^{00i} = \lambda_8^{ijk}
= \mu^{0i} = 0 ~.
\label{sym}
\end{equation}
Note that with this $R$-parity assignment, the inflationary potential
would be of the exact Starobinsky form given in \eqref{staropot}.
On the other hand, $\lambda_8^{0ij}$ couplings are allowed by this $\mathbb{Z}_2$
symmetry, but as we have seen in our previous numerical results, taking
these couplings of order unity does not adversely affect
inflation. Thus, this $R$-parity assignment would be sufficient for
obtaining the necessary strong segregation. We also note that this
$R$-parity is not broken by the vevs of $H$ and $\bar{H}$ since they are
$R$-parity even. Therefore, this $R$-parity is respected at low energies
unless $\phi_i$ acquire vevs.\footnote{
This $R$-parity assignment differs from that used in \cite{ehkno}. With this assignment
$R$-parity is exact unless some of the couplings in (\ref{sym}) are turned on. }

\section{Reheating Constraints}
\label{sec:reheating}

In this section, we study the neutrino mass structure in this model and
discuss its connection to reheating after inflation. As we see below,
neutrino mass terms are provided by Yukawa couplings in the
superpotential. In GUTs, these Yukawa couplings may be related to other
Yukawa couplings---thus, we first discuss Yukawa unification in our
model in Section~\ref{sec:yukawauni}. We then study the neutrino mass
structure in Section~\ref{sec:neutrinomass} and its connection to reheating
dynamics in Section~\ref{sec:singletdecay}.

\subsection{Yukawa unification}
\label{sec:yukawauni}

The Yukawa coupling terms in the low-energy effective theory of this
model are given by
\begin{equation}
 W_{\rm Yukawa} = f_u h_u Q\bar{u} + f_\nu h_u L\nu_R^c 
-f_d h_d Q \bar{d} -f_e h_d L \bar{e} ~,
\end{equation}
where we have suppressed generation indices for simplicity. In
ordinary SU(5) GUTs, we expect 
\begin{equation}
 f_d (M_{\rm GUT}) = f_e (M_{\rm GUT}) ~,
\label{eq:downlepyukuni}
\end{equation}
at the GUT scale. For the third generation (bottom and tau) Yukawa
couplings, this relation is satisfied at the ${\cal O}(10)$\% level. For
the first two generations, however, there are ${\cal O}(1)$
differences. Such deviations may be explained by means of
higher-dimensional operators suppressed by the Planck-scale
\cite{Ellis:1979fg} or higher-dimensional Higgs representations
\cite{Georgi:1979df} within the framework of SU(5) GUTs.

In the case of flipped ${\rm SU}(5) \times {\rm U}(1)$, on the other
hand, we have
\begin{equation}
 f_u (M_{\rm GUT}) = f_\nu (M_{\rm GUT}) ~,
\label{eq:yukawauniflip}
\end{equation}
as these two Yukawa couplings come from the same $\lambda_2$ term in
\eqref{Wgen}. In this case, the down-type Yukawa coupling $f_d$,
which is matched onto $\lambda_1$, is unrelated to the charged lepton
Yukawa coupling $f_e$, which originates from $\lambda_3$. Therefore, the
less successful prediction \eqref{eq:downlepyukuni}  for the first two
generations in ordinary SU(5) GUTs is not problematic in flipped ${\rm
SU}(5) \times {\rm U}(1)$ models. As we see below, even though we have
the unification condition \eqref{eq:yukawauniflip} for $f_\nu$, we can
explain the observed pattern of neutrino mass differences and mixing
angles by choosing $\lambda_6^{ia}$ and $\mu^{ab}$ appropriately.

\subsection{Neutrino masses}
\label{sec:neutrinomass}

Next, let us investigate the neutrino mass matrix in flipped ${\rm
SU}(5) \times {\rm U}(1)$. After $h$, $\bar{h}$, $H$, and $\bar{H}$
develop vevs, the Yukawa terms $\lambda_2$ and $\lambda_6$ lead to Dirac
mass terms for $\nu$, $\nu^c$, and $\tilde{\phi}_a$, where
$\tilde{\phi}_a$ denotes the fermionic component of $\phi_a$. 
If the singlet fields $\phi_a$ acquire vevs, the Higgsino may also mix with
right-handed neutrinos via the $\lambda_6^{ia}$ couplings, which results
in the $R$-parity violation. The $R$-parity violating effects may also
be induced by Higgsino-singlet mixing via the $\lambda^a_7$
couplings. Here, we focus on the following two cases where there is no (strong)
$R$-parity violation: (A) No singlet field develops a vev, $\langle
\phi_a \rangle = 0$, and the inflaton $S = \phi_0$ does not
participate in the neutrino mass generation. This setup is realized when
$\lambda_6^{i0} =\lambda_7^i = 0$ (though $\lambda_7^0$ is allowed). 
Note that this scenario requires at least four
singlets, our default assumption. (B) One of the $\phi_i$ fields
(denoted by $\phi_{i^\prime}$) acquires a non-zero vev.  If this field is responsible
for the $\mu$ term, its $R$-parity must be positive, and thus
does not couple to the neutrino sector. Instead of this
$\phi_{i^\prime}$, the inflaton $S$ plays a role in neutrino mass
generation. Thus only $\lambda_7^{i'}$ is non-zero and in particular, $\lambda_7^0 = 0$.
In the case of three singlets, since the $R$-parity of all three singlets must be negative, 
$\lambda_7 = 0$ for all three, and a GM term is necessary to produce a $\mu$ term.
Note that in this case, there is some $R$-parity violation due to the presence of 
both quadratic and cubic superpotential terms, though the $R$-parity violation is weakly
transmitted to the matter sector, and the lifetime of the lightest supersymmetric particle (LSP)
remains sufficiently long \cite{ENO8}.
In what follows, we study separately the neutrino mass structure and
reheating dynamics for these two scenarios (A) and (B).

\subsubsection{Scenario (A): inflaton decouples from the neutrino sector}

Here we assume $\langle \phi_a \rangle = 0$, which is achieved when
all of the $\mu^{ab}$ are much larger than the supersymmetry-breaking
scale. In this case, we need to introduce a Giudice--Masiero term to
obtain the MSSM Higgs $\mu$-term. We further assume 
\begin{equation}
 \lambda_6^{i0} = \lambda_7^i = \mu^{0i} = 0 ~,
\label{eq:conda}
\end{equation}
which is assured by the $R$-parity discussed in
Section~\ref{sec:rparity}. This prevents the fermionic component of
$\phi_0$ from mixing with neutrinos. The mass matrix for $\nu_i$,
$\nu^c_i$, and $\tilde{\phi}_i$ is then given by \cite{eln,Ellis:1992nq,Ellis:1993ks}
\begin{equation}
 {\cal L}^{(\nu)}_{\rm mass}=
-\frac{1}{2}
\left(
\begin{matrix}
{\nu}_i & {\nu}_i^c & \tilde{\phi}_i
\end{matrix}
\right) \left(
{\cal M}_\nu
\right)_{ij} \left(
\begin{matrix}
\nu_j \\ \nu_j^c \\ \tilde{\phi}_j
\end{matrix}
\right)
+ {\rm h.c.}~,
\end{equation}
with 
\begin{equation}
\left(
{\cal M}_\nu
\right)_{ij}\equiv
\begin{pmatrix}
 0 & \lambda_2^{ij}\langle \bar{h}_0\rangle & 0 \\
\lambda_2^{T ij}\langle \bar{h}_0\rangle & 0 & \lambda_6^{ij}\langle \tilde{\nu}_{\bar{H}}^c\rangle \\
0 & \lambda_6^{Tij}\langle \tilde{\nu}_{\bar{H}}^c\rangle & 2\mu^{ij}
\end{pmatrix}
~,
\label{eq:mnu}
\end{equation}
where we have used two-component notation. The mass matrix ${\cal
M}_\nu$ is a complex symmetric matrix and thus can be diagonalized with
a unitary matrix. By using $|\lambda_2^{ij} \langle \bar{h}_0 \rangle| \ll
|\lambda_6^{ij} \langle \tilde{\nu}_{\bar{H}}^c \rangle |, |\mu^{ij}|$, we
obtain the following mass matrix for the three light active neutrinos:
\begin{equation}
 {\cal M}_L \simeq \frac{2 \langle \bar{h}_0 \rangle^2}{\langle
  \tilde{\nu}_{\bar{H}}^c \rangle^2} \left[
\lambda_2 \left(\lambda_6^T\right)^{-1} \mu\, \lambda_6^{-1} \lambda_2^T
\right] ~.
\label{eq:ml}
\end{equation}
Thus, light neutrino masses can naturally be explained by the (double) seesaw
mechanism~\cite{Minkowski:1977sc,Georgi:1979dq}. Even though the structure of the
matrix $\lambda_2$ is related to the up-quark Yukawa matrix through the
unification relation \eqref{eq:yukawauniflip}, we still have a
sufficient number of degrees of freedom in the matrices $\lambda_6$ and
$\mu$, and thus can easily find a form of ${\cal M}_L$ that fits the
neutrino oscillation data. The couplings $\lambda_6^{ij}$ and $\mu^{ij}$
in general contain 
extra CP phases, and non-zero $\mu^{ij}$ cause lepton-number
violation. As a consequence, this model may explain baryon asymmetry of
the Universe via thermal leptogenesis \cite{fy,Giudice:2003jh}.\footnote{
As we discuss in Section~\ref{sec:infdecaya}, in this
scenario the inflaton does not decay directly into heavy neutrinos, leaving
thermal leptogenesis as a possibility. For this to occur, we need
a high reheating temperature, and thus the strong reheating case
discussed in Section~\ref{sec:strongreh} is favored in this scenario.}

Finally, the mass matrices of heavier states are given by
\begin{equation}
 {\cal M}_H = 
\begin{pmatrix}
 0 &  \lambda_6^{ij}\langle \tilde{\nu}_{\bar{H}}^c\rangle \\
 \lambda_6^{Tij}\langle \tilde{\nu}_{\bar{H}}^c\rangle & 2\mu^{ij}
\end{pmatrix}
~.
\label{eq:massheav}
\end{equation}
If furthermore, $ \lambda_6^{ij}\langle \tilde{\nu}_{\bar{H}}^c\rangle \ll \mu^{ij}$,
then the corresponding heavy mass eigenvalues are of order $
(\lambda_6^{ij}\langle \tilde{\nu}_{\bar{H}}^c\rangle)^2/\mu^{ij}$ and
$\mu^{ij}$.

\subsubsection{Scenario (B): The inflaton couples to the neutrino sector}

Next, we discuss the case where a combination of the singlet fields $\phi_i$,
called $\phi_{i^\prime}$, acquires a vev: $\langle\phi_{i^\prime}
\rangle \neq 0$. We assume that this singlet field does not have a
coupling to $F_i$, in order to suppress $R$-parity violation:
$\lambda_6^{i i^\prime} = 0$. In this case, the $\lambda_7^{i^\prime}$
term leads to the MSSM $\mu$-term, $\mu = \lambda_7^{i^\prime}
\langle\phi_{i^\prime} \rangle$. To obtain three massive active
neutrinos, we instead couple the inflaton field to $F_{i}$:
$\lambda_6^{i 0} \neq 0$. We then suppress the $\lambda_7^0$
coupling to avoid $R$-parity violation. 

For $i,j\neq i^\prime$, the neutrino mass matrix has the same structure
as ${\cal M}_\nu$ in \eqref{eq:mnu}, and thus light neutrino masses
for these generations are again given by
\eqref{eq:ml}~\footnote{Here, we neglect the effects of
$\lambda_6^{i0}$ and $\mu^{i0}$ for simplicity. The generalization to
non-zero $\lambda_6^{i0}$ and $\mu^{i0}$ is straightforward. We also
neglect mixing among generations, which is expected to be sizable
according to neutrino oscillation data, to simplify the expressions, but
the generalization is again straightforward. For more concrete
expressions, see Ref.~\cite{Ellis:1993ks}. }.
For $i=j=i^\prime$, on the other hand, the mass matrix is given by
\begin{equation}
 {\cal L}^{(i^\prime)}_{\rm mass}=
-\frac{1}{2}
\left(
\begin{matrix}
{\nu}_{i^\prime} & {\nu}_{i^\prime}^c & \tilde{S}
\end{matrix}
\right) 
\begin{pmatrix}
 0 & \lambda_2^{i^\prime i^\prime}\langle \bar{h}_0\rangle & 0 \\
\lambda_2^{i^\prime i^\prime}\langle \bar{h}_0\rangle & 0 &
 \lambda_6^{i^\prime 0}\langle \tilde{\nu}_{\bar{H}}^c\rangle \\
0 & \lambda_6^{i^\prime 0}\langle \tilde{\nu}_{\bar{H}}^c\rangle & m
\end{pmatrix}
 \left(
\begin{matrix}
\nu_{i^\prime} \\ \nu_{i^\prime}^c \\ \tilde{S}
\end{matrix}
\right)
+ {\rm h.c.}~,
\end{equation}
where $\tilde{S}$ denotes the fermionic partner of the inflaton $S$. The
light mass eigenvalue for this mass matrix is then given by
\begin{equation}
 m_{\nu_{i^\prime}} \simeq \frac{m\left(\lambda_2^{i^\prime
i^\prime}\langle \bar{h}_0\rangle\right)^2}{\left(\lambda_6^{i^\prime
0} \langle \tilde{\nu}^c_{\bar{H}} \rangle \right)^2} ~,
\label{eq:linumassesb}
\end{equation}
while the heavier eigenvalues have masses 
\begin{equation}
 m_{N_{i^\prime 1,2}} = \frac{1}{2}\left[
m \mp
\sqrt{\left(2 \lambda_6^{i^\prime 0}\langle
       \tilde{\nu}_{\bar{H}}^c\rangle\right)^2 + m^2}\,
\right] ~,
\label{eq:masseig12}
\end{equation}
where $N_{i^\prime 1}$ and $N_{i^\prime 2}$ are $\nu_{i^\prime}^c$-
and $\tilde{S}$-like states, respectively. For $m \ll \lambda_6 \langle
\tilde{\nu}_{\bar{H}}^c\rangle$, these two states form a pseudo-Dirac
state with mass $\lambda_6 \langle \tilde{\nu}_{\bar{H}}^c\rangle$ with
splitting of order $m$. It is interesting to note the role played by the
inflaton mass, $m$ for neutrino masses~\footnote{The parameter $m$ is
inflaton mass during inflation and reheating when the GUT symmetry
remains exact.  After GUT symmetry breaking the scalars associated with
the inflaton multiplet receive GUT scale masses proportional to 
$\lambda_6  \langle \tilde{\nu}^c_{\bar{H}} \rangle$.}.
The light (mostly left-handed) neutrino masses are proportional to the
inflaton mass, whilst the heavy state masses are split by the inflaton
mass.

The part of the superpotential relevant for the $\nu_{i^\prime}^c$
and ${S}$ couplings can be written as
\begin{equation}
 W = \lambda_2^{i^\prime j} \nu_{i^\prime}^c {L}_j h_u
+\lambda_6^{i\prime 0} \nu^c_{i^\prime} \nu_{\bar{H}}^c S
+ \frac{m}{2}S^2 ~.
\label{eq:wnus}
\end{equation}
Rotating the ${\nu}_{i^\prime}^c$ and ${S}$ fields into the mass
eigenstates: 
\begin{equation}
 \begin{pmatrix}
 {N}_{i^\prime 1} \\ {N}_{i^\prime 2}
 \end{pmatrix}
=
\begin{pmatrix}
 \cos \theta & \sin \theta \\
 -\sin \theta & \cos \theta 
\end{pmatrix}
\begin{pmatrix}
 \tilde{\nu}^c_{i^\prime} \\ S
\end{pmatrix}
~,
\end{equation}
with 
\begin{equation}
 \tan 2\theta = - \frac{2 \lambda_6^{i^\prime 0} \langle
  \tilde{\nu}^c_{\bar{H}} \rangle}{m} ~,
\end{equation}
the superpotential \eqref{eq:wnus} can then be expressed as
\begin{equation}
 W = 
\lambda_2^{i^\prime j} \left(\cos\theta N_{i^\prime 1} -\sin\theta
			N_{i^\prime 2}\right) {L}_j h_u
+ \frac{1}{2}m_{N_{i^\prime 1}} N_{i^\prime 1}^2
+ \frac{1}{2}m_{N_{i^\prime 2}} N_{i^\prime 2}^2
~,
\label{eq:winn12}
\end{equation}
where the masses $m_{N_{i^\prime 1,2}}$ are given in \eqref{eq:masseig12}.

As we see below, the neutrino mass structure in this scenario is
restricted by the constraint on the reheating temperature. We will
discuss the compatibility of this constraint with the observed neutrino
oscillation data in Section~\ref{eq:singletdecayb}.

\subsection{Singlet decays}
\label{sec:singletdecay}

Now we consider inflaton decay in the two scenarios discussed in the
previous subsection. In Scenario (A), the inflaton does not couple to
the neutrino sector, so at the tree level it can decay only into Higgs bosons and
Higgsinos. In Scenario (B), on the other hand, the inflaton $S$ does
couple to right-handed neutrinos but its coupling to the MSSM Higgs
fields is suppressed in order to evade $R$-parity violation. We will
find that there is a tight connection between neutrino masses and
reheating dynamics in this case.

\subsubsection{Scenario (A): inflaton decay into Higgs/Higgsino}
\label{sec:infdecaya}

Assuming \eqref{eq:conda}, the superpotential couplings
relevant to the inflaton decay are given by
\begin{equation}
 W_{S\, {\rm decay}} = 
\lambda_1^{ij} F_iF_jh + \lambda_2^{ij} F_i\bar{f}_j\bar{h} +
 \lambda_3^{ij}\bar{f}_i\ell^c_j h +
\lambda_7^0 S h\bar{h} + 3\lambda_8^{0ij} S \phi_{i} \phi_j 
+\frac{m}{2}S^2 ~.
\label{eq:wsdecay}
\end{equation}
If the mass eigenvalues of the mass matrix \eqref{eq:massheav} are
larger than the inflaton mass $m$, then the inflaton decay via the
couplings $\lambda_8^{0ij}$ is suppressed by a small light-heavy neutrino
mixing angle of ${\cal O}(\lambda_2 \langle \bar{h}_0 \rangle /(\mu, \lambda_6
\langle \tilde{\nu}_{\bar{H}}^c \rangle))$, and thus is negligible. The
$\lambda_7^0$ coupling gives rise to the inflaton-Higgs/Higgsino
interactions: 
\begin{equation}
 {\cal L}_{\rm int} = -\frac{\lambda_7^0}{\sqrt{2}} s \tilde{h}_u \tilde{h}_d 
- \frac{m^* \lambda_7^0}{\sqrt{2}} s h_u h_d + {\rm h.c.} ~,
\end{equation}
which yields the following singlet decay rate:
\begin{equation}
 \Gamma (s \to \tilde{h}_u \tilde{h}_d) = 
 \Gamma (s \to h_u h_d) \simeq
\frac{|\lambda_7^0|^2}{8\pi} |m| ~.
\end{equation}
The cross terms between the $\lambda_{1,2,3}$ and $\lambda_7$ terms in
\eqref{eq:wsdecay} also induce singlet-sfermion couplings. These
couplings give rise to either three-body decay or two-body decay
suppressed by the Higgs vev $\langle h^0 \rangle$. Hence, these sfermion
decay channels are sub-dominant.

As we discuss in Section~\ref{sec:reheat}, an upper limit on the inflaton decay
rate is given by the over-production of gravitinos, which restricts the
coupling $\lambda_{7}^0$ as~\footnote{The constant $\Delta \ge 1$ parametrizes any dilution of the gravitino relic density posterior to reheating, due to the entropy increase produced by the decay of a long-lived particle (see Section~\ref{sec:entropygrav}).}
\begin{equation}
 |\lambda_7^0| \lesssim 10^{-5}\Delta  ~,
\label{eq:conslam7}
\end{equation}
though it could be substantially smaller. 
Since the inflaton plays no role in the neutrino mass generation, this
limit has no implication for the neutrino mass structure, contrary to
Scenario (B) discussed below.

\subsubsection{Scenario (B): inflaton decay into neutrinos}
\label{eq:singletdecayb}

In this case, the $\lambda_7^0$ coupling is set to be zero to avoid
$R$-parity violation. Thus, the Higgs/Higgsino decay modes of the
inflaton are suppressed. Instead, the inflaton couples to the neutrino
sector through the $\lambda_6^{i^\prime 0}$ coupling, and thus can decay
into a lepton and a Higgsino, or a slepton and a Higgs boson. Other
decay channels are three-body decay processes or those dependent on a
small vev, $\langle \phi_{i^\prime} \rangle$ or $\langle h^0 \rangle$,
and are thus subdominant. 

The relevant interactions are readily obtained from the superpotential
\eqref{eq:winn12}, from which we evaluate the decay rates of the
$s$-like state ${\rm Re}(\sqrt{2} N_{i^\prime 2})$:
\begin{equation}\label{eq:stoLH}
 \Gamma (s \to L_j \tilde{h}_u) = \Gamma (s \to \tilde{L}_j h_u)
\simeq \frac{|\lambda_2^{i^\prime j} \sin \theta |^2}{8\pi} m_{N_{i^\prime 2}}
~. 
\end{equation}
We then obtain a constraint on $\sin \theta$ as in
\eqref{eq:conslam7}: 
\begin{equation}
 |\lambda_2^{i^\prime j} \sin \theta | \lesssim 10^{-5}\Delta ~.
\label{eq:conssinlam2}
\end{equation}

We now discuss the implication of this constraint for light neutrino
masses. We work on the basis where $\lambda_2$ is diagonalized. We
first consider the case $i^\prime = 3$, where $\lambda_2^{33} \simeq
m_t/\langle \bar{h}^0\rangle \simeq 1$. In this case, the constraint
\eqref{eq:conssinlam2} leads to 
\begin{equation}
  10^{-5}\Delta \gtrsim |\sin \theta | \simeq \left|
\frac{\lambda_6^{30} \langle \tilde{\nu}^c_{\bar{H}}\rangle}{m}
\right| ~.
\end{equation}
With \eqref{eq:linumassesb}, this bound gives
\begin{equation}
 m_{\nu_3} \gtrsim 10^{10} \cdot\frac{m_t^2}{m}\Delta^{-2} \sim 10~{\rm GeV}\, \Delta^{-2} ~,
\end{equation}
which, in the absence of significant entropy production, is much larger than the current experimental limit from
the Lyman $\alpha$ forest power spectrum obtained by BOSS in combination
with the Planck 2015 CMB data \cite{Palanque-Delabrouille:2015pga}:
$\sum_\nu m_\nu < 0.12$~eV. 

In the $i^\prime = 2$ case, although the bound is
relaxed by a factor of $10^{-4}$, the resultant neutrino mass value
is still above this limit. 
In the $i^\prime = 1$ case, however, $\lambda_2^{11} \simeq m_u
/\langle \bar{h}^0\rangle \simeq 10^{-5}$, and thus the constraint
\eqref{eq:conssinlam2} gives no limit on the mixing angle $\theta$. In
this case, the neutrino mass is given by
\begin{equation}
 m_{\nu_1} \simeq 10^{-9} \times 
\left(
\frac{m}{3\times 10^{13}\,{\rm GeV}}
\right)
\left(
\frac{|\lambda_6^{10}|}{10^{-3}}
\right)^{-2}
\left(
\frac{|\langle \tilde{\nu}^c_{\bar{H}} \rangle|}{10^{16}~{\rm GeV}}
\right)^{-2} ~{\rm eV}~,
\label{eq:mnu1}
\end{equation}
which evades the experimental limit.

Recent global fits to neutrino oscillation data give
\cite{Esteban:2016qun}
\begin{align}
 |\delta m^2|&\equiv |m_{\nu_2}^2 -m_{\nu_1}^2|\simeq 7.4\times
 |10^{-5}\,{\rm eV}^2~, \nonumber\\ 
 |\Delta m^2|&\equiv |m_{\nu_3}^2-(m_{\nu_2}^2+m_{\nu_1}^2)/2 |
\simeq 2.5\times 10^{-3}\,{\rm eV}^2~.
\end{align}
These values as well as the result in \eqref{eq:mnu1} indicate that,
unless $|\lambda_6^{10} \langle \tilde{\nu}^c_{\bar{H}} \rangle|$ is extremely
small, a Normal Hierarchy (NH) mass spectrum, {\it i.e.}, $m_{\nu_1}
\ll m_{\nu_2} < m_{\nu_3}$, is favored in this model. The other light
neutrino masses in this case are predicted to be 
\begin{align}
 m_{\nu_2} &\simeq |\delta m^2|^{\frac{1}{2}} \simeq 9 \times 10^{-3}~
 {\rm eV} ~, \nonumber \\
 m_{\nu_3} &\simeq |\Delta m^2|^{\frac{1}{2}} \simeq 5 \times 10^{-2}
 ~{\rm eV} ~.
 \label{numasses}
\end{align}
We can easily obtain these values by choosing appropriately $\mu$ and
$\lambda_6$ in \eqref{eq:ml}: reheating does not impose significant
restrictions for these two generations.

\section{Post-Inflation}
\label{sec:phasetr}

\subsection{Reheating}
\label{sec:reheat}

The temperature of the Universe following inflation depends on the inflaton decay rate, which we parameterize as
$\Gamma_{s}=|y|^2m/8\pi$. For case (A), $y = \sqrt{2} \lambda_7^0$, and for case (B),
$y = \sqrt{2} \lambda_2^{i^\prime j} \sin \theta$.   This decay rate is bounded by the upper limit on the 
density of gravitinos produced in the relativistic plasma arising from the inflaton decay products \cite{weinberg,elinn,nos,ehnos,kl,ekn,Juszkiewicz:gg,mmy,Kawasaki:1994af,Moroi:1995fs,enor,Giudice:1999am,bbb,kmy,stef,Pradler:2006qh,ps2,rs,kkmy,EGNOP}. 
Big-Bang Nucleosynthesis (BBN) imposes tight constraints on the decay rate of the inflaton for small gravitino masses \cite{kkmy,kmy,bbn,Kawasaki:1994af,kkm,ceflos,stef}.
However, if one assumes that the gravitino is sufficiently heavy to decay before BBN \cite{kkm,ceflos}, 
the dominant bound on its abundance comes from its contribution to the cold dark matter relic density. 
Assuming a present dark matter density $\Omega_{\rm cold}h^2=0.12$~\cite{planck15} and a standard thermal history with no post-reheating entropy production, we find the following constraint
on the inflaton decay coupling $y$~\cite{EGNOP}:
\beq
|y| < 2.7\times 10^{-5} \left( 1 + 0.56 \frac{m_{1/2}^2}{m_{3/2}^2} \right)^{-1}\left(\frac{100\,{\rm GeV}}{m_{\rm LSP}}\right)\, ,
\label{700}
\eeq
implying that $\Gamma_s \lesssim 900$~GeV for $m = 3 \times 10^{13}$ GeV. 
When the limit on $y$ is saturated, 
the relic density of the LSP is obtained by the non-thermal decay of the gravitino.
For smaller $y$, the LSP abundance from decay is reduced and other mechanisms
(such as freeze-out or coannihilations) must be operating so as to give the correct cold dark matter density.

Equivalently, in terms of the reheating temperature,
\beq\label{TrehC}
T_{\rm reh} \;\equiv\; \left(\frac{40}{g_{\rm reh}\pi^2}\right)^{1/4}\left(\Gamma_{s}M_P\right)^{1/2} \;\lesssim\; \left(\frac{915/4}{g_{\rm reh}}\right)^{1/4} (1.7\times 10^{10}\,{\rm GeV})\,,
\eeq
where $g_{\rm reh}$ is the number of relativistic degrees of freedom at $T_{\rm reh}$. {We use here the reference value $g_{\rm MSSM}=915/4$ instead of $g_{\text{SU}(5)\times \text{U}(1)}=1545/4$, 
as most of the difference is due to heavy fields, whose production is kinematically forbidden.} However, as is well known, the reheating temperature does not constitute an upper bound on 
the effective instantaneous temperature $T$, 
as higher effective temperatures can be reached during the reheating process \cite{ckr,Giudice:1999am,EGNOP}. Using the relation
\beq
T=\left(\frac{30 \rho_{\gamma}}{\pi^2g(T)}\right)^{1/4}\,,
\eeq
where $\rho_\gamma$ denotes the instantaneous energy density of the relativistic decay products,
an effective instantaneous temperature during reheating may be defined. 
This leads to a maximum temperature of the dilute plasma shortly after the start of inflaton decay, 
which may be written as
\beq\label{Tmax}
T_{\rm max} \;\simeq\; 0.74\left(\frac{\Gamma_{s}m M_P^2}{g_{\rm max}}\right)^{1/4} \;\lesssim\; \left(\frac{915/4}{g_{\rm max}}\right)^{1/4} (3.8\times 10^{12}\,{\rm GeV})\,,
\eeq
where the inequality is due to the gravitino production bound.  As discussed in~\cite{EGNOP}, 
because of the finite rate of the thermalization process, the actual maximum temperature of the Universe is in the range
$T_{\rm max}\gtrsim T > T_{\rm reh}$.

The previous constraints on the decay rate of the inflaton, and the maximum temperatures during and after reheating, change if there is an intermediate matter-dominated phase between the end of reheating and the end of the radiation-dominated era. We consider this effect in Section~\ref{sec:entropygrav}.

\subsection{Supercosmology and The GUT phase transition}

The maximum temperature during reheating is typically a few orders of
magnitude lower than the GUT scale. However, due to the stabilization of
$\tilde\nu_H^c, \tilde\nu_{\bar{H}}^c$ at their origins during
inflation, the Universe enters the reheating epoch in an $\text{SU} (5)
\times \text{U}(1)$ symmetric state. The eventual breaking of the symmetry takes place along an $F$- and $D$-flat direction of  the potential $\tilde{\nu}_H^c=\tilde{\nu}_{\bar{H}}^c\equiv \Phi$ and finite-temperature corrections to the effective potential must be taken into account.  As we will see,
the strong coupling behaviour of SU(5) at low temperatures help drive the transition \cite{supercosm,Campbell:1987eb}.

The leading-order running of the SU(5) gauge coupling is
\begin{equation}
\frac{1}{\alpha_5 (\mu)}  = \frac{1}{\alpha_{\text{GUT}}}
 -\frac{b_5}{2\pi} \ln\left(\frac{M_{\text{GUT}}}{\mu}\right) ~,
 \label{su5rg}
\end{equation}
where $b_5 = 15-(n_5 +3n_{10})/2$ with $n_{5}$ ($n_{10}$) the number of
${\bf 5}$ and $\overline{\bf 5}$ (${\bf 10}$ and $\overline{\bf 10}$)
multiplets. For the field content introduced in Section~\ref{sec:model},
$n_{5} = 5$ and $n_{10}=5$, and thus $b_5=5$.
As discussed above in (\ref{alphagut}),  $\alpha_{\rm GUT}\simeq 0.0374$ for 
$M_{\rm GUT}\simeq 1.2 \times 10^{16}\,{\rm GeV}$, and equation (\ref{su5rg}) implies that the SU(5) group
is asymptotically free, with the coupling in the unbroken phase becoming strong at a large energy scale 
$\Lambda_c \gg m_W$. Naively, this scale would be associated with the condition $g_5\sim 1$, 
corresponding to $\mu \sim 2 \times 10^8$~GeV, but symmetry-breaking
bilinear condensates may be formed above/below this threshold.  
These condensates acquire masses of order $\Lambda_c$ and effectively decouple from the low-energy theory. 
The nature of the condensates and their transformation properties can in principle be investigated
using a generalization of the so-called most attractive channel (MAC) hypothesis~\cite{Raby:1979my,Georgi:1981mh}. 
Schematically, lattice calculations have indicated that the exchange potential for the fermion bilinear may be 
sufficiently large for condensation to occur if~\cite{Kogut:1983sm}
\beq
g^2(\Lambda_c)\Delta C \equiv g^2(\Lambda_c)(C_c-C_1-C_2)\simeq 4,  \qquad \alpha_c \equiv \alpha(\Lambda_c) \simeq \frac{1}{\pi \Delta C} \, ,
\label{ourMAC}
\eeq
where $C_c$, $C_1$ and $C_2$ denote the  quadratic Casimirs of the composite channel 
and the two elementary supermultiplets, respectively~\footnote{Larger values of $\alpha_c$ are
estimated in the ladder approximation~\cite{sannino}. The differences between the various estimates of $\alpha_c$ 
stem from our incomplete understanding of strong
dynamics.}. More specifically, it is assumed that the channel that 
maximizes the effective coupling (the MAC) is the one that condenses first. Then, a second MAC condensate 
may form in the new broken phase, further reducing the symmetry group of the model. 
Successive MACs continue to condense until the non-Abelian gauge group is completely broken.

The MAC spectrum for the flipped model was studied within this approach in~\cite{Campbell:1987eb}, 
under the additional condition that supersymmetry remains unbroken during the MAC formation. 
We do not repeat that analysis here, but summarize the results.
All in all, the total number of light states in the strong-coupling phase is expected to be at most 
$4({\bf 1}_i) + 1 ({\bf 1}) + 14({\bf 1}_4) + 3({\bf 1}_j) + 3(\ell_i^c) =25$ where (as discussed in detail in~\cite{Campbell:1987eb}),
the ${\bf 1}_i$ are singlet fields arising from $\mathbf{10}_i \times \mathbf{\overline{10}}_{\bar H}$ condensation (where $i = 1, 2, 3, H$),
${\bf 1}$ is a singlet arising from $\mathbf{5}_h \times \mathbf{\bar{5}}_{\bar h}$ condensation,
and the ${\bf 1}_j$ are SU(4) singlets in the $\mathbf{\bar 5}_j$ matter representations.
Among these, 14 get masses $\propto\Phi$ and 11 do not, which is fewer than in \cite{Campbell:1987eb}, as the singlets $\phi_a$ 
are not included among the light states, as we assume here the presence of the bilinear couplings $\mu^{ab}$.
This is to be contrasted with the $\text{SU}(5)\times \text{U}(1)$-symmetric phase, which has 103 light superfield degrees of freedom, 
and the Higgs phase, in which 62 do not acquire masses $\sim \Phi$.

As a representative example of the net Casimir coefficient $\Delta C$ in a specific
MACs, we may take $\Delta C= \frac{24}{5}$ in (\ref{ourMAC}), which
gives $\alpha_c \simeq 0.0663$. 
Then, from Eq.~\eqref{su5rg} we have
\begin{equation}
 \Lambda_c \simeq M_{\text{GUT}} \exp \left[-\frac{2\pi}{b_5}
\left(\frac{1}{\alpha_{\text{GUT}}}- \frac{1}{\alpha_c} 
\right)\right] ~.
\label{eq:lambdacpred}
\end{equation}
Using our previous estimate that $\alpha(M_{\rm GUT}) = 0.0374$, we find that strong-coupling dynamics will be important for
\beq
\Lambda_c \simeq 4\times 10^{-7}\,M_{\rm GUT}\,.
\label{eq:lamcmac}
\eeq 
Using also our estimate $M_{\rm GUT} = 1.2 \times 10^{16}$~GeV, 
we then have $\Lambda_c \simeq 5 \times 10^{9}$~GeV.  Again using (\ref{ourMAC}), other MAC channels give
estimates of $\Lambda_c$ between $10^{8}$ and $10^{14}$ GeV~\footnote{For example,
$\Delta C = 36/5$ gives $\alpha_c =  0.0442$ and $\Lambda_c = 6 \times 10^{-3} M_{\rm GUT}$,
$\Delta C = 18/5$ gives $\alpha_c =  0.0884$ and $\Lambda_c = 4 \times 10^{-9} M_{\rm GUT}$,
and $\Delta C = 15/4$ gives $\alpha_c =  0.0849$ and $\Lambda_c = 7 \times 10^{-9} M_{\rm GUT}$.
See \cite{Campbell:1987eb} for more details.}.
As we will see in Section~\ref{eq:entropyrel}, if $T_R \gtrsim
\Lambda_c \gtrsim \sqrt{|m_\Phi| M_{\text{GUT}}} \sim 10^{10}$ GeV, the oscillation of the
$\Phi$ field occurs incoherently, which minimizes the entropy release
due to $\Phi$ decay. Eq.~\eqref{eq:lamcmac} 
suggests that it is plausible that $\Lambda_c$ falls into this
region, within the large current uncertainties.

\begin{figure}[t!]
\centering
    \scalebox{0.8}{\includegraphics{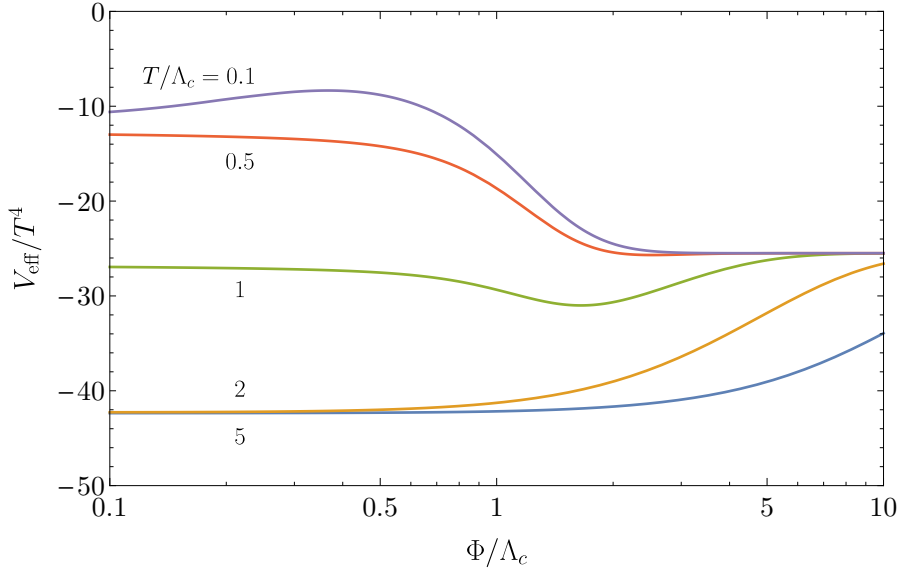}}
    \caption{\em The evolution with temperature of the effective potential $V_{\rm eff}(\Phi,T)$
    (\protect\ref{Veff}) in strongly-coupled $SU(5)\times U(1)$. 
        }
    \label{fig:FinTempV}
\end{figure}

The difference in the number of light degrees of freedom between the symmetric, 
strongly-coupled and Higgs phases of the theory is crucial for the onset of the 
$\text{SU}(5)\times \text{U}(1) \rightarrow  \text{SU}(3) \times
\text{SU}(2) \times \text{U}(1)$ phase transition, as we now show. 
From the one-loop temperature-dependent correction to the effective potential, we
have a contribution from 
light superfields that remain massless 
in the broken phases equivalent to that of an ideal ultrarelativistic
gas, {\it i.e.},
\beq
V_{\rm eff,\,light} = -\frac{\pi^2 T^4}{90}\, g \,,
\eeq
and the $\Phi$-independent heavy states will have negligible contributions. For the states
with $\Phi$-dependent masses, there are contributions to the chiral mass-squared matrices 
proportional to $|\lambda_{4,5,6}|^2\Phi^2$, and to $g^2C_a\Phi^2$ for the vector superfields. 
Under the assumption that $\lambda_{4,5,6},g_aC_a\sim\mathcal{O}(1)$ in the strong-coupling domain,
we may write a phenomenological fit to the temperature-dependent effective potential of the form
\beq
\label{Veff}
V_{\rm eff}(\Phi,T) \approx N_{\Phi}\frac{T^4}{2\pi^2} \sum_{\alpha=0,1} (-1)^{\alpha} \int_0^{\infty}dy\, y^2\,  \ln \left[1-(-1)^{\alpha}\exp\left(-\sqrt{y^2 + (\Phi/T)^2}\right)\right] \,,
\eeq
where $N_{\Phi}$ denotes the number of $\Phi$-dependent massive superfields in the corresponding regime. 
Fig.~\ref{fig:FinTempV} shows the resulting shape of the effective potential as a function of $\Phi$ when $T/\Lambda_c = \mathcal{O}(1)$. 
For definiteness, we have used a smooth (logistic function) interpolation for $g$ and 
$N_{\Phi}$ around the strong-coupling-transition 
scale $\Lambda_c$. In the topmost curve, a barrier that might trap $\Phi$ near the 
origin when $T<\Lambda_c$ is apparent. This effect may be an artifact of the approximations that we have considered, 
and we expect in any case that  strong-coupling and thermal effects would easily make an end run around any such barrier when $T\sim\Lambda_c$.

\subsection{Entropy release}
\label{eq:entropyrel}

Having verified how the phase transition takes place, we now estimate
the amount of entropy it releases. In what follows, we denote the decay
rate of the flat direction by $\Gamma_{\Phi}$, and the scale factor by
$a$. As we will discuss below, the amount of entropy release will be
dependent on whether it is possible or not for the flat direction $\Phi$
to undergo coherent oscillations after the completion of the phase
transition~\cite{Ellis:1988qe}.

One possibility is that reheating takes place at temperatures lower than
the strong-coupling scale, $T_{\rm reh}<\Lambda_c$. In this {\em weak
reheating} scenario, the $\text{SU}(5)\times \text{U}(1)$ gauge symmetry
is not restored after inflation, and the field $\Phi$ eventually reaches
its low-energy minimum and reheats the Universe through the coherent
decay of its oscillations. Disregarding non-renormalizable terms that
could lift the flat direction, field dependence in the zero-temperature
effective potential for $\Phi$ can only come from a
supersymmetry-breaking term $\sim m_{\Phi}^2\Phi^2$, where $m_{\Phi}^2$
is assumed to be negative. The energy stored
in the scalar field oscillations of $\Phi$ following the phase
transition may then be simply estimated as 
\beq\label{eq:rhophit}
\rho_{\Phi} \simeq |m_{\Phi}^2|\langle\Phi\rangle ^2 \left(\frac{a(t)}{a_{\Phi}}\right)^{-3}\,.
\eeq
where $\langle\Phi\rangle$ denotes the low-temperature vev of $\Phi$, responsible for the breaking of the GUT symmetry, and $a_{\Phi}$ denotes the size of the scale factor at the onset of $\Phi$-oscillations.

In contrast, for {\em strong reheating}, $T_{\rm reh} \gtrsim
\Lambda_c$, and the $\Phi$ field starts growing as the temperature falls
below $\Lambda_c$. This growth, however, will be driven by incoherent
fluctuations, in which there is a sizeable kinetic energy for the $\Phi$
field, $\dot{\Phi}\sim T^2$. For $\Lambda_c > (|m_{\Phi}|\Phi)^{1/2}$,
the incoherent component of $\Phi$ will dominate and destroy any
coherent contribution~\footnote{Even in the presence of coherence, the
condition for fast damping of the field oscillations, $|m_{\Phi}|\sim
H\sim \Lambda_c^2/M_P$, is not violated by more than an order of
magnitude, implying complete damping after a few oscillations.}. The
flat direction will then redshift as radiation until its temperature
decreases sufficiently to bring it to the non-relativistic regime,
during which it eventually decays and reheats the Universe. 

In the following subsections we determine the amount of entropy released by the decay of $\Phi$ in the weak and strong reheating scenarios, and determine their effect on the final baryon asymmetry.

\subsubsection{Weak reheating}

We consider first the weak reheating case, for which the GUT gauge
symmetry is not restored after inflation, and the $\Phi$ condensate
oscillates coherently about its low-energy minimum. Dependent on the
magnitudes of the decay rates of the inflaton $s$ and the flat direction
$\Phi$, the later can begin its oscillations and/or decay before or after
the completion of reheating. It is clear that a short-lived $\Phi$,
namely one which decays before the end of reheating, will not
contribute significantly towards the production of entropy, which
continues until the end of reheating. Therefore, to explore the
potential entropy injection produced by $\Phi$ decay, we need to
consider only the case in which $\Gamma_s>\Gamma_{\Phi}$. Let us assume
first that the flat direction starts oscillations during the
radiation-dominated era with $|m_{\Phi}|<\Gamma_s$. As the energy density
of the inflaton decay products at the end of reheating is
$\rho_{\gamma,\,{\rm reh}} \sim \Gamma_s^2M_P^2$~\cite{EGNO5,EGNOP}, we
can write at later times 
\beq\label{rhogamma}
\rho_{\gamma} \sim \Gamma_s^2M_P^2 \left(\frac{a(t)}{a_{\rm reh}}\right)
^{-4}\,. 
\eeq
With $a(t)\sim t^{1/2}$ during radiation domination and the Hubble
parameter given by $H_\gamma \sim \rho_\gamma^{1/2}/M_P$, and assuming
for simplicity the instantaneous onset of oscillations and decay of
$\Phi$ when $H(t_{\Phi}) \sim |m_{\Phi}|$ and $H(t_{d\Phi}) \sim
\Gamma_{\Phi}$, respectively, the ratio 
\beq\label{eq:ratio3a}
\left.\frac{\rho_{\Phi}}{\rho_{\gamma}}\right|_{d\Phi} \sim 
\frac{|m_{\Phi}^2|\langle\Phi\rangle
^2}{\Gamma_s^2M_P^2}\frac{a_{\Phi}^3a_{d\Phi}}{a_{\rm reh}^4} \simeq
\frac{|m_{\Phi}|^{1/2}\langle\Phi\rangle
^2}{\Gamma_{\Phi}^{1/2}M_P^2}\,, 
\eeq
will be smaller than one if the following constraint on the decay rate
of $\Phi$ is satisfied, 
\beq\label{case3a}
\frac{\Gamma_{\Phi}}{|m_{\Phi}|} >
\left(\frac{\langle\Phi\rangle}{M_P}\right)^{4} \gtrsim 2 \times
10^{-11} \,, 
\eeq
for $\langle \Phi \rangle \gtrsim 5 \times 10^{15}$ GeV as required by the
proton lifetime (see Eq.~\eqref{eq:protlim}).
If this occurs, a negligible amount of entropy will be released upon
$\Phi$ decay. When (\ref{case3a}) is not satisfied, the oscillations of
$\Phi$ will eventually dominate the energy density of the Universe. In
this case, from (\ref{eq:rhophit}) and (\ref{rhogamma}), we can compute
the scale factor $a_*$ at $\Phi$-radiation equality as 
\beq
 \frac{a_*}{a_{\Phi}} \simeq \left(\frac{M_P\Gamma_s}{|m_{\Phi}|
 \langle\Phi\rangle}\right)^2\left(\frac{a_{\rm reh}}{a_{\Phi}}\right)^4
 \simeq \left(\frac{M_P}{\langle\Phi\rangle}\right)^2\,. 
\eeq
With the Hubble parameter during $\Phi$ domination given by
\beq
H_{\Phi} \sim \frac{|m_{\Phi}|\langle\Phi\rangle}{
M_P}\left(\frac{a_{\Phi}}{a}\right)^{3/2}\,,  
\eeq
and the decay of the flat direction occurring at $\Gamma_{\Phi}\sim
H_{\Phi}$, we can re-calculate the ratio (\ref{eq:ratio3a}) for the flat
direction-dominated case as\footnote{
Here we have assumed that $\rho_\gamma$ goes as $a^{-4}$ during the
whole epoch. Generically speaking, during the coherent oscillation of
$\Phi$, the radiation originating from the decay of $\Phi$ might have
dominated that from inflaton decay. In this case, the Universe would
experience non-adiabatic expansion and $\rho_\gamma$ goes as
$a^{-3/2}$ \cite{Scherrer:1984fd}. However, it turns out that the
non-adiabatic expansion occurs only at a very low temperature because of
the small decay rate of $\Phi$ (see below), and thus this modification
would be negligible in our scenario. 
} 
\beq
\left.\frac{\rho_{\Phi}}{\rho_{\gamma}}\right|_{d\Phi} =
\left(\frac{a_{d\Phi}}{a_*}\right) =
\left(\frac{a_{d\Phi}}{a_{\Phi}}\right)
\left(\frac{a_{\Phi}}{a_*}\right) \sim
\left(\frac{|m_{\Phi}|\langle\Phi\rangle^4}{\Gamma_{\Phi}M_P^4}\right)^{2/3}
\,. 
\eeq
which is $>1$, consistent with the violation of the condition (\ref{case3a}).

Let us now assume that the oscillation of $\Phi$ starts during
reheating, $|m_{\Phi}|>\Gamma_s$. In this case we can interpolate between
the start of oscillations of the flat direction and the end of reheating
as 
\beq
\left(\frac{a_{\rm reh}}{a_{\Phi}}\right) \sim
\left(\frac{|m_{\Phi}|}{\Gamma_s}\right)^{2/3}\,. 
\eeq
If the Universe is dominated by radiation by the time of the decay of
$\Phi$, we will have 
\beq
\left.\frac{\rho_{\Phi}}{\rho_{\gamma}}\right|_{d\Phi} \sim
\frac{|m_{\Phi}^2| \langle\Phi\rangle
^2}{\Gamma_s^2M_P^2}\left(\frac{a_{\Phi}}{a_{\rm
reh}}\right)^3\left(\frac{a_{d\Phi}}{a_{\rm reh}}\right) \simeq
\frac{\Gamma_s^{1/2}\langle\Phi\rangle ^2}{\Gamma_{\Phi}^{1/2}M_P^2}\,. 
\eeq
Thus, no significant amount of entropy will be released  if the following constraint on the decay rate of $\Phi$ is satisfied,
\beq
\frac{\Gamma_{\Phi}}{\Gamma_s} > \left(\frac{\langle\Phi\rangle}{M_P}\right)^{4} \ga 2 \times 10^{-11} \,.
\eeq
Conversely, if the previous relation is violated, the flat direction oscillations will dominate the energy density until their decay. In this scenario, we can write 
\beq
\left.\frac{\rho_{\Phi}}{\rho_{\gamma}}\right|_{d\Phi} = \left(\frac{a_{d\Phi}}{a_*}\right) =  \left(\frac{a_{d\Phi}}{a_{\Phi}}\right) \left(\frac{a_{\Phi}}{a_*}\right) \sim \left(\frac{\Gamma_s \langle\Phi\rangle^4}{\Gamma_{\Phi}M_P^4}\right)^{2/3}>1  \,.
\eeq
With the entropy density in radiation given by $s_{\gamma}=4/3(g_{\rm reh}\pi^2/30)^{1/4}\rho_{\gamma}^{3/4}$, and a similar expression for the entropy density produced from $\Phi$ decays, we can summarize the amount of entropy released in both scenarios as
\beq\label{sphisg}
\Delta\;\equiv\;\left.\frac{s_{\Phi}}{s_{\gamma}}\right|_{d\Phi} \sim  \left(\frac{g_{d\Phi}}{g_{\rm reh}}\right)^{1/4}\left(\frac{\min[|m_{\Phi}|,\Gamma_s]\langle\Phi\rangle^4}{\Gamma_{\Phi}M_P^4}\right)^{1/2} \,.  
\eeq

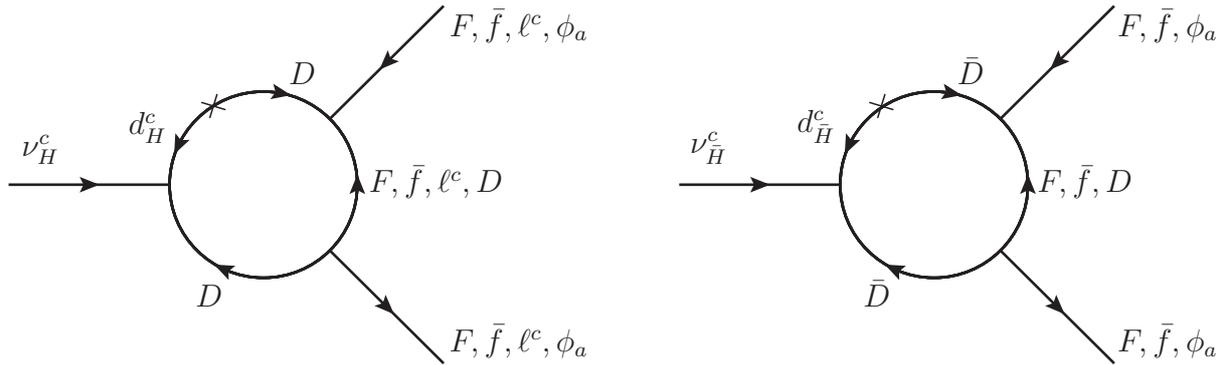
\begin{figure}[t] 
   \centering
   \fcolorbox{white}{white}{
   \begin{picture}(455,140) (0,-20)
    \SetWidth{1.0}
    \SetColor{Black}
    \Arc[arrow,arrowpos=1,arrowlength=5,arrowwidth=2,arrowinset=0.2](95,50)(35,0,360)
    \Arc[arrow,arrowpos=0.68,arrowlength=5,arrowwidth=2,arrowinset=0.2,flip](95,50)(35,0,360)
    \Arc[arrow,arrowpos=0.43,arrowlength=5,arrowwidth=2,arrowinset=0.2](95,50)(35,0,360)
    \Arc[arrow,arrowpos=0.22,arrowlength=5,arrowwidth=2,arrowinset=0.2,flip](95,50)(35,0,360)
    \Line[arrow,arrowpos=0.5,arrowlength=5,arrowwidth=2,arrowinset=0.2](0,50)(60,50)
    \Line[arrow,arrowpos=0.5,arrowlength=5,arrowwidth=2,arrowinset=0.2](162.175,117.175)(119.749,74.749)
    \Line[arrow,arrowpos=0.5,arrowlength=5,arrowwidth=2,arrowinset=0.2](119.749,25.251)(162.175,-17.175)
    \Text(5,58)[lb]{\Black{$\nu^c_H$}}
    \Text(165,104)[lb]{\Black{$F,\bar{f},\ell^c,\phi_a$}}
    \Text(165,-15)[lb]{\Black{$F,\bar{f},\ell^c,\phi_a$}}
    \Text(135,45)[lb]{\Black{$F,\bar{f},\ell^c,D$}}
    \Text(70,4)[lb]{\Black{$D$}}
    \Text(105,88)[lb]{\Black{$D$}}
    \Text(45,66)[lb]{\Black{$d^c_H$}}
    \Text(74,73)[lb]{\large{\Black{$\rput[lb]{28}{\times}$}}}
    \Line[arrow,arrowpos=0.5,arrowlength=5,arrowwidth=2,arrowinset=0.2](250,50)(310,50)
    \Line[arrow,arrowpos=0.5,arrowlength=5,arrowwidth=2,arrowinset=0.2](412.175,117.175)(369.749,74.749)
    \Line[arrow,arrowpos=0.5,arrowlength=5,arrowwidth=2,arrowinset=0.2](369.749,25.251)(412.175,-17.175)
    \Arc[arrow,arrowpos=1,arrowlength=5,arrowwidth=2,arrowinset=0.2](345,50)(35,0,360)
    \Arc[arrow,arrowpos=0.68,arrowlength=5,arrowwidth=2,arrowinset=0.2,flip](345,50)(35,0,360)
    \Arc[arrow,arrowpos=0.43,arrowlength=5,arrowwidth=2,arrowinset=0.2](345,50)(35,0,360)
    \Arc[arrow,arrowpos=0.22,arrowlength=5,arrowwidth=2,arrowinset=0.2,flip](345,50)(35,0,360)
    \Text(255,58)[lb]{\Black{$\nu^c_{\bar{H}}$}}
    \Text(415,104)[lb]{\Black{$F,\bar{f},\phi_a$}}
    \Text(415,-15)[lb]{\Black{$F,\bar{f},\phi_a$}}
    \Text(385,45)[lb]{\Black{$F,\bar{f},D$}}
    \Text(320,4)[lb]{\Black{$\bar{D}$}}
    \Text(355,88)[lb]{\Black{$\bar{D}$}}
    \Text(295,66)[lb]{\Black{$d^c_{\bar{H}}$}}
    \Text(324,73)[lb]{\large{\Black{$\rput[lb]{28}{\times}$}}}
  \end{picture}
}
\caption{\it Diagrams contributing to a $D$-term generating the decay
 rate (\ref{eq:phidecr}) for the $\Phi$ field in the flipped
 $SU(5)\times U(1)$ GUT model.
}
   \label{fig:phidec}
\end{figure}

As was discussed in~\cite{Campbell:1987eb}, the decay rate of the flat direction occurs 
via the effective $D$-term diagrams shown in Fig.~\ref{fig:phidec}, which lead to the decay rate
\beq\label{eq:phidecr}
\Gamma_{\Phi} \simeq
\frac{9\lambda_{1,2,3,7}^4}{2048\pi^5}\left(\frac{|m_{\Phi}| m^2_{F,\bar{f},\ell^c,\tilde{\phi}_a}}{\langle\Phi\rangle^2}\right)\,.
\eeq
The gravitino production constraint implies that $\Gamma_s\lesssim 900\,{\rm GeV}$ (see (\ref{700})) and, if we assume that the effective mass
in the flat direction is heavier than the weak scale, the entropy ratio (\ref{sphisg}) evaluates to
\begin{align}
\Delta & \simeq  1.5 \times 10^8\, \lambda_{1,2,3,7}^{-2} \left(\frac{g_{d\Phi}}{g_{\rm reh}}\right)^{1/4}\left(\frac{\langle\Phi\rangle}{5\times10^{15}\,{\rm GeV}}\right)^3\bigg(\frac{(10\,{\rm TeV})^3}{m_{F,\bar{f},\ell^c,\tilde{\phi}_a}^2 |m_{\Phi}|}\bigg)^{1/2} \left(\frac{\Gamma_s}{900\,{\rm GeV}}\right)^{1/2}  \simeq  \nonumber \\
&
\hspace{-0.2cm}
6.2 \times 10^{7}\, \lambda_{1,2,3,7}^{-2} \left(\frac{g_{d\Phi}}{g_{\rm reh}}\right)^{1/4} 
\hspace{-0.2cm}
\left(\frac{\langle\Phi\rangle}{5\times10^{15}\,{\rm GeV}}\right)^3
\hspace{-0.2cm}
\bigg(\frac{(10\,{\rm TeV})^3}{m_{F,\bar{f},\ell^c,\tilde{\phi}_a}^2 |m_{\Phi}|}\bigg)^{1/2} 
\hspace{-0.2cm}
\left(\frac{y}{10^{-5}}\right)  
\hspace{-0.2cm}
\left(\frac{m}{3 \times 10^{13}\,{\rm GeV}}\right)^{1/2}  \,. \label{entropyfactor}
\end{align}
A large release of entropy could be problematic if it overly dilutes the
baryon asymmetry. But we should bear in mind that 1) the estimate in
(\ref{entropyfactor}) is proportional to $y$ and so may be significantly
smaller if $y$ is small; 2) the matter-antimatter asymmetry may be quite
large initially, {\it e.g.}, if generated via the Affleck--Dine
mechanism~\cite{Affleck:1984fy}, in which case some dilution could be
acceptable or even welcome. Such dilution can also relax the
gravitino over-production problem as we see in Section~\ref{sec:entropygrav}.

\subsubsection{Strong reheating}
\label{sec:strongreh}

When the temperature of the relativistic plasma following reheating is $T_{\rm max}>T_{\rm reh}\sim \Lambda_c$, the energy density of the flat direction $\Phi$ will be dominated by incoherent fluctuations of energy $\mathcal{O}(\Lambda_c)$ as the phase transition to the broken symmetry phase takes place. As was discussed earlier, these fluctuations will erase the coherent component of $\Phi$. Eventually, the interactions that lifted the energy of $\Phi$ to the plasma temperature will cease to maintain it in equilibrium, and the flat direction will decouple at $T_{\rm dec}\lesssim \Lambda_c$, maintaining a profile that evolves as $T_{\Phi}=T_{\rm dec}(a_{\rm dec}/a)=T ( g(T)/g_{\rm dec} )^{1/3}$, where $T$ is the temperature of the radiation background. At a later time, this temperature will fall below $|m_{\Phi}|$, $\Phi$ will become non-relativistic, and the Universe will be matter-dominated until the decay of $\Phi$ at $T_{ d\Phi}$. Equating the Hubble rate $H=\sqrt{\rho_{\Phi}/3M_P^2}$, with $\rho_{\Phi}=\zeta(3) |m_{\Phi}|T_{\Phi}^3/\pi^2$, to the decay rate (\ref{eq:phidecr}) one obtains
\beq
T_{ d\Phi} \simeq 2\times 10^{-3} \lambda_{1,2,3,7}^{8/3}
\left(\frac{|m_{\Phi}|m^4_{F,\bar{f},\ell^c,\tilde{\phi}_a}M_P^2}{\langle\Phi\rangle^4}\right)^{1/3}\,. 
\eeq
Upon the decay of $\Phi$, the Universe will be once again dominated by radiation, with a temperature now given by
\beq
T_{\rm reh}' = \left(\frac{40}{g_{d\Phi} \pi^2}
\right)^{1/4}(\Gamma_{\Phi}M_P)^{1/2}  \simeq 5\times 10^{-3} g_{
d\Phi}^{-1/4} \lambda_{1,2,3,7}^{2} \left(\frac{|m_{\Phi}| m^2_{F,\bar{f},\ell^c,\tilde{\phi}_a}M_P }{\langle\Phi\rangle^2}\right)^{1/2}\,,
\eeq
where we have neglected the delay arising from the conversion of the heavy supersymmetric decay products into the truly relativistic Standard Model particles. For $\langle \Phi\rangle \simeq 5\times 10^{15}\,{\rm GeV}$, $\lambda_{1,2,3,7}\sim 1$ and $m_{\Phi,F,\bar{f},\ell^c,\tilde{\phi}_a}\gtrsim 10\,{\rm TeV}$, this temperature is $T_{\rm reh}'\gtrsim 1\,{\rm MeV}$, remarkably around what is needed to re-start nucleosynthesis. The amount of entropy released by the decay of $\Phi$ can therefore be estimated as
\begin{align}
\label{strongdelta}
\Delta &\;=\; \frac{g_{d\Phi}T_{\rm reh}^{'3}}{\left.(g(T)T^3)\right|_{d\Phi}} \;=\; \frac{g_{d\Phi}T_{\rm reh}^{'3} }{g_{\rm dec} T_{d\Phi}^3}\\ \label{entropyfactor2}
&\;\simeq\; 8\times 10^3\, \lambda_{1,2,3,7}^{-2} \left(\frac{g_{d\Phi}}{43/4}\right)^{1/4}\left(\frac{915/4}{g_{\rm dec}}\right) \left(\frac{\langle \Phi\rangle}{5\times 10^{15}\,{\rm GeV}}\right) \bigg(\frac{ 10\,{\rm TeV} }{m_{F,\bar{f},\ell^c,\tilde{\phi}_a}^2/|m_{\Phi}| }\bigg)^{1/2}\,.
\end{align}
Therefore, in the strong reheating scenario the entropy dilution is reduced by $\mathcal{O}(10^4)$ with respect to weak reheating,
providing more leeway for the initial asymmetry-generating mechanism.

\subsubsection{Entropy production and the gravitino bound on reheating}
\label{sec:entropygrav}

A late injection of entropy would dilute any previously produced relics, such as gravitinos or their decay products. In particular,
the gravitino yield $Y_{3/2}\equiv n_{3/2}/n_{\gamma}$ would be reduced by a factor of $\Delta^{-1}$. 
Since in the absence of a secondary matter-dominated era driven by $\Phi$, the yield at late times is given approximately by~\cite{EGNOP}
\beq
Y_{3/2}(T) \simeq 0.0036\left(1+0.56\frac{m_{1/2}^2}{m_{3/2}^2}\right)\left(\frac{\Gamma_s}{M_P}\right)^{1/2}\,,
\eeq
the extra dilution would weaken the decay rate constraint imposed by the LSP relic density by a factor of $\Delta^2$, thus allowing a higher reheat temperature by a factor of $\Delta$. 
More specifically, (\ref{700}) would now become
\beq\label{newymax}
|y| < 2.7\times 10^{-5} \Delta \left( 1 + 0.56 \frac{m_{1/2}^2}{m_{3/2}^2} \right)^{-1}\left(\frac{100\,{\rm GeV}}{m_{\rm LSP}}\right)\, ,
\eeq
and $T_{\rm reh}\lesssim \Delta\, (10^{10}\,{\rm GeV})$. Thus, the strong reheating condition $T_{\rm reh}\gtrsim \Lambda_c$
would automatically be allowed by the late decay of the flaton $\Phi$. 

An additional effect of the entropy increase (\ref{entropyfactor}) or (\ref{entropyfactor2}) would be a shift in the number of $e$-folds of 
inflation after the pivot scale $k_*$ crosses the horizon, due to the non-standard thermal history~\cite{Easther:2013nga,Adshead:2010mc}. 
In the slow-roll approximation,
the number of $e$-folds to the end of inflation can be expressed as~\cite{LiddleLeach,MRcmb,planck15}
\beq
N_* = \ln\left(\frac{\rho_{\rm reh}^{1/4}a_{\rm reh}}{\sqrt{3}a_0H_0}\right) - \ln\left(\frac{k_*}{a_0H_0}\right) + \frac{1}{4}\ln\left(\frac{V_*^2}{M_P^4\rho_{\rm end}}\right) + \frac{1-3w_{\rm int}}{12(1+w_{\rm int})}\ln\left(\frac{\rho_{\rm reh}}{\rho_{\rm end}}\right) \ ,
\label{howmany}
\eeq
where $a_0$ and $H_0$ are the present cosmological scale factor and Hubble expansion rate, respectively, 
$V_*$ is the inflationary energy density at the reference scale, 
$\rho_{\rm end}$ and $\rho_{\rm reh}$ are the energy densities at the end of inflation and after reheating, and 
$w_{\rm int}$ is the $e$-fold average of the equation-of-state parameter during the thermalization epoch. For 
the standard thermal history (STH), the first term evaluates to $66.9- \frac{1}{12}\ln g_{\rm reh}$, as entropy is assumed to be conserved after reheating. 
In our case of an intermediate matter-dominated era, it can be rewritten as
\beq\label{Nstar1}
\ln\left(\frac{\rho_{\rm reh}^{1/4}a_{\rm reh}}{\sqrt{3}a_0H_0}\right) = \ln\left(\frac{\rho_{d\Phi}^{1/4}a_{d\Phi}}{\sqrt{3}a_0H_0}\right) + \ln\left(\frac{\rho_{\rm reh}^{1/4}a_{\rm reh}}{\rho_{d\Phi}^{1/4}a_{d\Phi}}\right)\,,
\eeq
where now the first term of (\ref{Nstar1}) can be evaluated assuming entropy conservation, and the second term is directly related to the dilution factor $\Delta$. 
We obtain for the number of $e$-folds the following expression,
\beq\label{Nstar2}
N_* = N_*^{\rm STH} - \frac{1}{3}\ln\Delta\,.
\eeq
At first sight, the reduction of $N_*$ due to the late entropy injection might appear to put Starobinsky-like inflation under stress, 
as the Planck data disfavors $N_*\lesssim 50\,(44)$ at the 68\% (95\%) CL. However, the physical range for $N_*$ depends on
the underlying particle model, since the last term in (\ref{howmany}) depends implicitly on the decay rate of the inflaton. 
For Starobinsky-like inflation, it can be evaluated in the perturbative regime as~\cite{EGNO5}
\beq
\frac{1-3w_{\rm int}}{12(1+w_{\rm int})}\ln\left(\frac{\rho_{\rm reh}}{\rho_{\rm end}}\right) \;\simeq\; \frac{1-3w_{\rm int}}{12(1+w_{\rm int})}\left(2\ln\left(\frac{\Gamma_s}{m}\right) + {\rm const.}\right)\,,
\eeq
where $w_{\rm int}\simeq 0.782/\ln(2m/\Gamma_s)$. In the STH case,
the gravitino upper bound for $\Gamma_s$ constrains $N_*$ to be less than $N_*^{\rm max}\simeq 53.3$. 
However, in the case of intermediate $\Phi$-domination, this maximum value is reduced due to the modified $e$-fold expression (\ref{Nstar2}), 
and also increased because of the weakened gravitino bound (\ref{newymax}). These two effects combine to give
\beq
\Delta N_*^{\rm max} \simeq -4\times 10^{-3} \ln \Delta \,.
\eeq
Therefore, the favored range for $N_*$ is practically unchanged for strong reheating when one accounts for the increased decay rate limit. In the weak reheating regime, the maximum value for $N_*$ cannot be reached within perturbation theory, as for $|y|\lesssim 1$ the number of $e$-folds is limited to $N_*\lesssim 49$ with $\Delta \simeq 5\times 10^8$, the value of the entropy factor obtained by taking $\min[|m_{\Phi}|,\Gamma_s]=|m_{\Phi}|$ in (\ref{sphisg}). Hence, for weak reheating, the inflationary predictions lie outside the 1$\sigma$ Planck bounds.

\subsubsection{Baryon asymmetry}

Finally, let us investigate the generation of the baryon asymmetry in this class of models.
In Scenario (A), as we noted earlier, at the tree level the inflaton decays primarily to
Higgs bosons and Higgsinos, and there is no decay to neutrinos. 
At one-loop, through the exchange of a heavy right-handed neutrino,
there is the possibility of a lepton-number-violating decay to two neutrinos.
However, in that case, in order to obtain a net lepton asymmetry one must consider the interference
between one-loop diagrams and their two-loop corrections, greatly suppressing the final lepton
asymmetry. 

However, in the case of strong reheating in Scenario (A), it may be possible to 
produce thermally the right-handed neutrinos, though this is possible only if 
the reheating temperature is comparable to the right-handed neutrino mass.
We recall however, that in Scenario (A) the right-handed neutrino mass
is of order $\lambda_6^{ij} \langle \tilde{\nu}_{\bar{H}}^c\rangle$ (for
$\tilde{\nu}_{\bar{H}}^c \rangle \sim \mu^{ij}$) and, from
(\ref{TrehC}), the reheating temperature is $T_R \lesssim 8.7 \times
10^{14}\, |\lambda_7^0|~ {\rm GeV} \lesssim 8.7 \times 10^9\, \Delta
~{\rm GeV}$. Thus we would require 
\beq
\lambda_6^{ij} \lesssim 1.7 \times 10^{-6} \left(\frac{5 \times 10^{15} {\rm GeV}}{\langle \tilde{\nu}_{\bar{H}}^c\rangle} \right) \Delta\,.
\eeq
Taking into account the entropy production factor in (\ref{strongdelta}), this is a viable path towards producing the
final baryon asymmetry.

Generating the baryon asymmetry is more straightforward in Scenario (B), in which the inflaton decays to $Lh$ in much the same way that 
right-handed neutrinos decay in standard out-of-equilibrium leptogenesis models \cite{fy}. 
At the end of inflation, the  lepton and baryon
asymmetries can be related directly to the reheat temperature, $T_R$, by
\cite{dl,nos,cdo,eln,noz2} 
\beq
\frac{n_B}{s}  \sim \frac{n_L}{s}  \sim  \frac{\epsilon}{\Delta} f \frac{n_s}{T_R^3}
\sim \frac{\epsilon}{\Delta} f \frac{T_R}{m} \, ,
\eeq
where $n_s$ is the number density of inflatons at the time of their decay, and
$f$ is the branching fraction into $Lh$, ($f\sim 1$ for Scenario (B)). 
The amount of C and CP is given by \cite{Ellis:1993ks, Covi:1996wh} 
\begin{equation}
 \epsilon \simeq -\frac{3}{4\pi}\frac{1}
{\left(U_{\nu^c}^\dagger (\lambda_2^D)^2 U_{\nu^c}\right)_{11}}
 \sum_{i=2,3} {\rm Im} \left[\left(U_{\nu^c}^\dagger (\lambda_2^D)^2 U_{\nu^c}\right)^2_{i1}\right] 
\frac{m}{M_i} ~,
\end{equation} 
where $U_{\nu^c}$ is a mixing matrix associated with the diagonalization of
$\tilde{\nu}_i^c$ and $S$ in the basis where $\lambda_2$ is
diagonalized to $\lambda_2^D$ \cite{Ellis:1993ks}, and 
we assume for simplicity that the second and third generation heavy neutrino
states (two states for each generation) have similar masses $M_2$ and
$M_3$, and $m \ll M_i$. Generically, we expect $(U_{\nu^c}^\dagger (\lambda_2^D)^2 U_{\nu^c})_{11} \sim (\lambda_2^D)_{33}^2 \sim m_t/\langle
\bar{h}^0\rangle \sim 1$ is the largest entry in
$(\lambda_2^D)^2$. This gives
\begin{equation}
 |\epsilon| \sim 7\times 10^{-3} \times \left(
\frac{m}{3 \times 10^{13}~\text{GeV}}
\right)
\left(
\frac{M_i}{10^{15}~\text{GeV}}
\right)^{-1} \times \delta ~.
\label{eps}
\end{equation} 
where $\delta$ denotes an ${\cal O}(1)$ factor that depends on the
amount of CP violation in the matrix $U_{\nu^c}^\dagger
(\lambda_2^D)^2 U_{\nu^c}$. 
From Eq.~\eqref{TrehC}
\begin{align}\notag
T_R \;&\approx\; 0.07  |y| \left(\frac{915/4}{g_{\rm reh}}\right)^{1/4}  (m M_P)^{1/2}\\
\;&\approx\; 6 \times 10^{14} |y| \left(\frac{915/4}{g_{\rm reh}}\right)^{1/4} \left(\frac{m}{3 \times 10^{13} {\rm GeV}}\right)^{1/2} ,
\end{align}
we have
\beq
\frac{n_B}{s}  \sim 20  \frac{\epsilon}{\Delta} f |y| \left(\frac{915/4}{g_{\rm reh}}\right)^{1/4} \left(\frac{3 \times 10^{13} {\rm GeV}}{m}\right)^{1/2}\,.
\eeq

In the weak reheating scenario, the entropy dilution factor is given by (\ref{entropyfactor}), and leads to
\beq
\frac{n_B}{s} \sim 7.3 \times 10^{-12} \epsilon f  \lambda_{1,2,3,7}^{2} \left(\frac{43/4}{g_{d\Phi}}\right)^{1/4} \left(\frac{5\times10^{15}\,{\rm GeV}}{\langle\Phi\rangle}\right)^{3} \bigg(\frac{m_{F,\bar{f},\ell^c,\tilde{\phi}_a}^2 |m_{\Phi}|}{(10\,{\rm TeV})^3}\bigg)^{1/2} \left(\frac{3 \times 10^{13} {\rm GeV}}{m}\right)\,,
\eeq
from which it is clear that the enormous amount of dilution will lead to an insufficient asymmetry. 
Note that this estimate is now independent of the inflaton coupling $y$. 

However, in the strong reheating regime, (\ref{entropyfactor2}) implies that
\begin{align} \notag
\frac{n_B}{s} &\simeq  2.7\times 10^{-8} \,\epsilon f \lambda_{1,2,3,7}^{2} \left(\frac{43/4}{g_{d\Phi}}\right)^{1/4}\left(\frac{915/4}{g_{\rm reh}}\right)^{1/4}\left(\frac{g_{\rm dec}}{915/4}\right) \left(\frac{y}{10^{-5}}\right)\\ \label{nbssr}
&\qquad \qquad \times  \left(\frac{5\times 10^{15}\,{\rm GeV}}{\langle \Phi\rangle}\right) \bigg(\frac{m_{F,\bar{f},\ell^c,\tilde{\phi}_a}^2/|m_{\Phi}| }{ 10\,{\rm TeV} }\bigg)^{1/2} \left(\frac{m}{3 \times 10^{13} {\rm GeV}}\right)^{-1/2}\,.
\end{align}
Substituting the expression (\ref{eps}) for $\epsilon$ and taking $M_{2,3} \sim \lambda_6^2 \langle
\Phi \rangle^2/\mu \sim \lambda_6^2 \langle \Phi \rangle$ from (\ref{eq:massheav}) with $\langle 
\Phi \rangle \sim \mu$, 
we have
\begin{align} \notag
\frac{n_B}{s} &\simeq 3.8\times 10^{-11} \,\delta f \lambda_{1,2,3,7}^{2} \lambda_6^{-2} \left(\frac{43/4}{g_{d\Phi}}\right)^{1/4}\left(\frac{915/4}{g_{\rm reh}}\right)^{1/4}\left(\frac{g_{\rm dec}}{915/4}\right) \left(\frac{y}{10^{-5}}\right)\\ 
&\qquad \qquad \times  \left(\frac{5\times 10^{15}\,{\rm GeV}}{\langle \Phi\rangle}\right)^2 \bigg(\frac{m_{F,\bar{f},\ell^c,\tilde{\phi}_a}^2/|m_{\Phi}| }{ 10\,{\rm TeV} }\bigg)^{1/2} \left(\frac{m}{3 \times 10^{13} {\rm GeV}}\right)^{1/2}\,.
\end{align}
Thus, if the product $\delta f  \lambda_{1,2,3,7}^{2}/\lambda_6^2 $ is of order $2.2$,
we obtain the correct baryon asymmetry. Moreover, this estimate for the asymmetry increases significantly if the weakened gravitino production bound (\ref{newymax}) is saturated, in which case we would have $n_B/s \simeq  8.4\times 10^{-7} \,\delta f \lambda_{1,2,3,7}^{2}/\lambda_6^2$.

Finally, we note that previously we had argued against the strong reheating case,
on the basis of a potential overdensity of flatinos, $\tilde{\Phi}$ \cite{ehkno}. However,
there it was presumed that the only sources for the flatino mass were radiative\cite{dt},
making it likely that the flatino was a long-lived LSP.  However, as we have argued earlier,
the flaton and flatino may receive significant mass contributions 
from either a GM term or a higher-dimensional superpotential term
needed to lift the flat direction (or both). In such a case, we would not expect the flatino
to be the LSP. The thermal LSP density may be less than the observed cold dark matter
if there is significant entropy production, but the correct non-thermal density 
from gravitino decays could be obtained when the reheating bound is saturated.

\section{Summary and Discussion}
\label{sec:summary}

We have discussed in this paper the scope for constructing models of
cosmological inflation based on a flipped $\text{SU}(5)\times
\text{U}(1)$ GUT model within the framework of no-scale
supergravity. These two model ingredients are each attractive in their
own rights, since flipped $\text{SU}(5) \times \text{U}(1)$ avoids the
problem of proton stability that plagues many GUT models by
incorporating a minimal and elegant missing-partner mechanism, and
no-scale supergravity avoids the cosmological issues of generic
supergravity models by ensuring a positive semi-definite effective
potential with asymptotically-flat directions that are suitable for
accommodating Starobinsky-like inflation. Moreover, both flipped
$\text{SU}(5) \times \text{U}(1)$ and no-scale supergravity emerge
naturally in models of string compactification.

Within this no-scale flipped $\text{SU}(5)\times \text{U}(1)$ framework,
we have focused on realizations of inflation via a superpotential
resembling (\ref{WZW}), which can yield predictions for the CMB
observables $(n_s, r)$ that resemble those of the Starobinsky model. The
minimal flipped $\text{SU}(5) \times \text{U}(1)$ model contains 4
singlet fields that mix, in general, and we have studied the
circumstances under which one of these could be the inflaton
field. Generically, one may consider a scenario in which the inflaton
eigenstate is hierarchically lighter than the other singlet mass
eigenstates, or a scenario in which there is no such mass hierarchy. In
both scenarios, we have studied the constraints on the couplings of the
model for it to lead to Starobinsky-like predictions. Typically, we find
predictions for the tensor-to-scalar ratio $r$ that are within a factor
${\cal O}(2)$ of the Starobinsky prediction, but the predictions for
$n_s$ are much more sensitive to the model parameters, as seen in
Fig.~\ref{fig:nsrcurve}, and measurements of $n_s$ provide the tightest
CMB constraints on them, as seen in Fig.~\ref{fig:Planck1}.

It is important, when evaluating the no-scale flipped
$\text{SU}(5)\times \text{U}(1)$ predictions, to take into
account the cosmological evolutions of all the Standard Model singlet
fields, including the $\tilde{\nu}^c$ components of the {\bf 10}
representations $F$ of matter fields, and their analogues in the ${\bf
10} + \mathbf{\overline{10}}$ Higgs representations, as seen in
Figs.~\ref{fig:sol1_1} to \ref{fig:sol2_2}. These effects were taken
into account numerically in deriving the model parameter constraints
shown in Fig.~\ref{fig:Planck1}. As we have discussed, the upper limits
on model parameters could be respected naturally by postulating a
symmetry argument for `segregation' between the inflaton and the other
singlet fields.

Neutrino masses and mixing provide another relevant set of constraints
on the no-scale flipped $\text{SU}(5) \times \text{U}(1)$ framework,
which depend whether the inflaton decouples from the neutrino sector. If
it does, the inflationary implications for the neutrino mass matrix are
not important, whereas if the inflaton does not decouple there are
interesting model indications in favour of a Normal Hierarchy of the
light neutrino masses, with predictions for the masses of the
eigenstates, see (\ref{numasses}).

We have also addressed the evolution of the Universe after inflation,
discussing the post-inflationary reheating, which imposes a constraint
on the inflaton decay coupling via the upper limit on the density of
supersymmetric relic particles produced by the decays of gravitinos. We
have also discussed the GUT phase transition, building upon a previous
MAC analysis of the breaking pattern of the flipped $\text{SU}(5) \times
\text{U}(1)$ gauge group. A final set of issues that we have studied in
this paper was the amount of entropy release and the baryon
asymmetry. This would have diluted the baryon asymmetry, and may be
substantial in weak reheating scenarios, see (\ref{entropyfactor}),
providing potentially an important constraint on the couplings
responsible for the decays of the singlet fields in our
model~\footnote{On the other hand, the cosmological baryon asymmetry
might have been generated in a different way, {\it e.g.}, via the Affleck--Dine
mechanism.}. The entropy release would be considerably smaller in strong
reheating scenarios, leading to a smaller dilution of the initial
matter-antimatter asymmetry and facilitating the possibility that it was
generated in the decays of heavy neutrinos~\cite{fy}.

The overall conclusion of our work is that the ambitious no-scale
flipped $\text {SU}(5) \times \text{U}(1)$ framework is capable of
satisfying the many different types of constraints ranging from CMB
measurements to neutrino masses, the dark matter density and the
generation of the cosmological baryon asymmetry. The particular line we
have followed is based on superpotentials resembling (\ref{WZW}), and it
should be emphasized that this is not the only option for obtaining
successful Starobinsky-like predictions for the CMB
observables. Nevertheless, the consistency of our framework with the
available constraints, coupled with the facts that both no-scale
supergravity and flipped $\text{SU}(5) \times \text{U}(1)$ emerge
naturally in models of string compactification, suggests that it may
provide a good avenue for linking a wide range of particle and
cosmological phenomenology to an underlying string model.

\section*{Acknowledgements}

The work of J.E. was supported in part by the UK STFC via the research
grant ST/J002798/1. The work of D.V.N. was supported in part by the DOE
grant DE-FG02-13ER42020 and in part by the Alexander~S.~Onassis Public
Benefit Foundation. The work of K.A.O. was supported in part by
DOE grant DE-SC0011842 at the University of Minnesota. The work of
N.N. was supported by the Grant-in-Aid for Scientific Research
(No.17K14270).

\end{document}